\documentclass[longauth]{aa} 

\usepackage{graphicx}
\usepackage{txfonts}
\usepackage[colorlinks=true,citecolor=cyan]{hyperref}
\begin{document}

   \title{Density distributions, magnetic field structures and fragmentation in high-mass star formation}


 \author{H.~Beuther
        \inst{1}
        \and
        C.~Gieser
        \inst{2}
        \and
        J.D.~Soler
        \inst{3}
        \and
        Q.~Zhang
        \inst{4}
        \and
        R.~Rao
        \inst{4,5}
        \and
        D.~Semenov
        \inst{1}
        \and
        Th.~Henning
        \inst{1}
        \and
        R.~Pudritz
        \inst{6,1}
        \and
        T.~Peters
        \inst{7}
        \and
        P.~Klaassen
        \inst{9}
        \and
        M.\ T.\ Beltr\'an
        \inst{10}
        \and
        A.~Palau
        \inst{11}
        \and
        T.~M\"oller
        \inst{9}
        \and
        K.G.~Johnston
        \inst{12}
        \and
        H.~Zinnecker
        \inst{13}
        \and
        J.~Urquhart
        \inst{14}
        \and
        R.~Kuiper
        \inst{15}
        \and
        A.~Ahmadi
        \inst{16}
        \and
        \'A. S\'anchez-Monge
        \inst{17,18}
        \and
        S.~Feng
        \inst{19}
        \and
        S.~Leurini
        \inst{20}
        \and
        S.E.~Ragan
        \inst{21}
}
   \institute{$^1$ Max Planck Institute for Astronomy, K\"onigstuhl 17,
     69117 Heidelberg, Germany, \email{name@mpia.de}\\
     $^2$ Max-Planck-Institute for Extraterrestrial Physics, Gie\ss enbachstrasse 1, 85748 Garching\\
     $^3$  Instituto di Astrofisicae Planetologia Spaziali (IAPS), INAF, Via Fossodel Cavaliere 100, 00133 Roma, Italy\\
     $^4$ Harvard-Smithsonian Center for Astrophysics, 60 Garden Street, Cambridge MA 02138, USA\\
     $^5$ Institute of Astronomy and Astrophysics, Academia Sinica, 11F of Astronomy-Mathematics Building, AS/NTU No.1, Sec. 4, Roosevelt Road, Taipei 10617, Taiwan, Republic of China\\
     $^6$ Department of Physics and Astronomy, McMaster University, Hamilton, ON L8S 4M1, Canada\\
     $^7$ Max-Planck-Institut f\"{u}r Astrophysik, Karl-Schwarzschild-Str. 1, D-85748 Garching, Germany\\
     $^8$ I.~Physikalisches Institut, Universit\"at zu K\"oln, Z\"ulpicher Str.~77, 50937 K\"oln, Germany\\
     $^9$ UK Astronomy Technology Centre, Royal Observatory Edinburgh, Blackford Hill, Edinburgh EH9 3HJ, UK \\
     $^{10}$ INAF -Osservatorio Astrofisico di Arcetri, Largo E.~Fermi 5, 50125 Firenze, Italy\\
     $^{11}$ Instituto de Radioastronom\'ia y Astrof\'isica, Universidad Nacional Aut\'onoma de M\'exico, Antigua Carretera a P\'atzcuaro 8701, Ex-Hda. San Jos\'e de la Huerta, 58089 Morelia, Michoac\'an, M\'exico\\
     $^{12}$  School of Physics \& Astronomy, E.C. Stoner Building, The University of Leeds, Leeds LS2 9JT, UK\\
     $^{13}$ Universidad Autonoma de Chile, Providencia, 7500912 Santiago de Chile, Chile\\
     $^{14}$ Centre for Astrophysics and Planetary Science, University of Kent, Canterbury, CT2 7NH, UK\\
     $^{15}$ Fakult\"at für Physik, Universit\"at Duisburg-Essen, Lotharstr. 1, D-47057 Duisburg, Germany\\
     $^{16}$ Leiden Observatory, Leiden University, PO Box 9513, 2300 RA Leiden, The Netherlands\\
     $^{17}$ Institut de Ci\`encies de l'Espai (ICE, CSIC), Carrer de Can Magrans, s/n, E-08193 Bellaterra, Barcelona, Spain\\
     $^{18}$  Institut d'Estudis Espacials de Catalunya (IEEC), Barcelona, Spain\\
     $^{19}$ Department of Astronomy, Xiamen University, Xiamen, Fujian 361005, People’s Republic of China\\
     $^{20}$ INAF  Osservatorio Astronomico di Cagliari, Via della Scienza 5, 09047 Selargius (CA),Italy\\
     $^{21}$ School of Physics and Astronomy, Cardiff University, Queen’s Buildings, The Parade, Cardiff CF24 3AA, UK
   }

   \date{Version of \today}

  \abstract
   {The fragmentation of high-mass star-forming regions depends on a variety of physical parameters, including the density, magnetic field and turbulent gas properties.}
   {We evaluate the importance of the density and magnetic field structures in relation to the fragmentation properties during high-mass star formation.}
   {Observing the large pc-scale Stokes $I$ mm dust continuum emission with the IRAM 30\,m telescope and the intermediate-scale ($<$0.1\,pc) polarized submm dust emission with the Submillimeter Array toward a sample of 20 high-mass star-forming regions allows us to quantify the dependence of the fragmentation behaviour of these regions depending on the density and magnetic field structures.}
   {Based on the IRAM\,30\,m data, we infer density distributions $n\propto r^{-p}$ of the regions with typical power-law slopes $p$ around $\sim$1.5. There is no obvious correlation between the power-law slopes of the density structures on larger clump scales ($\sim$1\,pc) and the number of fragments on smaller core scales ($<$0.1\,pc). Comparing the large-scale single-dish density profiles to those derived earlier from interferometric observations at smaller spatial scales, we find that the smaller-scale power-law slopes are steeper, typically around $\sim$2.0. The flattening toward larger scales is consistent with the star-forming regions being embedded in larger cloud structures that do not decrease in density away from a particular core. Regarding the magnetic field, for several regions it appears aligned with filamentary structures leading toward the densest central cores. Furthermore, we find different polarization structures with some regions exhibiting central polarization holes whereas other regions show polarized emission also toward the central peak positions. Nevertheless, the polarized intensities are inversely related to the Stokes $I$ intensities, following roughly a power law slope of $\propto S_I^{-0.62}$. We estimate magnetic field strengths between $\sim$0.2 and $\sim$4.5\,mG, and we find no clear correlation between magnetic field strength and the fragmentation level of the regions.
   Comparison of the turbulent to magnetic energies shows that they are of roughly equal importance in this sample. The mass-to-flux ratios range between $\sim$2 and $\sim$7, consistent with collapsing star-forming regions.} 
   {Finding no clear correlations between the present-day large-scale density structure, the magnetic field strength and the smaller-scale fragmentation properties of the regions, indicates that the fragmentation of high-mass star-forming regions may not be affected strongly by the initial density profiles and magnetic field properties. 
   However, considering the limited evolutionary range and spatial scales of the presented CORE analysis, future research directions should include density structure analysis of younger regions that better resemble the initial conditions, as well as connecting the observed intermediate-scale magnetic field structure with the larger-scale magnetic fields of the parental molecular clouds.}
   
   \maketitle
%

\section{Introduction}
\label{intro}

Most high-mass stars are known to form in clusters, however, it is not yet clear to what extent different physical processes contribute to the fragmentation of the parental star-forming regions. Is thermal Jeans fragmentation the main mechanism (e.g., \citealt{palau2013,palau2014,palau2015}) or does turbulence play a significant role (e.g., \citealt{wang2014})? How important are the initial conditions, in particular the density structure (e.g., \citealt{girichidis2011})? Can thermal feedback suppress fragmentation (e.g., \citealt{krumholz2009})? How does the magnetic field influence the fragmentation of the cluster-forming regions (e.g., \citealt{commercon2011,peters2011,commercon2022,myers2013,myers2014,zhang2014b,federrath2015,beuther2018,beuther2020b,palau2021})?

To study the fragmentation of high-mass star-forming regions, we have embarked on the IRAM NOEMA (Northern Extended Millimeter Array, formerly Plateau de Bute Interferometer, PdBI) large program CORE\footnote{\url{http://www.mpia.de/core}} that investigates 20 well-known high-mass star-forming regions in the northern hemisphere at high spatial resolution ($\sim 0.3''-0.4''$) in the 1.3\,mm continuum and spectral line emission \citep{beuther2018b}. The sample is selected to have high luminosities $>10^4$\,L$_{\odot}$, distances less than 6\,kpc, and high declination to ensure good uv-coverage with northern hemisphere interferometers. While the continuum data are used for the fragmentation analysis, the spectral line data allow us to investigate the turbulent, kinematic and physical properties, as well as the chemical characteristics of the regions \citep{beuther2018b,gieser2021,ahmadi2023}.

\begin{table*}[htb]
\begin{center}
\caption{CORE Sample \citep{beuther2018b}.}
\label{sample}
\begin{tabular}{lrrrrrr}
  \hline
  \hline
                &      &                &     &     & & \\
  Source & R.A.~(J2000) & Dec.~(J2000) & $v_{\rm{lsr}}$ & $D$ & $L$ & \#cores \\
  & (h:min:sec) & ($^o:':''$) & (km\,s$^{-1}$) & (kpc) & ($10^4$L$_{\odot}$) & \\
  \hline
  IRAS23151+5912  & 23:17:21.01 & +59:28:47.49 & -54.4 & 5.7 & 10.0 & 5   \\
  IRAS23033+5951  & 23:05:25.00 & +60:08:15.49 & -53.1 & 3.5 & 1.2  & 4   \\
  IRAS23385+6053  & 23:40:54.40 & +61:10:28.20 & -50.2 & 4.9 & 1.6  & 3   \\
  W3(H$_2$O)$^a$  & 02:27:04.60 & +61:52:24.73 & -48.5 & 2.0 & 3.0  & 7   \\
  W3IRS4          & 02:25:31.22 & +62:06:21.00 & -42.8 & 2.0 & 6.0  & 6   \\
  IRAS21078+5211  & 21:09:21.64 & +52:22:37.50 & -6.1 & 1.5 & 1.3   & 20  \\
  AFGL2591        & 20:29:24.86 & +40:11:19.40 & -5.5 & 3.3 & 20.0  & 3   \\
  G75.78+0.34     & 20:21:44.03 & +37:26:37.70 & -0.5 & 3.8 & 7.0   & 4   \\
  S87 IRS1        & 19:46:20.14 & +24:35:29.00 & 22.0 & 2.7 & 3.3   & 11  \\
  G084.9505-00.691& 20:55:32.47 & +44:06:10.10 & -34.6 & 5.5 & 1.3  & 8   \\
  G094.6028-01.797& 21:39:58.25 & +50:14:20.90 & -43.6 & 4.9 & 4.3  & 4   \\
  G100.3779-03.578& 22:16:10.35 & +52:21:34.70 & -37.6 &3.7 & 1.7   & 20  \\
  G108.7575-00.986& 22:58:47.25 & +58:45:01.60 & -51.5 & 4.3 & 1.3  & 3   \\
  G138.2957+01.555& 03:01:31.32 & +60:29:13.20 & -37.5 & 2.9 & 1.4  & 3   \\
  G139.9091+00.197& 03:07:24.52 & +58:30:48.30 & -40.5 & 3.2 & 1.1  & 2   \\
  S106            & 20:27:26.77 & +37:22:47.70 & -1.0 & 1.4 & 4.0   & 2   \\
  CepAHW2         & 22:56:17.98 & +62:01:49.50 & -10.0 & 0.7 & 2.5  & 2   \\
  NGC7538IRS9     & 23:14:01.68 & +61:27:19.10 & -57.0 & 2.7 & 4.5  & 9   \\
  NGC7538IRS1$^a$ & 23:13:45.36 & +61:28:10.55 & -57.3 & 2.7 & 8.0   & 1   \\
  NGC7538S$^a$    & 23:13:44.86 & +61:26:48.10 & -56.4 & 2.7 & 1.5   & 6   \\
  \hline
  \hline
\end{tabular}
~\\
Notes: $^a$Archival SMA data from \citet{chen2012,frau2014,palau2021}.
\end{center}
\end{table*}

Investigating the 1.3\,mm continuum data of the sample, \citet{beuther2018b} distinguished a large variety in fragmentation properties ranging from sources that are dominated by a single massive core to those that fragment into as many as 20 cores (independent of distance). Since the sample is selected uniformly with respect to their evolutionary phase -- they are all high-mass protostellar objects (HMPOs, indicators are discussed in \citealt{beuther2018b}), some also harboring hot molecular cores and ultracompact H{\sc ii} regions -- evolutionary effects are unlikely to explain the results. Hence, the diversity from highly fragmented to barely fragmented regions is a real outcome of this study \citep{beuther2018b}.

The challenge is to isolate which of the above physical processes dominate the fragmentation diversity. Conducting a minimum spanning tree analysis, \citet{beuther2018b} find that typical nearest neighbor separations are below the thermal Jeans fragmentation scale, indicating that turbulence cannot explain the observed core separations. Since all regions are in similar evolutionary stages, thermal feedback is also unlikely to cause the fragmentation differences. Two possible explanations are differences in the magnetic field and/or the density structure of the parental gas clump. We also investigated whether the interferometric spatial filtering varies for regions with different fragmentation properties because steep density profiles should result in less missing flux than flatter density distributions. However, no considerable spatial filtering differences have been found within the sample \citep{beuther2018b}. In the following, we refer to clumps as the parental star-forming regions on pc-scales and to cores as the fragments on sub-0.1\,pc scales.

In this study, we now investigate the density distribution of the parental gas clumps and the magnetic field structures of the regions. While the density structures are studied by means of large-scale single-dish 1.2\,mm dust continuum mapping of the regions with the IRAM\,30\,m telescope, the magnetic fields are investigated with polarization observations of the dust emission with the Submillimeter Array (SMA) at 875\,$\mu$m. Overall parameters for the investigated CORE sample are presented in Table~\ref{sample}.

\subsection{Density structure} 

Density distributions $\rho$ of star-forming regions can be described via radial $r$ power law distributions like $\rho\propto r^{-p}$.
To first order, flatter density distributions may result in more fragments whereas steeper density distributions favor less fragmentation (e.g., \citealt{girichidis2011}). Observational studies of low-mass cores typically found density distributions with $p$ varying between 1.5 and 2, resembling finite-sized Bonnor-Ebert spheres (e.g., \citealt{motte2001,alves2001}). While early single-dish studies in the high-mass regime of ultracompact H{\sc ii} regions indicated potentially shallower slopes that may have favored logatropic equations of state (e.g., \citealt{vandertak2000,hatchell2000}), later studies of younger evolutionary stages again found density distributions with exponents $p$ mostly between 1.5 and 2 \citep{beuther2002a,mueller2002,hatchell2003,palau2014}. Hence, these studies indicated that the density distributions between low- and high-mass star-forming regions should not be very different. Since then, (very) few observational studies have focused on clump-scales to constrain the density structures of the whole cluster-forming regions. Although the interstellar medium is known to be filamentary (e.g., \citealt{andre2014,molinari2016b}), investigating structures on clump scales still allows us to derive the physical properties of individual (high-mass) star-forming regions. On the theoretical side, \citet{shu1977} already determined that the singular isothermal sphere solution results in density profiles with $p=2$. Analytic and numerical modeling showed that $p=2$ is an attractor, i.e., that initially flatter density distributions approach $p=2$ during the star formation process (e.g., \citealt{naranjo-romero2015,gomez2021}). 

\begin{table*}[htb]
\begin{center}
\caption{Observational and fit parameters}
\label{obs}
\begin{tabular}{l|rrrrrrr|rrrr}
  \hline
  \hline  
 & \multicolumn{7}{c}{IRAM30m} &\multicolumn{4}{c}{SMA}\\
Source & rms & $m_i$ & $p_i$ & $p^b$ & $S_{\rm{peak}}^c$ & $N_{\rm{H2}}$ & $n_{\rm{mean}}$ & rms($I$) & rms($Q$) & rms($U$) & Beam \\
 & $\left(\frac{\rm mJy}{\rm beam}\right)$ & & & & $\left(\frac{\rm mJy}{\rm beam}\right)$ & ($10^{23}$cm$^{-2}$) & ($10^5$cm$^{-3}$) & $\left(\frac{\rm mJy}{\rm beam}\right)$ & $\left(\frac{\rm mJy}{\rm beam}\right)$ & $\left(\frac{\rm mJy}{\rm beam}\right)$ & $('')$ \\
  \hline
  IRAS23151+5912   & 5 & 1.1 & 1.5 & 2.12 &  283 & 0.6 & 1.1  & 5.0  & 1.7 & 1.9 & $4.1\times 2.6$  \\
  IRAS23033+5951   & 7 & 1.1 & 1.5 & 2.16 &  671 & 1.5 & 2.0  & 10.0 & 2.2 & 2.2 & $4.1\times 2.6$  \\
  IRAS23385+6053   & 6 & 1.0 & 1.4 & 1.97 &  334 & 0.8 & 0.9  & 5.5  & 3.1 & 4.3 & $4.9\times 2.2$  \\
  W3(H$_2$O)$^a$   & 12& 1.3 & 1.7 & 1.95 & 3355 & 7.6 & 21.1 & 24.1 & 3.4 & 3.4 & $1.5\times 1.4$  \\
  W3IRS4           & 7 & 0.5 & 0.9 & 2.08 & 1372 & 3.1 & 8.6  & 15.0 & 1.5 & 1.1 & $4.2\times 2.3$  \\
  IRAS21078+5211   & 5 & 1.3 & 1.7 & 1.70 & 1541 & 3.5 & 12.9 & 4.5  & 2.3 & 2.3 & $2.0\times 1.7$  \\
  AFGL2591         & 4 & 1.2 & 1.6 & 1.97 & 1321 & 3.0 & 5.2  & 4.7  & 1.5 & 2.0 & $2.4\times 1.5$  \\
  G75.78+0.34      & 7 & 1.2 & 1.6 & 1.96 & 1459 & 3.3 & 4.8  & 4.7  & 2.1 & 2.7 & $2.4\times 1.5$  \\
  S87 IRS1         & 12& 0.8 & 1.2 & 2.08 &  790 & 1.8 & 4.5  & 6.4  & 2.7 & 2.9 & $2.2\times 1.7$  \\
  G084.9505-00.691 & 8 & 1.2 & 1.6 & 1.81 &  200 & 0.5 & 0.5  & 1.6  & 0.2 & 0.7 & $ 2.0\times 1.9$ \\
  G094.6028-01.797 & 7 & 1.3 & 1.7 & 1.85 &  300 & 0.7 & 0.9  & 1.7  & 0.8 & 0.8 & $2.0\times 2.0$  \\
  G100.3779-03.578 & 7 & 1.3 & 1.7 & 1.84 &  199 & 0.4 & 0.7  & 2.0  & 1.6 & 1.5 & $2.1\times 1.7$  \\
  G108.7575-00.986 & 5 & 0.5 & 0.9 & 2.30 &  455 & 1.0 & 2.6  & 2.8  & 0.8 & 0.8 & $2.2\times 1.9$  \\
  G138.2957+01.555 & 16& 1.1 & 1.5 &      &  773 & 1.7 & 3.3  & 7.3  & 2.3 & 1.9 & $ 3.5\times 2.6$ \\
  G139.9091+00.197 & 6 & 0.8 & 1.2 &      &  397 & 0.9 & 1.6  & 3.9  & 1.6 & 1.8 & $3.4\times 2.6$  \\
  S106             & 5 & 1.1 & 1.5 &      & 1097 & 2.5 & 10.6 & 6.8  & 1.9 & 2.3 & $2.0\times 1.9$  \\
  CepAHW2          & 4 & 1.0 & 1.4 & 2.16 & 3732 & 8.4 & 66.9 & 8.5  & 1.3 & 2.0 & $2.2\times 1.9$  \\
  NGC7538IRS9      & 8 & 1.0 & 1.4 & 2.23 &  788 & 1.8 & 3.7  & 7.1  & 3.1 & 6.4 & $5.0\times 2.2$  \\
  NGC7538IRS1$^a$  & 8 & 1.1 & 1.5 &      & 4284 & 9.7 & 19.9 & 21.1 & 10.4& 10.6& $2.3\times 2.0$  \\
  NGC7538S$^a$     & 8 & 1.1 & 1.5 &      & 2690 & 6.1 & 12.5 & 11.1 & 6.3 & 3.1 & $2.0\times 1.6$  \\
  \hline
  \hline
\end{tabular}
\end{center}
Notes: $m_i$ and $p_i$ are the intensity and density power law indices.\\
$^a$Archival SMA data from \citet{chen2012,frau2014,palau2021}.\\
$^b$ Small-scale density power law index based on the NOEMA data from \citet{gieser2021}, assuming a temperature power law index $q=0.4$.\\
$^c$ The peak flux densities $S_{\rm{peak}}$ are corrected for free-free emission based on: W3(H$_2$O) \citep{wyrowski1999}, W3IRS4 \citep{tieftrunk1997}, AFGL2591 \citep{vandertak2005b}, S87IRS1 \& S106 \citep{kurtz1994}, G094 \citep{skinner1993}, G139 \citep{manjarrez2012}, CepA \citep{torrelles1996}, NGC7538IRS1 \citep{sandell2009}.
\end{table*}

While the density structures of fragmented cores for the CORE sample have been studied in detail by \citet{gieser2021}, their large-scale density structure from the parental clumps is not well constrained yet (with the exceptions of IRAS\,23033 and IRAS\,23151 studied in \citealt{beuther2002a}). Studying now the large-scale distributions, we can directly set them into context with the observed fragmentation properties (e.g., number of fragments, separation between fragments) as well as with the small-scale density structures reported in \citet{gieser2021}.

\subsection{Magnetic field structure} 
\label{intro_magnetic}

Six of the CORE sample regions have been observed at single-dish resolution ($20''$) with the SCUBA polarimeter SCUPOL \citep{matthews2009}. However, all these data show comparably weak polarization signal toward the Stokes $I$ peak positions (so-called polarization holes), mainly caused by unresolved magnetic field structures in the single-dish beam (cf.~W3(H$_2$O) in \citealt{matthews2009} versus \citealt{chen2012}). Thus, higher spatial resolution observations are imperative. For three of the regions (W3(H$_2$O), NGC7538IRS1 and NGC7538S), SMA magnetic field studies have already been published \citep{chen2012,frau2014,palau2021}. While NGC7538IRS1 shows barely any fragmentation in the CORE study (\citealt{beuther2018b}, Table \ref{sample}) and high magnetic field strengths on the order of $\sim$2\,mG in the SMA data \citep{frau2014}, being consistent with reduced gas fragmentation (e.g., \citealt{commercon2011,peters2011,commercon2022,myers2013,myers2014}), W3(H$_2$O) and NGC7538S show intermediate levels of fragmentation (7 and 6 fragments, respectively, Table \ref{sample}) with again magnetic field strengths in the mG regime \citep{chen2012,palau2021}. Therefore, with only three regions of the CORE sample observed in the past, the effect of magnetic fields remains inconclusive. \citet{zhang2014b} observed a sample of intermediate- to high-mass star-forming regions with the SMA and inferred that magnetic fields are indeed important during the star-formation process. Recently, \citet{palau2021} compared the same magnetic field data to the fragmentation properties of these regions, and they found a tentative correlation between the number of fragments and the mass-to-flux ratio. Summaries of the current state of high-resolution interferometric studies of star-forming regions can be found in \citet{hull2019} or \citet{liu2022}. Clearly this is only the beginning of an exciting avenue to follow. Therefore, to further investigate the fragmentation and magnetic field properties in context, complementary high-spatial-resolution density and magnetic field data are needed. Since the $0.3''-0.4''$ mm continuum data for the CORE sample have already been acquired, we now complement these with the magnetic field information from the SMA.

\begin{figure*}[htb]
\includegraphics[width=0.99\textwidth]{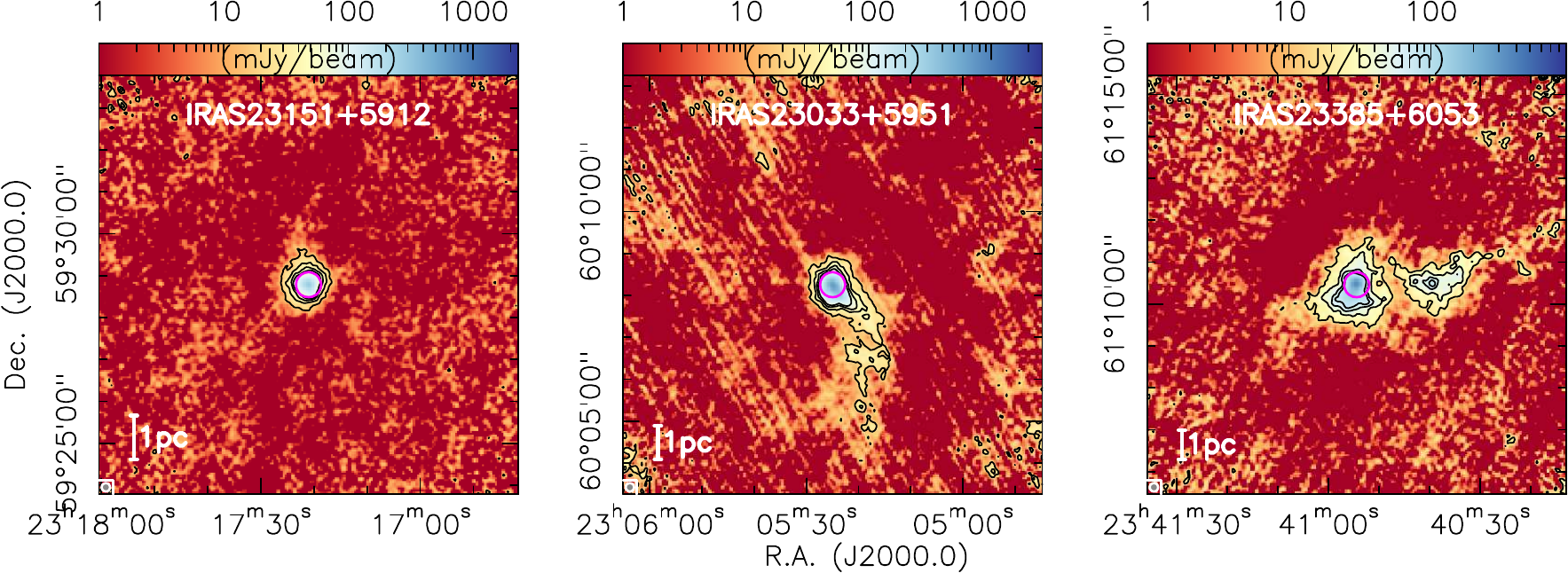}
\caption{NIKA2 1.2\,mm dust continuum images toward three example regions of the CORE sample. The color-scale shows the flux densities, and the contours are always from $3\sigma$ to $12\sigma$ in $3\sigma$ steps. The $12''$ beam and a 1\,pc scale-bar are shown at the bottom left of each panel. The magenta circles outline the $\sim 36''$ primary beam size of the corresponding SMA observations. The remaining maps are shown in Figs.~\ref{nika2_2} and \ref{nika2_3}.}
\label{nika2_1} 
\end{figure*} 

\section{Observations}
\label{observations}

\subsection{IRAM 30\,m telescope: Large-scale density structure}
\label{30m}

All 20 CORE regions were observed with New IRAM KIDs Array 2 (NIKA2, \citealt{adam2018,perotto2020}\footnote{Detailed descriptions of NIKA2 can be found at https://publicwiki.iram.es/Continuum/NIKA2/Main\#NIKA2\_Homepage.}) on the IRAM 30\,m telescope between December 2019 and February 2020 (project 143-19). NIKA2 consists of 2900 kinetic inductance detectors (KIDs) operating at 150 and 260\,GHz. Since the main goal of this study is to derive reliable intensity profiles (and from that density profiles) of the central star-forming regions, the inherent spatial filtering in single-dish continuum observations has to be considered (e.g., \citealt{perotto2020}, see also Sect.~\ref{density}). Therefore, we followed the IRAM recommendation to observe extended map sizes of roughly $9.8'$ per region. While smaller maps were also observed, here we use only the extended maps.

Data reduction was conducted with the PIIC software package (Pointing and Imaging In Continuum\footnote{PIIC handbook: http://www.iram.fr/~gildas/dist/piic.pdf}). Since all regions have extended emission, for best recovery of spatial structures, we extensively tested various source structures as well as mapping parameters within PIIC. For that purpose, we first created artificial sources with different power-law profiles. These were then smoothed to the $12''$ beam of NIKA2 at 1.2\,mm wavelengths and run through the PIIC pipeline. We explored, for example, how much source masks affect the retrieval of the intensity profiles. We found that for our application the best results were obtained without applying any source masks since the emission is extended with complex substructures such as dense clumps and filaments. Therefore, for our final dataset no source masks were defined and we applied baselines of order one. At most 35 iterations were used, but we always stopped further iterations when the peak flux density difference between the previous two iterations was less than 0.1\%. For more details about source structure recovery, see also Sect.~\ref{density}.

NIKA2 always observes simultaneously at 1.2\,mm and 2\,mm wavelengths. For our study here, we require the highest spatial resolution. Therefore, we only present the 1.2\,mm data. The angular resolution of the final maps is $12''$ and the $1\sigma$ rms varies between 4 and 16\,mJy\,beam$^{-1}$ (Table \ref{obs}).

\subsection{Submillimeter Array (SMA): Magnetic field structure}
\label{obs_sma}

The CORE sample was observed with the Submillimeter Array \citep{ho2004} over a period of several years. The first few regions were observed in the winter term 2018/2019 and the last sources in September 2021. Altogether, our observing campaign covered 17 CORE regions while the remaining three CORE regions were already observed with the SMA in earlier campaigns (Sect.~\ref{intro_magnetic} and Table \ref{sample}). Typically, we observed two regions together in track-sharing mode. Using the compact configuration in the 875\,$\mu$m wavelength band, a spatial resolution between $2''$ and $4''$ can be achieved. The final synthesized beams for all regions are listed in Table \ref{obs}.

The target sources were always observed in loops together with gain calibrators that are also sensitive to the polarization. Typical loop lengths were 15\,minutes. Bandpass and flux calibrations were conducted either at the beginning or end of the tracks. The bandpass calibrator was also used to derive the instrumental polarization of the Array. The SMA observations for the 17 newly observed targets cover in total 16\,GHz of spectral bandwith, 8\,GHz in the lower and upper sideband, respectively. For details of the remaining three regions (W3(H$_2$O), NGC7538IRS1 and NGC7538S) we refer to the respective publications by \citet{chen2012}, \citet{frau2014} and \citet{palau2021}. The exact frequency coverage of the new observations was 338.85 to 346.88\,GHz in the lower sideband and 354.84 to 362.87\,GHz in the upper sideband, respectively. The native spectral resolution was 140\,kHz per channel and we binned the data to a spectral resolution of 559\,kHz, corresponding at 345\,GHz to a velocity resolution of $\sim$0.5\,km\,s$^{-1}$. The continuum data were created by collapsing the entire bandpass after flagging the strong CO(3-2) line. While the polarized Stokes $Q$ and $U$ data exhibit barely any line emission beyond the CO(3--2) line, the Stokes $I$ continuum emission can have some additional line contamination, in particular toward the peak positions of the hot cores like NGC7538IRS1, W3(H$_2$O) or AFGL2591. However, past analysis of hot cores typically found line contamination even toward the peak positions of less than 10-20\% (e.g., \citealt{beuther2017b}), and it is even lower for the more extended filamentary structures. Hence, line contamination does not significantly affect our results. The typical rms in Stokes $I$, $Q$ and $U$ continuum images is a few mJy\,beam$^{-1}$, a bit higher for Stokes $I$ because of larger sidelobes during the cleaning of the strong submm Stokes $I$ continuum emission. All rms values are listed in Table~\ref{obs}.

\begin{figure*}[htb]
\includegraphics[width=0.33\textwidth]{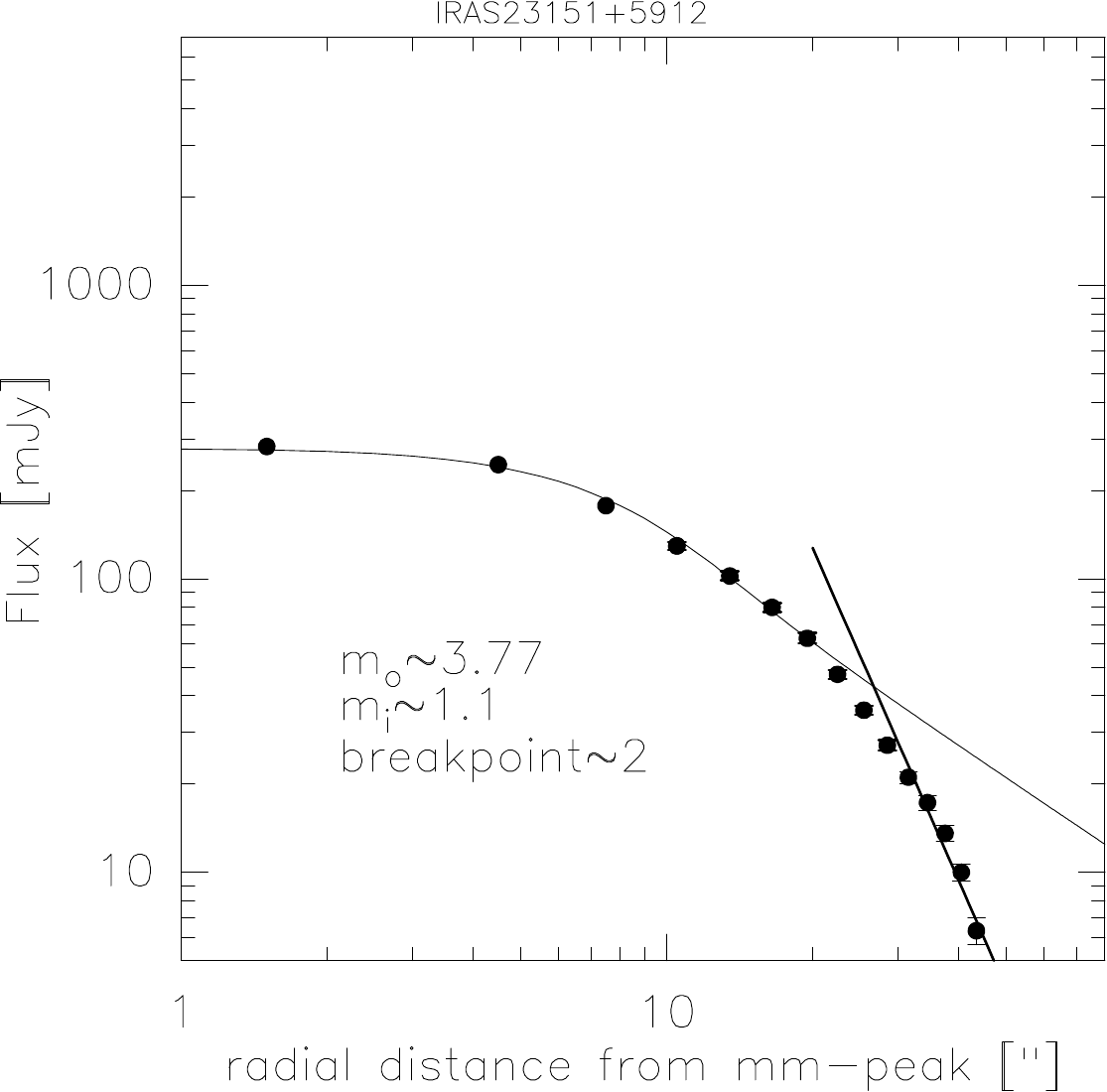}
\includegraphics[width=0.33\textwidth]{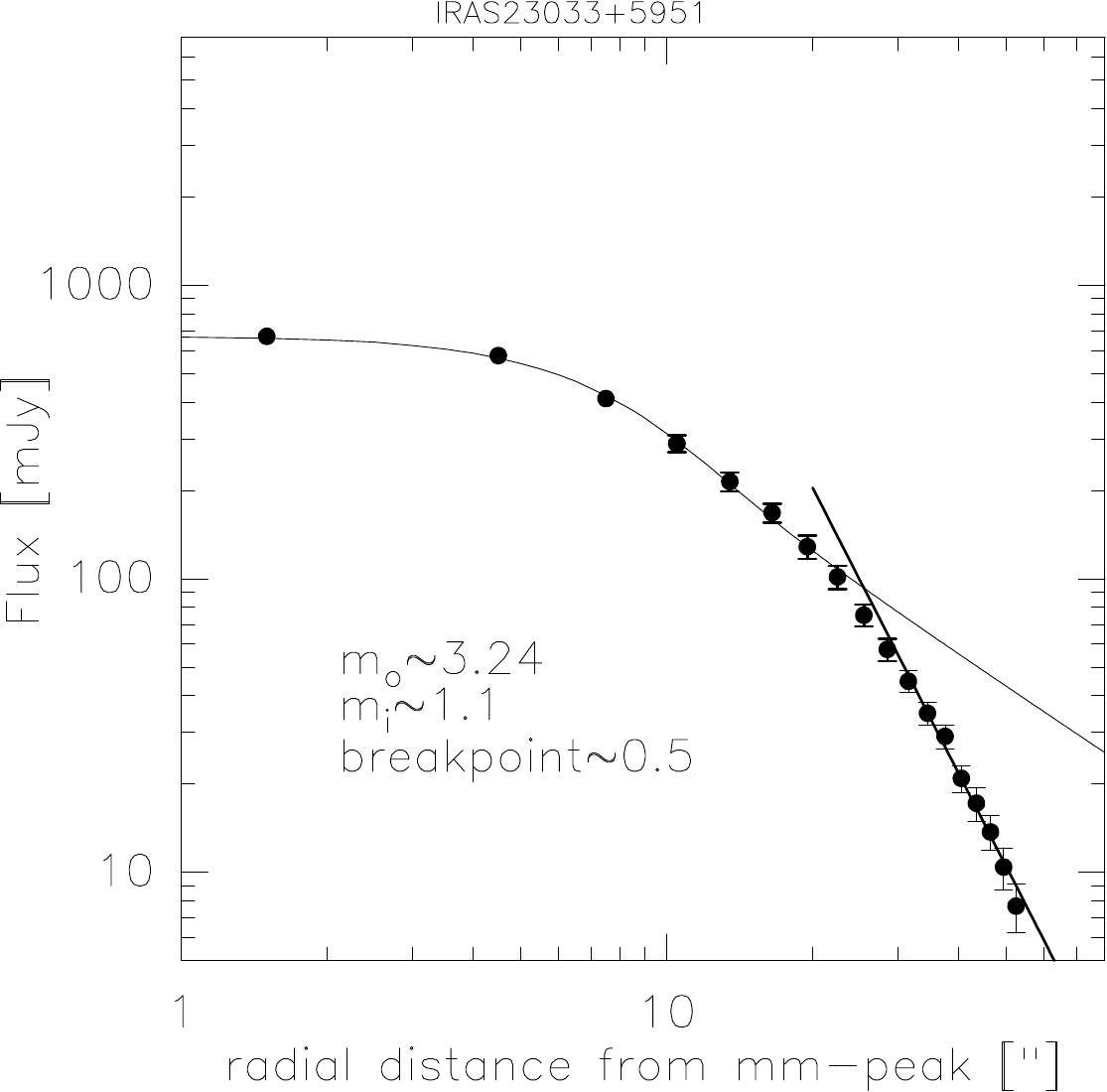}
\includegraphics[width=0.33\textwidth]{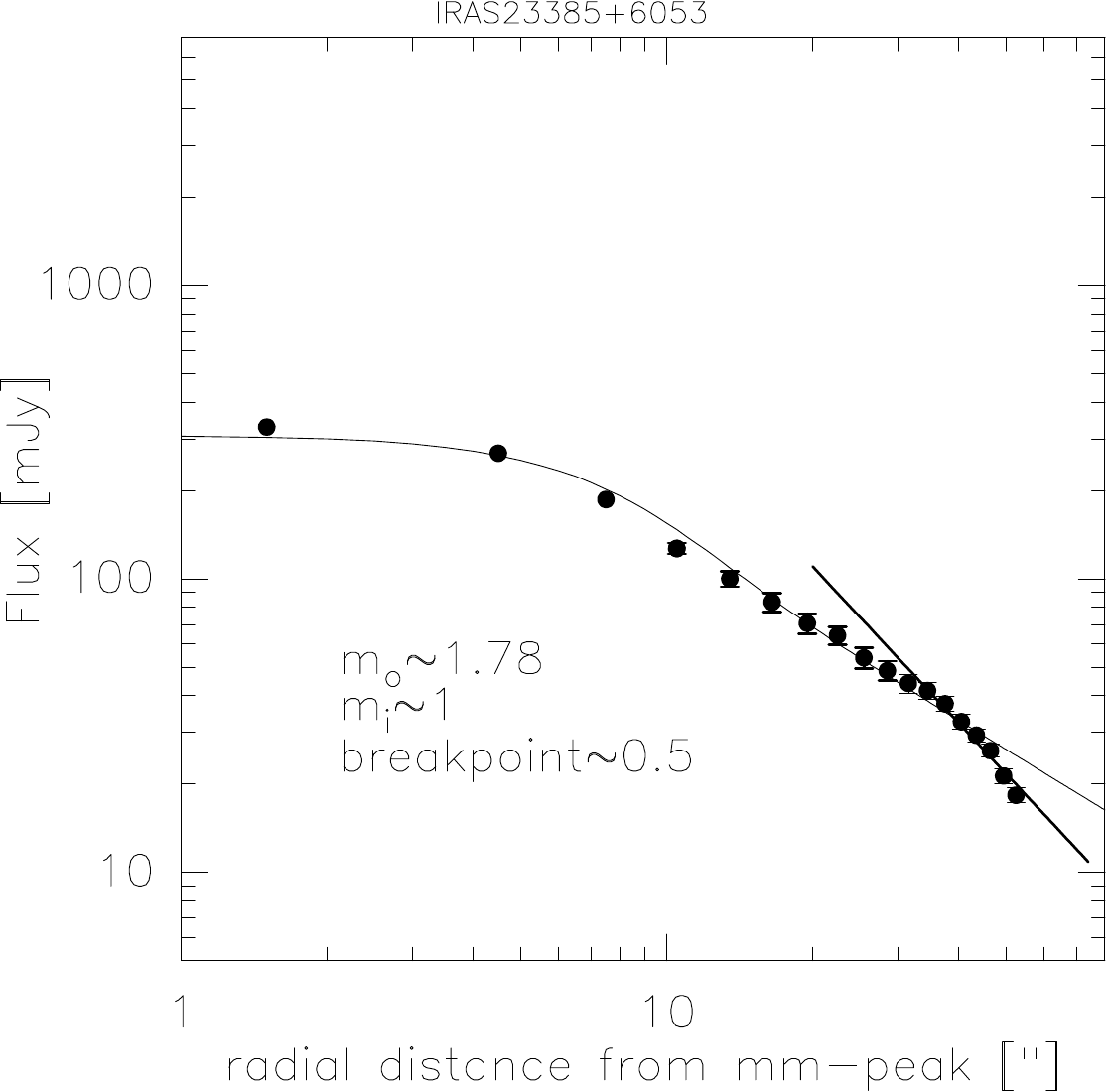}
\caption{Radial intensity profiles derived for the main 1.2\,mm dust continuum sources in the CORE sample. The power-law slopes of the inner and outer fits ($m_i$ and $m_o$) as well as the inner breakpoint in arcsecond are labeled in each panel. See main text for fitting details. The beam size of $12''$ corresponds to radial distances of $6''$.  The remaining fits are presented in Figs.~\ref{profiles_2} and \ref{profiles_3}.}
\label{profiles_1} 
\end{figure*} 

\section{Results}

While the density structure and magnetic field analysis relies on two very different datasets, we start the presentation of results on the larger scales with the single-dish dust continuum observations in Sect.~\ref{density}, and then zoom into the more central region obtained with the SMA observations in Sect.~\ref{magnetic}.

\subsection{Intensity and inferred density profiles}
\label{density} 

The main goal of the single-dish 1.2\,mm continuum mapping of the CORE sample is to derive intensity profiles, and based on these to derive density profiles of the parental gas clumps of the CORE region. Typical mean densities for such regions based on mm continuum data are in the regime of $10^5$\, cm$^{-3}$ (e.g., \citealt{beuther2002a}). To obtain reliable source structures from single-dish continuum mapping, comparatively large maps are needed to account for the inherent spatial filtering during continuum mapping (see also Sect.~\ref{30m}). In such scanning map approaches (e.g., \citealt{adam2018}), correlated noise between KIDs (kinetic inductance detectors) has to be considered, and the largest mapped scales are regarded as emission-free and used to subtract the background (e.g., Sect.~4.4 in \citealt{perotto2020}). Therefore, large-scale emission is filtered out in the final map. This approach can also affect source structures. Hence, the larger the maps, the more reliable the measured source structures are (e.g., \citealt{motte1998,beuther2002a,hatchell2003,perotto2020,rigby2021}). As outlined in Sect.~\ref{30m}, the mapping was conducted with map sizes of $\sim 9.8'$, and the final maps are presented in Figs.~\ref{nika2_1}, \ref{nika2_2} and \ref{nika2_3}. The magenta circles in these figures outline the primary beam size of the corresponding SMA observations of $36''$ discussed in the following Sect.~\ref{magnetic}. The primary beam size of the original NOEMA 1.3\,mm CORE observations was $22''$ \citep{beuther2018b}.

These larger maps outline the general environment of the regions and reveal quite a diversity. While some of the star-forming regions are rather isolated (e.g., IRAS\,23151+5912, Fig.~\ref{nika2_1}, or G100.3779-3.578, Fig.~\ref{nika2_2}), others are part of large-scale structures (e.g., W3IRS4, G75.78+0.34, Fig.~\ref{nika2_2}, or the NGC7538 complex, Fig.~\ref{nika2_3}). While these large-scale structures are interesting in themselves and deserve a separate analysis, here we are focusing only on the intensity and density structures of the main central regions marked by the magenta ellipses in Figs.~\ref{nika2_1}, \ref{nika2_2} and \ref{nika2_3}. Nevertheless, all these 1.2\,mm continuum maps are provided in electronic form via CDS (Centre de Donnees astronomiques de Strasbourg) for further analysis.

For our intensity profile analysis, we follow largely the approach described in \citet{beuther2002a} for a sample of 69 high-mass protostellar objects (HMPOs). The two dimensional intensity distributions are a convolution of the intrinsic source intensity with the $12''$ beam, and it is also affected by the scanning mapping technique (see discussion above). Regarding the convolution with the beam, theoretical work and simulations have shown that the beam convolution does not significantly affect scales larger than the beam (e.g., \citealt{adams1991,motte2001}). Structures smaller than the beam size cannot be spatially resolved but the data still give the integrated continuum flux densities of the inner regions.

\begin{figure}[htb]
\includegraphics[width=0.49\textwidth]{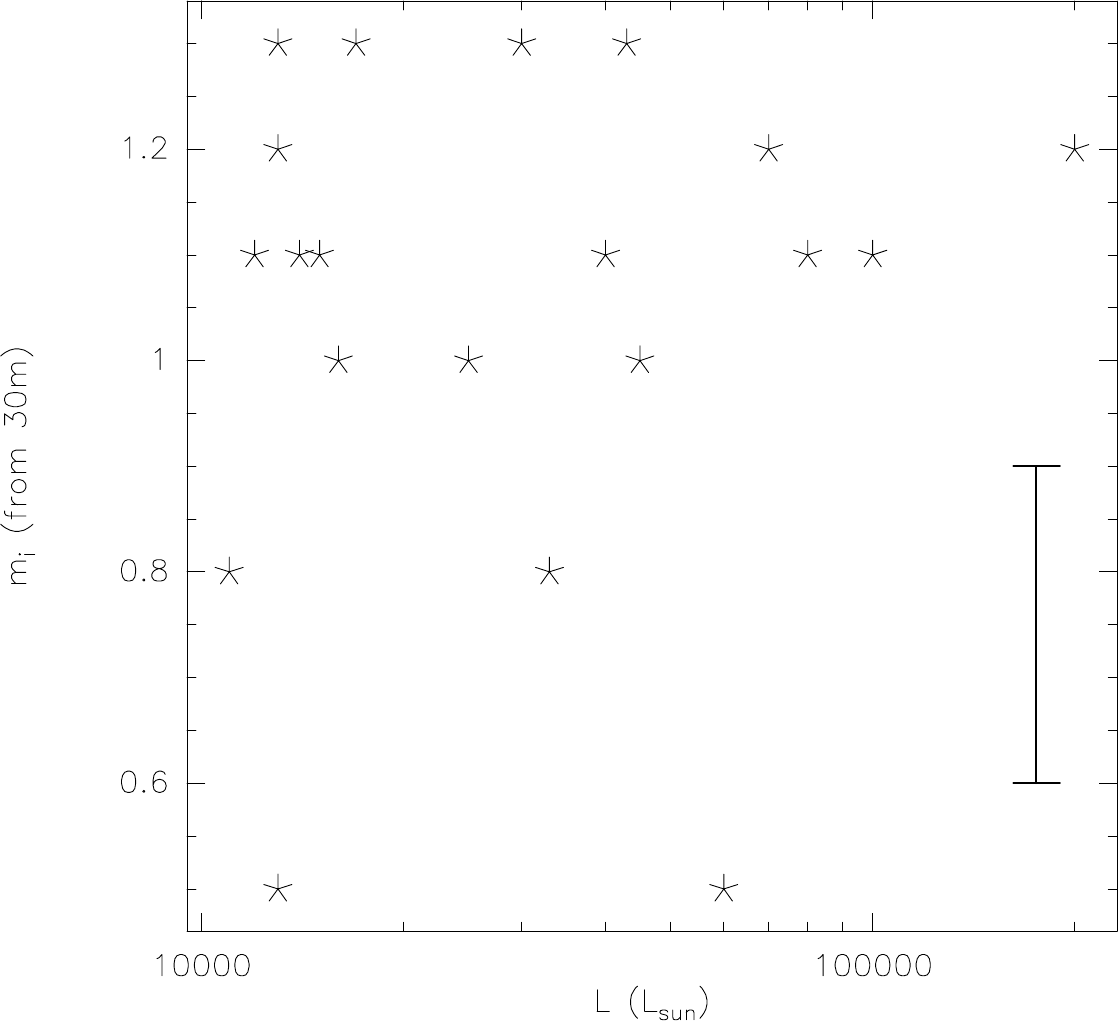}
\caption{Intensity power law index $m_i$ plotted versus the luminosity of the region (Table \ref{sample}). A representative error bar for $m_i$ is shown at the bottom-right.}
\label{m_L} 
\end{figure} 

Since most central regions resemble to first order spherical-like structures (Figs.~\ref{nika2_1}, \ref{nika2_2} and \ref{nika2_3}), we derive the radial intensity profiles in circular annulli of $3''$ steps starting at the continuum peak position. Figures \ref{profiles_1}, \ref{profiles_2} and \ref{profiles_3} present the derived intensity profiles. As already seen in the data of the 69 regions presented in \citet{beuther2002a}, outside the spatial resolution limit of $\sim 12''$, the profiles typically exhibit broken power-law slopes, steepening to the outside. This steepening to the outside is most likely an observational artifact of the spatial filtering caused by the scanning maps conducted with NIKA2, and the subsequent background subtraction during the calibration. The steepening typically occurs around $\sim 30''$ from the center. Following \citet{beuther2002a}, we fit the intensity profiles with an inner power-law slope $I\propto r^{-m_i}$ out to a radius of $30''$ and an outer power-law slope $I\propto r^{-m_o}$ from $30''$ to $54''$. Since the integrated flux toward the center is finite, the inner power-law $I\propto r^{-m_i}$ has to break somewhere, and we model that with an inner flat region. For the combination of $m_i$ and the inner breakpoint we create models, smooth them with a Gaussian beam of $12''$, and fit $m_i$ and inner breakpoint simultaneously to the data. 
That inner flat but unresolved region typically corresponds roughly to the area where fragmentation is observed in the higher-resolution interferometric NOEMA and SMA data. Hence, with the single-dish continuum observations, we trace the density structure of the paternal gas clump and its potential implication for the fragmentation observed at smaller spatial scales with NOEMA and the SMA. In total, our model fitting contains three free parameters $m_i$, $m_o$ and the inner breakpoint. In practice, the outer profile $m_o$ is directly fitted to the obtained profiles since the beam at those scales is negligible. 

\begin{figure}[htb]
\includegraphics[width=0.49\textwidth]{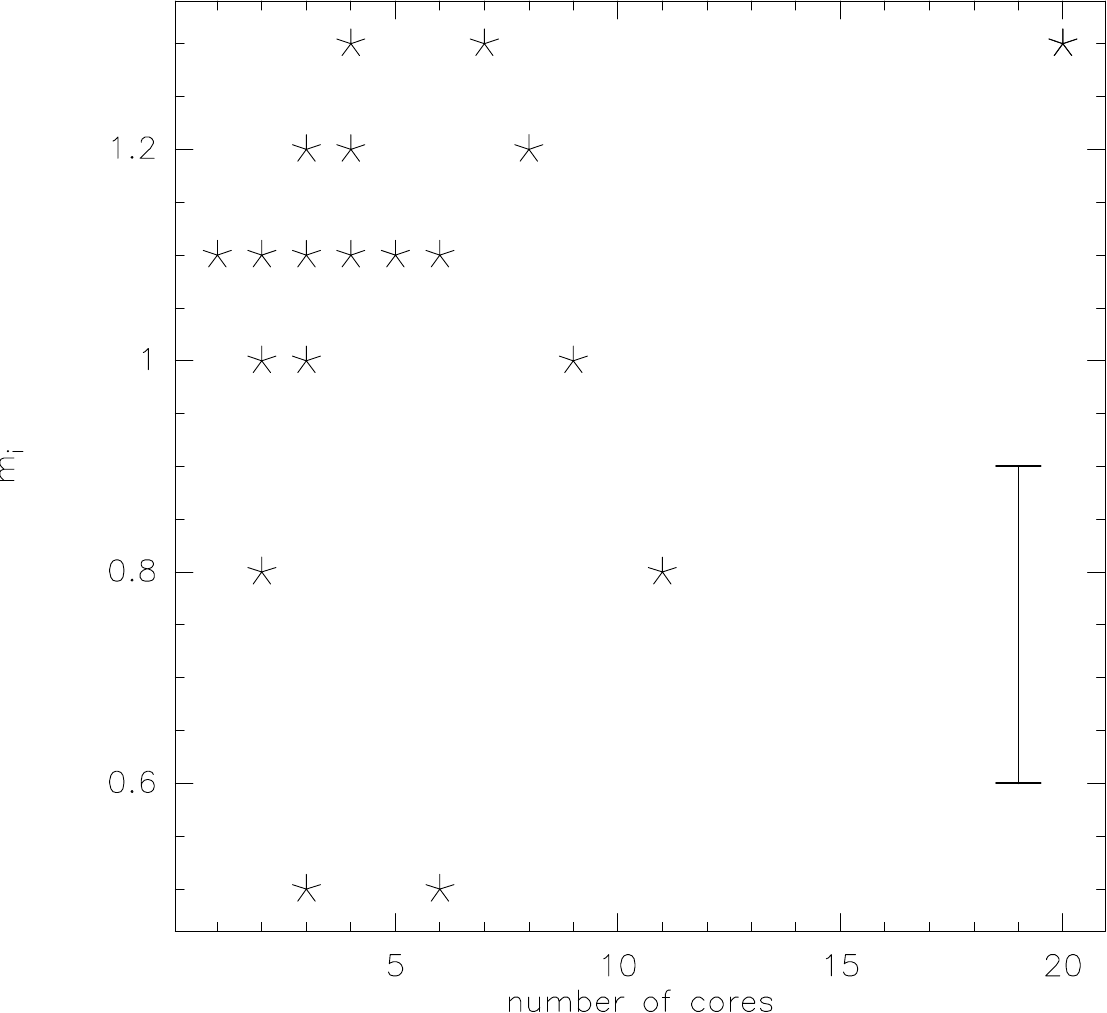}
\caption{Intensity power law index $m_i$ plotted versus the number of cores (Table \ref{sample}) from \citet{beuther2018b}. The point for 20 cores and a slope of 1.3 consists of two sources. A representative error bar for $m_i$ is shown at the bottom-right.}
\label{num_cores_mi} 
\end{figure} 

As outlined in Sect.~\ref{30m}, by conducting tests with artificially created sources, convolving them with the $12''$ beam, and injecting typical noise parameters, we explored how much the fitting deviates from the model profiles. For test inputs of $I\propto r^{-m}$ with $m$ equal to 2, 1.5 and 1.0, the fitting resulted in corresponding values of $m_i$ of 2.1, 1.7 and 1.3, respectively. The fits to the artificially created profiles also steepen at larger radii, similar to our observed data and confirming the spatial filtering of the scanning mapping approach with NIKA2. While the fitted $m_i$ of the steeper profile is barely affected by the correlated noise and spatial filtering, flatter profiles have increased deviations. This is reasonable since for flatter initial profiles, more larger-scale flux is filtered out in the observations, which steepens the fitted profiles. Based on these tests, we estimate the uncertainties of $\Delta m_i$ to $\sim$0.3 (similar results were also found in the past, e.g., \citealt{motte2001}). Since the outer profiles suffer more from such correlated noise and filtering effects, and the inner breakpoint is anyway unresolved, in the following analysis, only the inner power law $m_i$ is considered. All resulting fits are presented in Figures \ref{profiles_1}, \ref{profiles_2} and \ref{profiles_3}. Furthermore, Table \ref{obs} lists the fitted values for the inner power-law profiles $m_i$. The range of fitted inner intensity profiles $m_i$ between 0.5 and 1.3 is slightly narrower than for the larger HMPO sample of 69 regions in \citet{beuther2002a} where $m_i$ values were found in the range 0.4 to 2.1. Two regions are in common between our CORE sample and the \citet{beuther2002a} study (IRAS\,23151 and IRAS\,23033), and while we derive here values of $m_i$ of 1.1 for both sources, \citet{beuther2002a} found $m_i$ of 1.3 for the same regions. Considering that these observations are taken with two entirely different continuum instruments decades apart, the similarity of profiles, consistent within the error budget, is reassuring. 
Furthermore, only 4 regions or 20\% of the sample exhibit extremely flat intensity profiles with $m_i$ below 1.0. The remaining 80\% of the sample cluster in the narrow range of intensity power-law profiles between 1.0 and 1.3.

\begin{figure}[htb]
\includegraphics[width=0.49\textwidth]{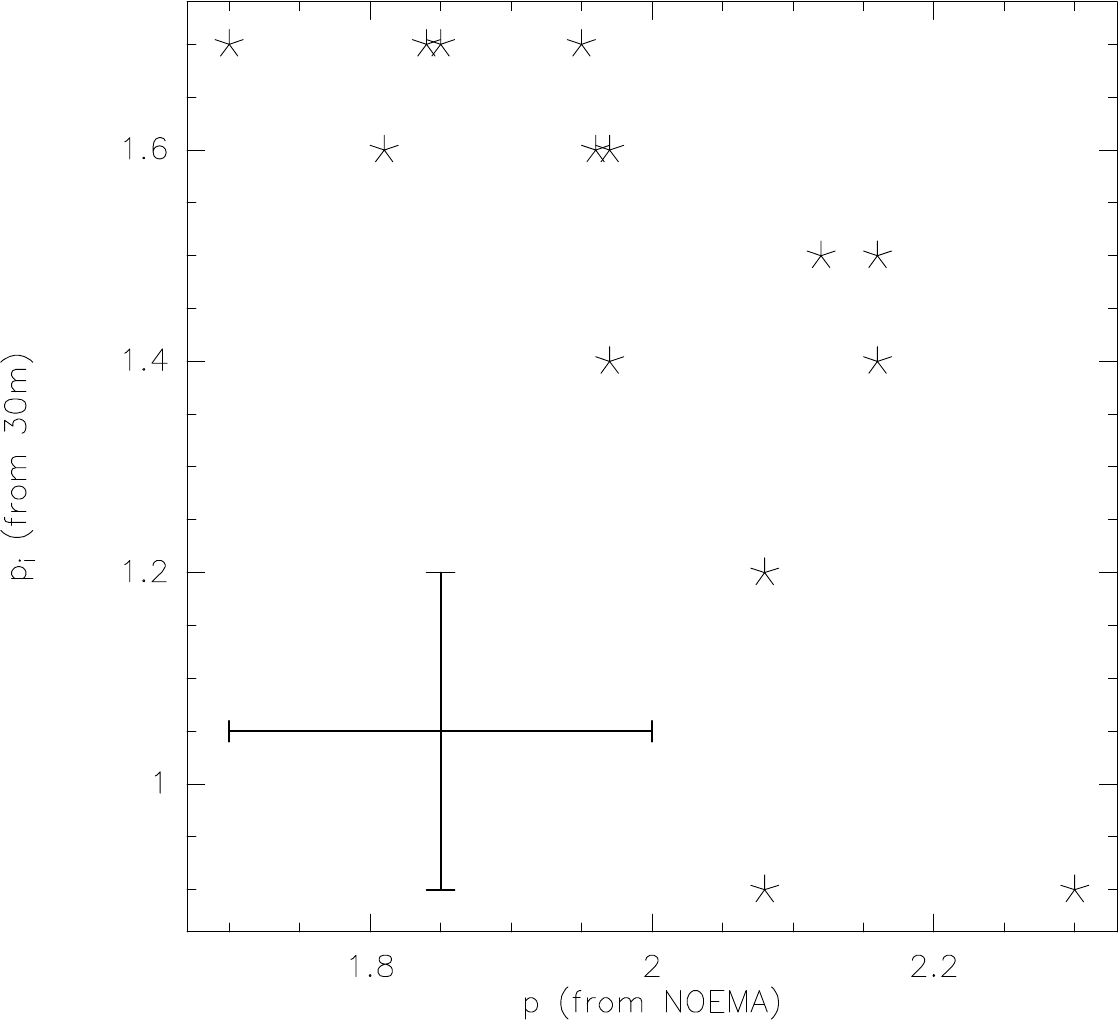}
\caption{Comparison of density distributions measured on large scale with the IRAM\,30\,m telescope ($p_i$) with the small-scale density power-law index $p$ derived from the NOEMA data by \citet{gieser2021}.  Representative error bars are shown at the bottom-left.}
\label{p_pi} 
\end{figure} 

Figure \ref{m_L} compares the fitted intensity profiles to the luminosities of the regions (Table \ref{sample}). While the data may indicate a weak tendency that the steeper profiles could be related to more luminous regions, considering the errorbars and the scatter, no reliable correlation between these two quantities can be discerned. This is consistent with the recent compilation of density indices from the literature by \citet{gomez2021} who also did not find significant differences between samples of low- and high-mass star-forming regions. We also checked whether the intensity profile slopes may be related to the distances of the sources, but no correlation between these parameters can be identified.

\begin{figure*}[htb]
\includegraphics[width=0.33\textwidth]{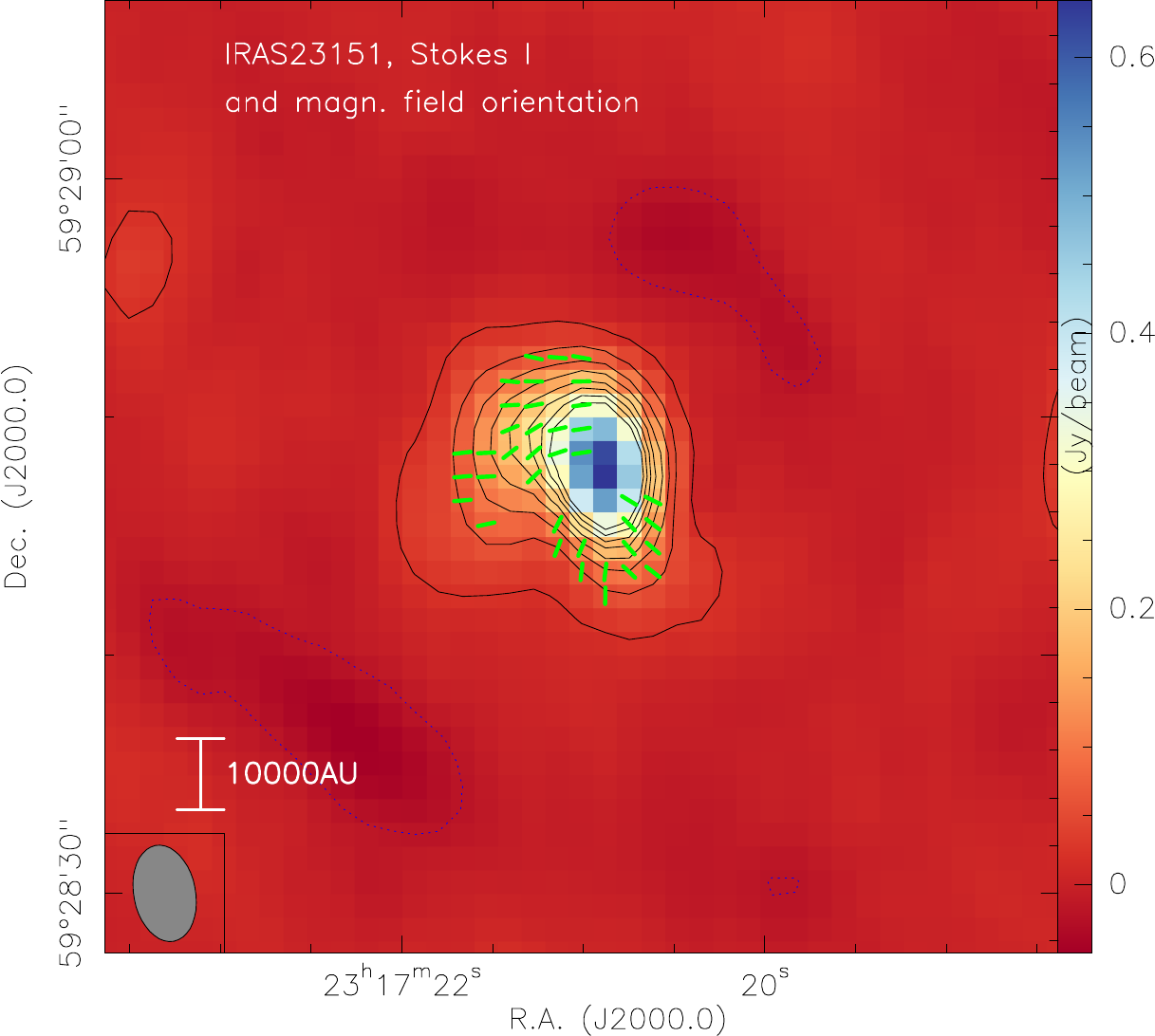}
\includegraphics[width=0.33\textwidth]{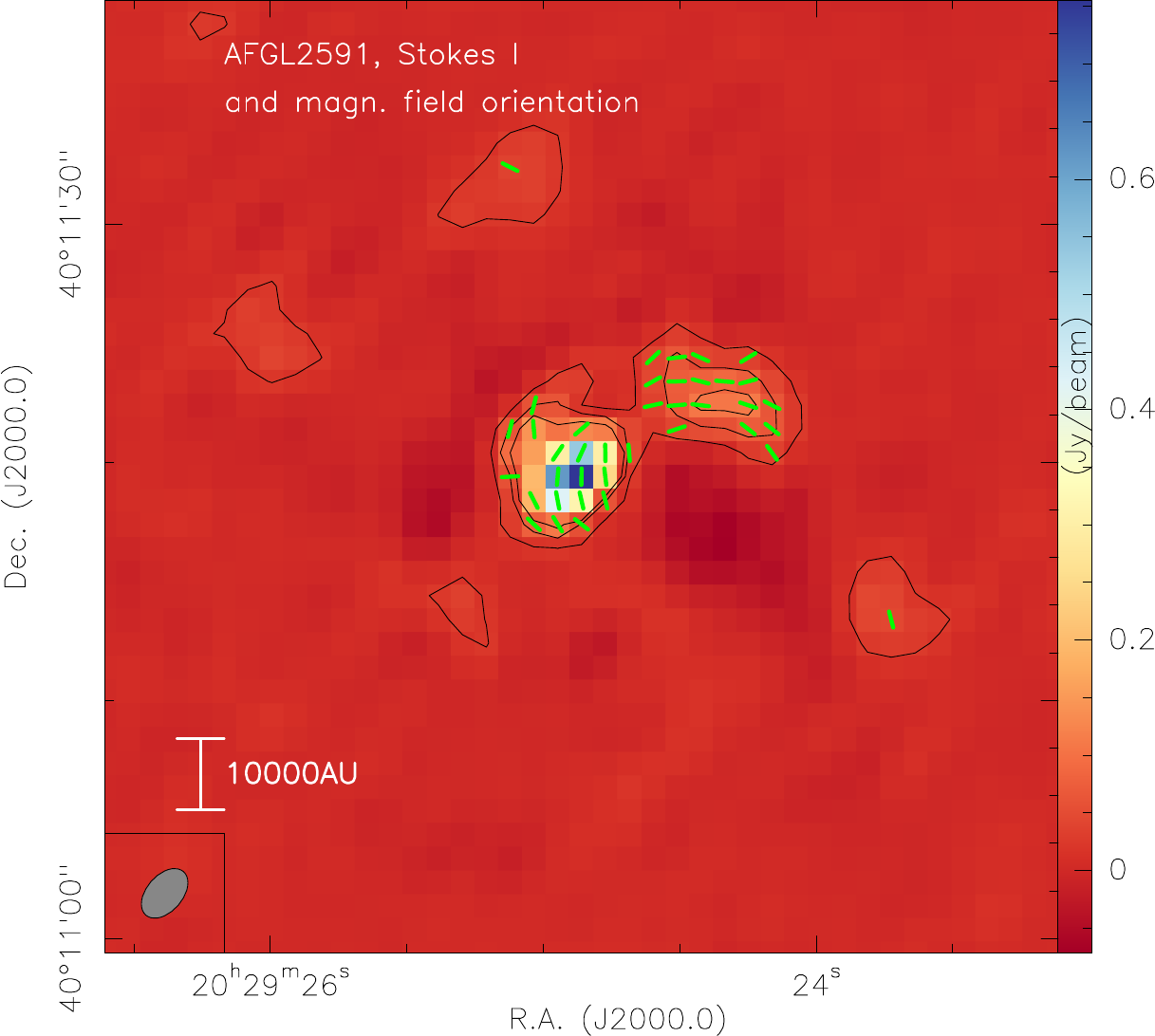}
\includegraphics[width=0.33\textwidth]{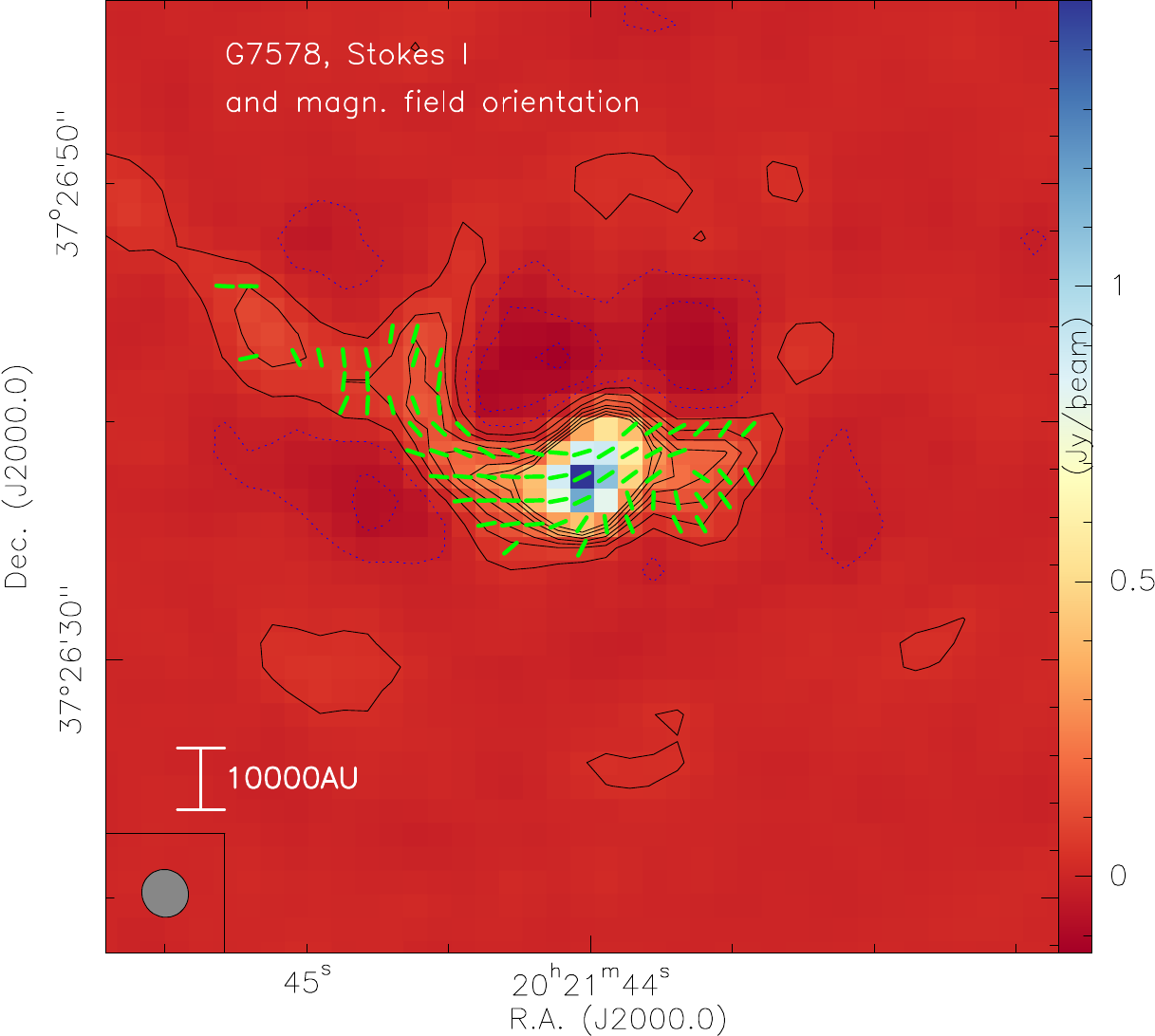}
\caption{Example SMA polarization maps. The color-scale presents the Stokes $I$ total intensity data. The contours show the same data starting at the $4\sigma$ level and continue in $8\sigma$ steps. The green constant-length line segments present the magnetic field orientation (polarization angles rotated by 90\,deg) derived from the linearly polarized continuum data above the $2\sigma$ level (independent of the polarization fraction). The synthesized beam and a linear scale bar are shown at the bottom of each panel. The corresponding maps for the remaining sample are presented in Figs.~\ref{sma_pol2} and \ref{sma_pol3}.}
\label{sma_pol1} 
\end{figure*} 

More importantly, how do the fitted intensity profiles relate to the fragmentation properties of the regions? Although the fragmentation in the CORE interferometer data is observed over a primary beam size of $22''$ and the single-dish intensity profiles cover larger radii of $\sim$60$''$, the difference between barely fragmented to highly fragmented regions may show signatures in the large-scale intensity distributions. In the high-spatial-resolution NOEMA data, low-fragmented regions typically show emission concentrated in the center of the field, whereas highly fragmented regions are well distributed over the size of the primary beam \citep{beuther2018b}. Hence, although the absolute number of fragments depends on the size of the primary beam, the sensitivity and the spatial resolution (all similar for the entire CORE sample, \citealt{beuther2018}), the observed variations between barely fragmented to highly fragmented regions is real. To investigate the relation between large-scale intensity distribution and number of cores per region, Figure \ref{num_cores_mi} shows the power-law indices $m_i$ against the number of cores identified in \citet{beuther2018b} within each region. While the regions are covering a range of distances (Table \ref{sample}), we compare the intensity structures to the number of fragments within the same regions. Therefore, the distance is not critical for this direct comparison. 

No clear trend is visible here, and one cannot identify any relation of the single-dish intensity profile with the number of cores. Although the data point with 20 cores and an intensity index $m_i$ of 1.3 appears like an outlier in Figure \ref{num_cores_mi}, we point out that this location is occupied twice. Both highly fragmented regions with 20 identified cores (IRAS20178 and G100.3779) exhibit the same steep intensity distribution.  With two regions like that, an outlier "rejection" argument seems less likely. This is interesting also when comparing their large-scale distributions in Figure \ref{nika2_2}. While IRAS\,21078 is part of a more extended region with filamentary large-scale structures, G100.3779 appears more isolated. However, there may also exist more extended low-intensity structures in G100.3779 that could be below our sensitivity. Looking at the whole sample, the large-scale intensity distributions seems to be uncorrelated with the number of observed cores or fragments. We will return to this point in Sect.~\ref{discussion}.

The millimeter continuum emission profiles can also be used to estimate the underlying density profiles. Here, we concentrate only on the inner profiles $m_i$. Following, \citet{adams1991}, \citet{motte2001} and \citet{beuther2002a}, the mm intensity profile is related to the density profile via:

\begin{eqnarray}
m = -1 + p + Qq + \epsilon_f
\label{eq_m}
\end{eqnarray}

\noindent with the intensity profile power-law index $m$, the density profile $n\propto r^{-p}$ and the temperature profile $T\propto r^{-q}$. $Q$ is a temperature and frequency dependent correction factor that is $\sim$1.2 at 30\,K and 1.2\,mm wavelength \citep{adams1991,beuther2002a}. The de-projection term $\epsilon_f$ takes into account that the relation between intensity, density and temperature profile was originally derived for infinite power law distributions, but our regions typically have finite sizes \citep{yun1991}. However, $\epsilon_f$ is rather small, and \citet{motte2001} have estimated that $\epsilon_f$ should typically be around 0.1. Regarding the temperature distribution $T\propto r^{-q}$, for centrally heated regions like those in our CORE sample, radiative transfer calculations typically find power-law indices $q$ around 0.4 (e.g., \citealt{emerson1988,vandertak2000c}). Furthermore, \citet{gieser2021} measured the temperature distributions in the CORE sample from H$_2$CO and CH$_3$CN emission line data, and they found an average temperature power-law index $q=0.4\pm 0.1$. Hence, we will use that $q$ value for the estimates of the density distributions based on the single-dish 1.2\,mm data. Using the discussed parameters in Eq.~ \ref{eq_m}, the power-law density index $p$ can be approximated as:

\begin{eqnarray}
    p \approx m + 1 -1.2\times 0.4 -0.1 \approx m+0.4
\end{eqnarray}

\noindent The corresponding estimated inner density power-law indices $p_i$ for all regions are also listed in Table \ref{obs}. 
With Gaussian error propagation, the error $\Delta p_i$ can be approximated as $\Delta p_i\approx \sqrt{\Delta m_i^2+\Delta q^2}$. With $\Delta m_i\approx 0.3$ and $\Delta q\approx 0.1$ we get $\Delta p_i\approx 0.32\approx 0.3$.
Similar to the intensity profiles, 80\% of the regions exhibit large-scale density distributions with a narrow power-law exponent range between 1.4 and 1.7, clustering around 1.5. Only four regions or 20\% of the sample have flatter profiles closer to power-law indices of 1.0. 

\begin{figure}[htb]
\includegraphics[width=0.49\textwidth]{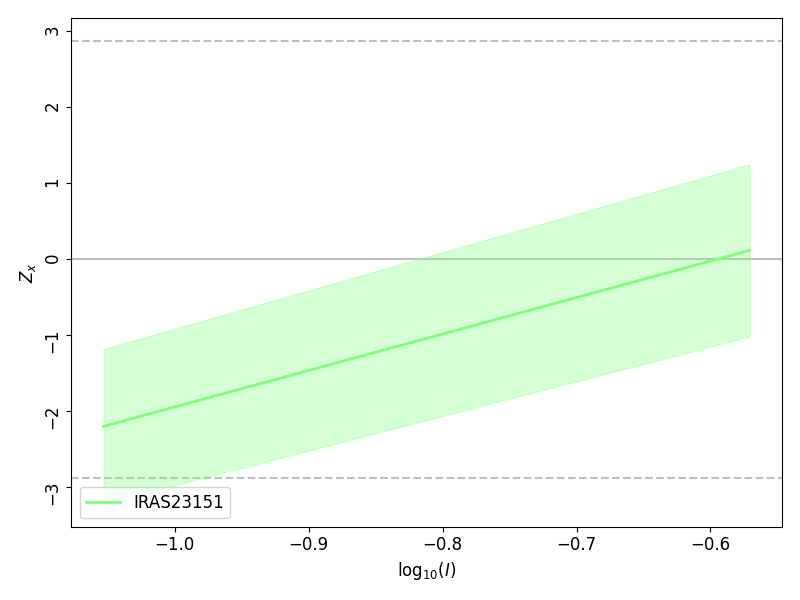}
\includegraphics[width=0.49\textwidth]{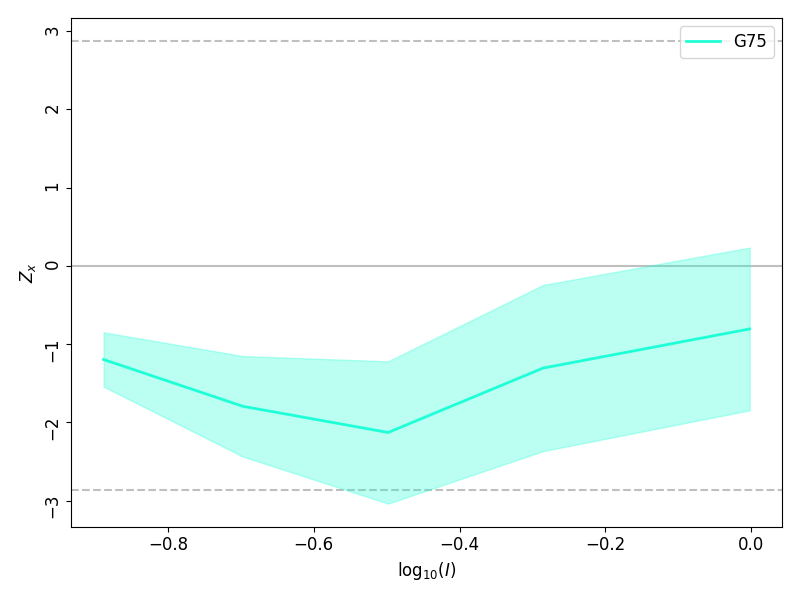}
\caption{Histogram of oriented gradient (HRO) plots for two example regions, IRAS23151 (top) and G75.78 (bottom). The X-axis shows the Stokes $I$ intensity, and the Y-axis the projected Rayleigh statistic $Z_x$ (also known as $V$). Positive $Z_x$ values correspond to largely parallel orientation and negative $Z_x$ to largely perpendicular orientations between magnetic field and gas column density. The dashed lines at $\sim$|2.87| correspond to $3\sigma$ significance in circular statistics.}
\label{HRO} 
\end{figure} 

We now compare the density power-law distributions derived on large-scales from the single-dish data with the corresponding small-scale density structures estimated from the NOEMA data in \citet{gieser2021}. They estimate the density profiles by fitting the interferometric visibilities directly in the uv-plane with a power-law slope $\alpha$. This $\alpha$ relates to the power law slopes $p$ and $q$ of the density and temperature distributions as $\alpha = p + q -3$ (e.g., \citealt{adams1991,looney2003,gieser2021}). For a consistent comparison with our data, we derive the small-scale density power-law slope $p$ from their $\alpha$ using again a temperature power-law slope of $q=0.4$ (Table \ref{obs}). We approximate the same $\Delta p\approx 0.3$ for these interferometrically derived small-scale density profiles.

While the large-scale density profiles cluster around a power-law slope of 1.5, the small-scale density power-law slopes have a mean value of 2.0 \citep{gieser2021}.  Although the error bars are large on these measurements, on average there seems to be a steepening of the density profiles from large to small spatial scales. Figure \ref{p_pi} presents the direct comparison of the large- and small-scale density structures. While for many regions, the difference between large-scale and small-scale profiles is rather small, interestingly, the two flattest large-scale profiles $p_i$ below 1.0 both exhibit comparably steep small-scale profiles above 2.0. Figure \ref{p_pi} also gives the impression of increasing small-scale $p$ with decreasing large-scale $p_i$. While this trend is only tentative considering the given uncertainties in the density profiles, we note that \citet{gieser2023} found in a recent analysis of ALMA data toward high-mass star-forming regions in different evolutionary stages a similar flattening of the density profiles from small core to large clump spatial scales. The density profiles are further discussed in Sect.~\ref{discussion}.

In addition to the intensity and density profiles, we can also use the 1.2\,mm continuum data to estimate column densities and mean densities towards the peak position. This will be important for the magnetic field analysis following below. Table \ref{obs} lists the peak flux densities $S_{\rm{peak}}$ derived towards the 20 regions. For those regions that encompass also an ultracompact H{\sc ii} region, the $S_{\rm{peak}}$ values are corrected for free-free emission based on the studies listed in footnote $c$ of Table \ref{obs}. Following \citet{hildebrand1983} or \citet{schuller2009}, assuming optically thin dust emission, a mean dust temperature of 25\,K, a gas-to-dust mass ratio of 150 \citep{draine2003}, and a dust opacity $\kappa\approx 1.0$cm$^2$g$^{-1}$ at 1.2\,mm \citep{ossenkopf1994}, we can estimate the mean column densities in the regions where also the polarized dust emission is detected (see following Section). The derived column densities range between $\sim 4\times 10^{22}$ and almost $10^{24}$\,cm$^{-2}$ as listed in Table \ref{obs}. As a next step, assuming a spherical structure of the parental star-forming clump, we can use these column densities to estimate the mean densities by dividing the peak column densities by the corresponding linear extend of the $12''$ beam. These mean densities are listed in Table \ref{obs} and range between $\sim 0.5\times 10^5$ and several times $10^6$\,cm$^{-3}$. 

\subsection{Polarization and magnetic field properties}
\label{magnetic}

\subsubsection{Alignment of magnetic field and submm continuum emission}

Polarized submm continuum emission is detected toward the entire CORE sample. Figures~\ref{sma_pol1}, \ref{sma_pol2} and \ref{sma_pol3} show the polarized emission angles rotated by 90\,deg, outlining the plane-of-the-sky magnetic field structure, overlaid on the Stokes $I$ 875\,$\mu$m continuum emission. In many regions we are able to map the magnetic field structure toward extended, often filamentary structures. Visual analysis indicates that the magnetic field structure is often aligned with filamentary structures leading toward the central source. 

A way to quantify the relative orientation between the magnetic field and the gas column density structure is the histogram of relative orientations (HRO), introduced for astrophysical magnetic field studies by \citet{soler2013}. For comparable methods, see also \citet{li2013b} or \citet{liu2022}. The HRO method measures the relative orientation between the plane-of-the-sky magnetic field component and the corresponding column density structure using its gradient. In our case, we measure the relative orientations of the magnetic field vectors against the gradient vectors of the Stokes $I$ submm continuum emission, which traces the optically thin dust emission and hence gas column density (e.g., \citealt{hildebrand1983}). Figure \ref{HRO} presents the corresponding results for two examples where the Y-axis quantifies the relative orientation between the magnetic field and continuum intensity gradient ($Z_x$, also known as projected Rayleigh statistic $V$) and the X-axis shows the Stokes $I$ 875\,$\mu$m continuum emission. Positive $Z_x$ values correspond to preferentially parallel orientation[s] and negative $Z_x$ to largely perpendicular orientations between the magnetic field and gas column density. The null hypothesis in this approach implies a uniform distribution. $Z_X$ values around |1.64| and |2.57| correspond to rejections of the null hypothesis with probabilities of 5\% and 0.5\%, and $Z_X\approx |2.87|$ corresponds roughly to a $3\sigma$ limit (e.g., \citealt{soler2022,batschelet1972}). Assuming the same temperature, increasing Stokes $I$ intensities correspond to increasing column densities.


Although the visual inspection of the data indicates that on larger scales of filamentary structures, the magnetic field appears aligned with some of the filaments (e.g., AFGL2591 or G75.78 in Fig.~\ref{sma_pol1}), the HRO analysis is less straight forward. While for IRAS\,23151 negative $Z_X$ values at low Stokes $I$ intensities are consistent with the magnetic field vectors roughly perpendicular to the densest inner regions (see also Fig.~\ref{sma_pol1}), even that is not highly significant. And in most other cases, the Z$_X$ values lie around 0. While this HRO analysis does not exclude alignment or misalignment as indicated by the visual analysis, the number of polarization vectors is typically too low for a proper statistical analysis as was done for example on larger scales with the Planck data by, e.g., \citet{soler2013} or \citet{jow2018}. 

Nevertheless, the visual analysis above can be interpreted in a framework where gravity dominates the dynamics of the gas flows, and the gas is channeled along filamentary structures toward the main gravitational centers where the most massive protostars are forming. Magneto-hydrodynamic simulations result in comparably aligned structures (e.g., \citealt{klassen2017,gomez2018}). At even higher spatial resolution, similar results have been found by \citet{beuther2020b}, \citet{sanhueza2021} or \citet{cortes2021}. 

\begin{figure}[htb]
\includegraphics[width=0.49\textwidth]{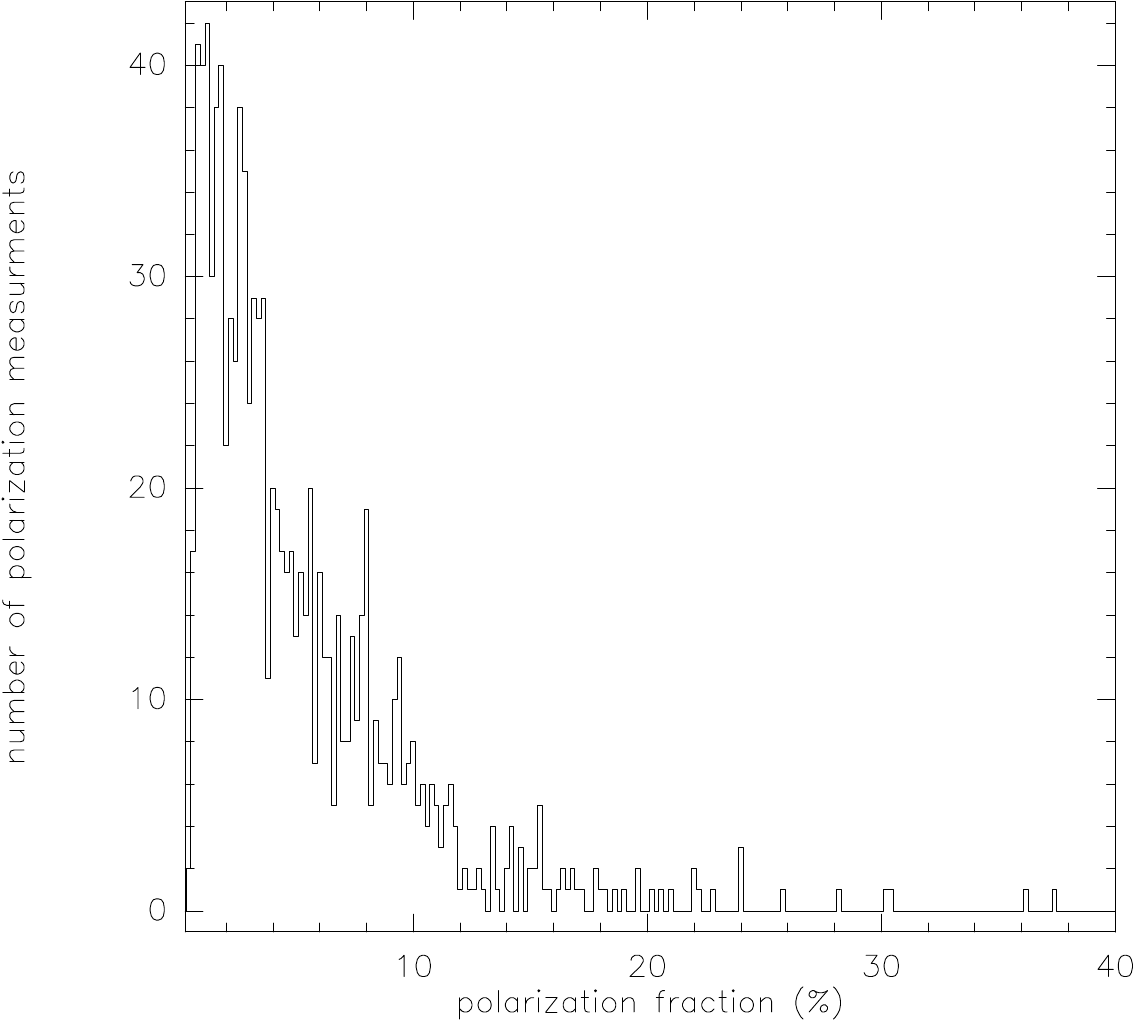}
\caption{Histogram of polarization fractions for the whole sample.}
\label{poli_histo} 
\end{figure} 

\begin{figure}[htb]
\includegraphics[width=0.49\textwidth]{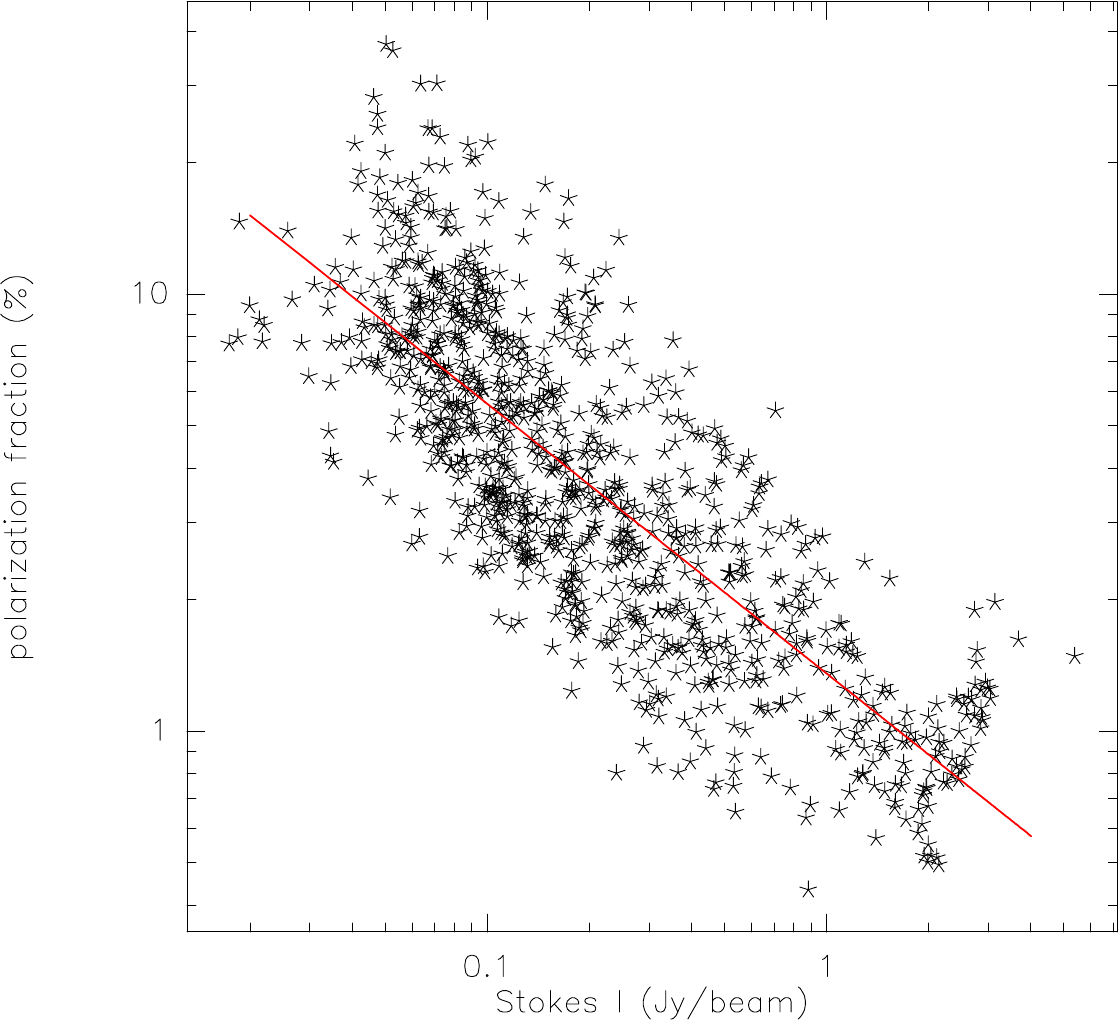}
\caption{Polarization fraction plotted against the Stokes $I$ total intensity. Uncertainties for Stokes $I$ are around the calibration uncertainty of $\sim$10\%. Uncertainties on the polarization fraction are $\sim$20\%. The red line shows a power-law fit to the data with the polarization fraction ($pol_{\rm frac}$) proportional to the Stokes $I$ intensity ($S_I$): $pol_{\rm frac} \propto S_I^{-0.62}$.}
\label{poli_stokesI} 
\end{figure} 

\subsubsection{Polarization fraction and relation to Stokes $I$}

A different quantity to analyze is the polarization fraction, i.e., the fraction between linearly polarized and Stokes $I$ intensities. Based on the typical flux calibration uncertainty of $\sim$10\% and potential line contamination toward the hot core peak positions discussed in section \ref{obs_sma}, we estimate the uncertainty on the polarization fraction to $\sim$20\%. Appendix Figure \ref{poli} presents the maps of polarization fractions for the entire CORE sample, and in Figure \ref{poli_histo} we show a histogram of the polarization fractions for all regions. More than half of the observed polarization fractions are below 4\%, and 88 percent are below 10\% polarization fraction.

Over the last decades so-called polarization holes are discussed that may indicate a decrease of fractional polarized intensity toward the Stokes $I$ peak intensities. (e.g., \citealt{dotson1996,fiege2000c,henning2001,matthews2002,wolf2003c,girart2006,tang2009a,liu2013,hull2014,soam2018,fernandez-lopez2021}). Physical reasons for measured lower polarization fractions at increased column densities range from unresolved magnetic field structures that may be smoothed out by too low angular resolution to less efficient radiative torque alignment of the dust grains at high densities and high optical depth (e.g., \citealt{lazarian2007,hull2014,soam2018,girart2018}). Inspecting the CORE sample, we see various structures (Figs.~\ref{sma_pol1}, \ref{sma_pol2}, \ref{sma_pol3} and \ref{poli}). While we do detect polarized emission towards many of the continuum peak positions (e.g., IRAS\,23033, G75.78 or NGC7538IRS1), toward other regions, we find barely any polarized emission toward the 875\,$\mu$m peak positions (e.g., IRAS\,23151, NGC7538IRS9, NGC7538S, see Figure \ref{poli}). Quantifying this behaviour further, Figure \ref{poli_stokesI} shows the polarization fraction plotted against the Stokes $I$ total intensity at 875\,$\mu$m wavelength. One clearly see that the polarization fraction decreases with increasing Stokes $I$ intensities. Even for sources where the polarized emission is detected towards the peak position, the fractional polarized intensities still decrease with increasing Stokes $I$ intensities (Fig.~\ref{poli}). Fitting a power-law between the fractional polarized intensities ($pol_{\rm frac}$) and the Stokes I intensities ($S_I$), one finds a relation as $pol_{\rm frac} \propto S_I^{-0.62}$. This slope is similar to the value of $-0.72$ recently reported for the high-mass region G5.89$-$0.39 \citep{fernandez-lopez2021}.

\begin{figure}[htb]
\includegraphics[width=0.49\textwidth]{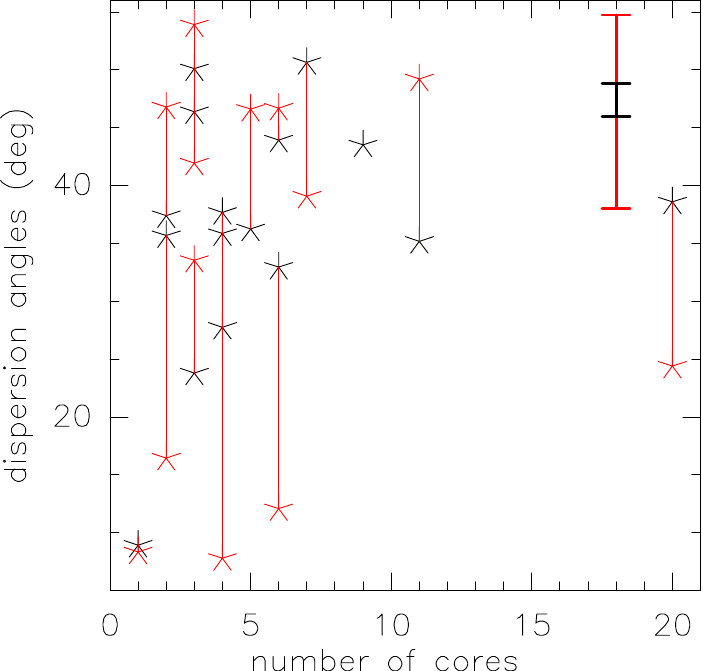}
\caption{Plot of dispersion angles versus number of cores. The black stars correspond to the directly measured $\sigma_{\psi}$ whereas the red stars correspond to the $b(0)$ value measured with the structure function analysis. Approximate error bars are shown at the top right in black and red for the $\sigma_{\psi}$ and $b(0)$ estimated magnetic field strengths, respectively.}
\label{num_cores_disp} 
\end{figure} 

\begin{figure}[htb]
\includegraphics[width=0.49\textwidth]{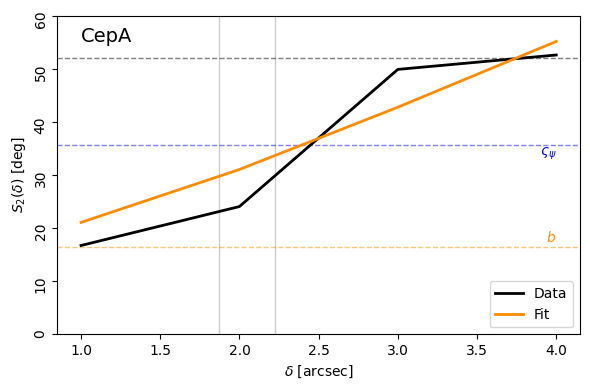}
\caption{Example structure function $S_2(l)$ for CepA. The black line shows the data, and the orange line presents the fit done for length scales above the beam size which is marked by the two vertical lines (Major and minor beam axis). The orange dashed line shows the intercept $b(0)$ and the blue dashed line shows the dispersion angles derived directly from the data as discussed above. The dashed line at $\sim$52\,deg shows the extreme for a random distribution.}
\label{structure_function_cepa} 
\end{figure} 

\subsubsection{Magnetic field and fragmentation}

Most importantly, we want to investigate how the magnetic field relates to the fragmentation of these high-mass star-forming regions. Figure \ref{num_cores_disp} presents the dispersion angles of polarized emission, i.e., of the magnetic field, plotted against the number of cores or fragments from \citet{beuther2018b}. Following Appendix D in \citet{planckXXXV}, we estimate the dispersion of polarization angles $\sigma_{\psi}$ directly from the Stokes $Q$ and $U$ images via 

\begin{equation}
\sigma_{\psi}=\sqrt{\left<(\Delta\psi)^{2}\right>},
\label{sigma_psi}
\end{equation}

\noindent with

\begin{equation}
\Delta\psi=0.5\times\arctan\left(\frac{Q\left<U\right>-\left<Q\right>U}{Q\left<Q\right>-\left<U\right>U}\right)
\end{equation}

\noindent and $\left<...\right>$ denotes the average values in the selected areas of the maps. To estimate reliable dispersion angles, we impose a threshold of the polarized emission to be detected in at least nine pixel, corresponding roughly to one beam size. This way, we can estimate the dispersion angle $\sigma_{\psi}$ for 16 regions (Table \ref{polarization_parameters}).

\begin{figure*}[htb]
\includegraphics[width=0.99\textwidth]{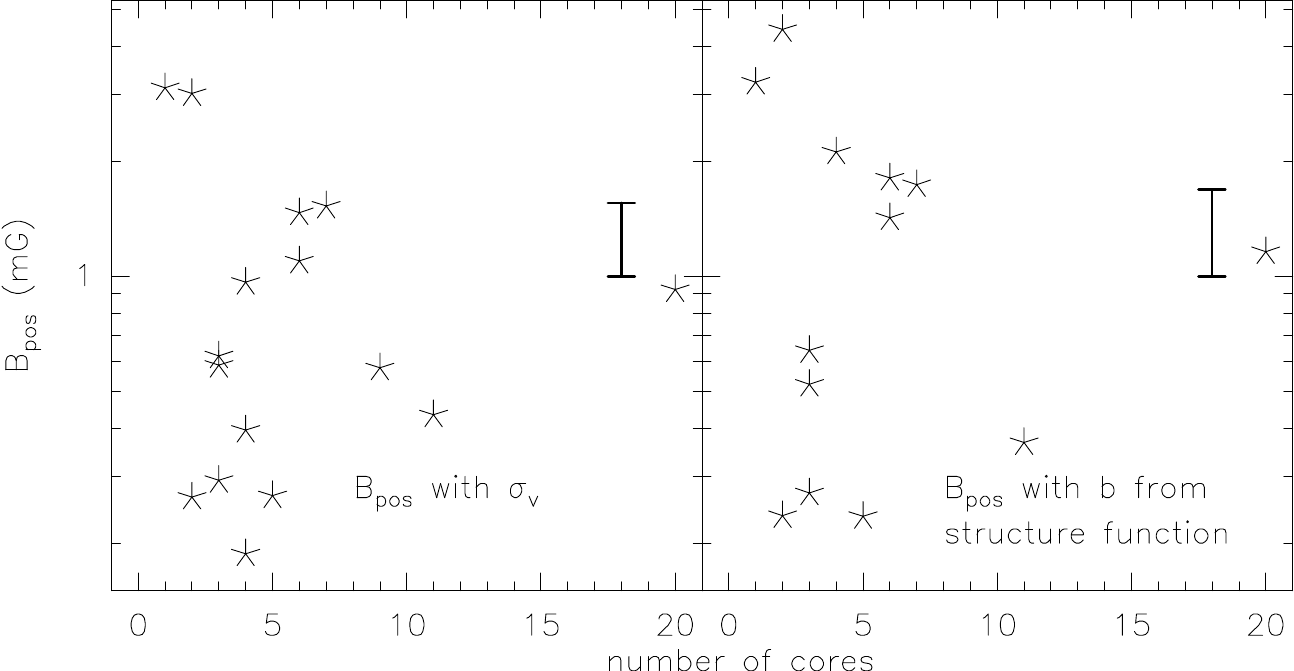}
\caption{Plot of magnetic field $B_{\rm pos}$ versus number of cores. The left panel uses the directly measured $\sigma_{\psi}$ whereas the right panel uses the $b(0)$ value measured with the structure function analysis. Approximate mean errors are shown at the right of each panel.}
\label{num_cores_magn_field} 
\end{figure*} 

To better account for the fact that not the entire angle dispersion distribution may be attributed to magneto-hydrodynamic waves and turbulence, \citet{houde2009} and \citet{hildebrand2009} developed a different approach to estimate the approximate angle dispersion associated with the magnetic field via the second order structure function analysis $S_2(l)$ where $l$ is a length scale. This structure function $S_2(l)$ measures the differences between polarization angles depending on the distances $l$ between them. The structure function analysis allows the estimation of the angle dispersion while reducing the potential contributions of large-scale turbulence \citep{hildebrand2009}. In practice, the angle dispersion $\sigma_{\psi}$ is estimated by fitting the square of the structure function with polynomial of second order $S^2_2(l)=b(l)+a_2l^2$ above length scales corresponding to the beam size (see Fig.~\ref{structure_function_cepa} for an example). The intercept of $b(0)$ is then an alternative measure for the dispersion angle $\sigma_{\psi}$. We are able to derive reasonable structure functions for 13 regions. The results are listed in Table \ref{polarization_parameters} and shown as red markers in Fig.~\ref{num_cores_disp}. In many cases, the measured $b(0)$ is below the directly inferred dispersion angles $\sigma_{\psi}$ from Eq.~ \ref{sigma_psi}. However, some regions also exhibit $b(0)$  larger than $\sigma_{\psi}$. These are typically at high angle dispersion values, close to $\sim$52\,deg that corresponds to random distributions. In that regime, measuring reliable dispersion values clearly is a difficult undertaking.

Even without estimating quantitative magnetic field strengths, already the measurement of the dispersion angles $\sigma_{\psi}$ allows a qualitative estimate of the magnetic field strengths. Larger values of $\sigma_{\psi}$ indicate smaller magnetic field strength (see Eq.~\ref{dcf} below). Using this qualitative estimate, we plot the dispersion angle $\sigma_{\psi}$ against the number of cores in Figure \ref{num_cores_disp}. This Figure shows an interesting feature in the way that the measurements cluster in the top-left half of the plot with no entries in the bottom-right of the figure. In other words, we indeed see large dispersion angles $\sigma_{\psi}$ dominating but not strictly correlated with the number of cores. However, the emptiness of the bottom-right part of Figure \ref{num_cores_disp} implies that a larger number of cores is not seen with low dispersion angles $\sigma_{\psi}$. Or formulated with respect to the magnetic field, a large number of fragments is not found for large magnetic field values. This indicates that the magnetic field can indeed prevent the parental gas clump from fragmenting, consistent with simulations (e.g., \citealt{commercon2011,peters2011,myers2013,myers2014,commercon2022}).

Quantifying the magnetic field strength and following \citet{davis1951} and \citet{chandrasekhar1953}, in the so-called DCF method the angle dispersion $\sigma_{\psi}$ is inversely related to the magnetic field in the plane of the sky. If the magnetic field is closely coupled ("frozen") to the gas, and the dispersion of the local magnetic field orientation angles is caused by transverse and incompressible Alfven waves, the magnetic field in the plane of the sky can be estimated via

\begin{equation}
B_{\rm pos}^{\rm{DCF}}=\sqrt{4\pi\rho}\frac{\sigma_{v}}{\sigma_{\psi}},
\label{dcf}
\end{equation}

\noindent with $\rho$ the gas density and $\sigma_{v}$ the 1D velocity dispersion of the gas. 
The accuracy of the DCF method has been discussed frequently (e.g., \citealt{ostriker2001,heitsch2001,liu2021,liu2022,skalidis2021a,skalidis2021b}), and different correction factors to Eq.~\ref{dcf}, typically around 0.5, have been proposed (e.g., \citealt{ostriker2001,liu2022}). Recently, \citet{skalidis2021a} proposed a method that depends on the squareroot of the angle dispersion $\sqrt{\sigma_{\psi}}$ and takes into account compressible motions in the interstellar medium. In the following, we estimate the magnetic field strength following that approach \citep{skalidis2021a}:
 
\begin{equation}
B_{\rm pos}=\sqrt{2\pi\rho}\frac{\sigma_{v}}{\sqrt{\sigma_{\psi}}}.
\label{st}
\end{equation}

While the central densities estimated at the highest resolution ($\sim 0.3-0.4''$) of the CORE data ranges between $10^6$ and $10^8$\,cm$^{-3}$ \citep{beuther2018b}, the polarized emission of the SMA data at lower angular resolution ($\sim 3''$, Table \ref{obs}) is observed more in the environmental structures at lower densities. Therefore, for the density $\rho$ we use the mean densities derived from the single-dish 1.2\,mm data at $12''$ resolution in Sect.~\ref{density} (Table \ref{obs}), roughly encompassing the areas of detected polarized emission in the SMA data. Furthermore, we use a mean molecular weight $\mu =2.8m_p$ with $m_p$ the proton mass. The 1D velocity dispersion is estimated from the single-dish IRAM 30\,m C$^{18}$O(2--1) data \citep{beuther2018b,mottram2020} over the beam size of $\sim 11''$, typically encompassing large parts of the detected polarized emission (Figs.~\ref{sma_pol1}, \ref{sma_pol2}, \ref{sma_pol3}). The full width half maximum $\Delta v$ is given in Table \ref{polarization_parameters} (typically a few km\,s$^{-1}$), and the corresponding 1D velocity dispersion is $\sigma_{v}=\Delta v/\sqrt{8\rm{ln}(2)}$. 

The  estimated magnetic field strengths $B_{\rm pos}(\sigma_{\psi})$ and $B_{\rm pos}(b)$, using either the direct angle dispersion $\sigma_{\psi}$ or the structure function intercept $b(0)$,  vary roughly between $\sim$0.2 and $\sim$4.5\,mG (Fig.~\ref{num_cores_magn_field} left and right panels, respectively, and Table \ref{polarization_parameters}).
The mean error shown in Fig.~\ref{num_cores_magn_field} is estimated from Gaussian error propagation with a relative error on the number density of a factor 2 (could easily be higher, even up to a factor 10, e.g., \citealt{liu2021}), the measured line width error of $\sim$0.1\,km\,s$^{-1}$ and the dispersion angle error of $\sim$3 and $\sim$17\,deg from the direct $\sigma_\psi$ and structure function intercept $b(0)$, respectively. 
In addition to that, one should keep in mind that the absolute error of the Davis-Chandrasekhar-Fermi method could be even larger as discussed in \citet{planckXXXV} or \citet{liu2021}. One caveat is that it is difficult to determine whether the angle dispersion is only caused by magneto-hydrodynamic (MHD) waves and turbulence. Furthermore, the angle dispersion is an average measurement within the beam along the line of sight, potentially reducing the measured projected dispersion again. Therefore, the absolute derived magnetic field strength should be considered as order-of magnitude estimate, but does not capture the entire complexity of the magnetic field structure. However, since the assumptions are the same for our entire sample, the relative differences between the sources are considered more reliable. 
We will discuss possible dependencies between magnetic field and fragmentation as well as further insights about the magnetic field in Sect.~\ref{discussion_magn_field}.


\begin{table*}[htb]
\begin{center}
\caption{Polarization and magnetic field parameters}
\label{polarization_parameters}
\begin{tabular}{lrrrrrr}
  \hline
  \hline
Source & $\sigma_{\psi}$ & $b(0)$ & $\Delta v$ & $B_{\rm pos}(\sigma_\psi)$ & $B_{\rm pos}(b)$ & $M/\Phi_B$ \\
 & (deg) & (deg) & (km\,s$^{-1}$) & (mG) & (mG) \\
  \hline
  IRAS23151+5912    & 36 & 47 & 2.8 & 0.27 & 0.23 & 1.8 \\
  IRAS23033+5951    & 28 & -- & 2.7 & 0.40 & --   & 2.9 \\
  IRAS23385+6053    & 46 & 54 & 3.9 & 0.29 & 0.27 & 2.0 \\
  W3(H$_2$O)$^a$    & 51 & 39 & 4.3 & 1.53 & 1.74 & 3.8 \\
  W3IRS4            & 33 & 12 & 3.9 & 1.10 & 1.81 & 2.1 \\
  IRAS21078+5211    & 39 & 24 & 2.9 & 0.92 & 1.16 & 2.9 \\
  AFGL2591          & 50 & 42 & 3.3 & 0.59 & 0.64 & 3.9 \\
  G75.78+0.34       & 38 & 8  & 4.9 & 0.97 & 2.12 & 2.6 \\
  S87 IRS1          & 35 & 49 & 2.2 & 0.43 & 0.37 & 3.1 \\
  G084.9505-00.691  & -- & -- & 3.7 & --   & --   & --  \\
  G094.6028-01.797  & 36 & -- & 2.1 & 0.19 & --   & 2.7 \\
  G100.3779-03.578  & -- & -- & 1.7 & --   & --   & --  \\
  G108.7575-00.986  & -- & -- & 3.3 & --   & --   & --  \\
  G138.2957+01.555  & 24 & 34 & 3.0 & 0.62 & 0.52 & 2.1 \\
  G139.9091+00.197  & -- & -- & 1.8 & --   & --   & --  \\
  S106              & 37 & 47 & 0.9 & 0.26 & 0.24 & 7.1 \\
  CepAHW2           & 36 & 16 & 4.0 & 3.02 & 4.45 & 2.1 \\
  NGC7538IRS9       & 44 & -- & 3.6 & 0.58 & --   & 2.3 \\
  NGC7538IRS1       & 9  & 8  & 3.8 & 3.12 & 3.23 & 2.3 \\
  NGC7538S          & 44 & 47 & 5.0 & 1.47 & 1.43 & 3.1 \\
  \hline
  \hline
\end{tabular}
\end{center}
Notes: $\sigma_{\psi}$ and $b(0)$ are the angle dispersions measured directly from the data and via the structure function analysis \citep{houde2009}. $B_{\rm pos}(\sigma_\psi)$ and $B_{\rm pos}(b)$ are the corresponding magnetic field strength estimates. $\Delta v$ is measured from the C$^{18}$O(2--1) IRAM 30\,m data with an $11''$ beam. $M/\Phi_B$ is the mass-to-flux ratio.
\end{table*}

\section{Discussion}
\label{discussion}

\subsection{Density structures and fragmentation}

Although only tentative, we find a trend of flatter density power-law slopes on large scales compared to those derived with NOEMA on smaller spatial scales (Fig.~\ref{p_pi}). Interestingly, \citet{gieser2023} found a similar trend in recent ALMA observations of high-mass star-forming regions in different evolutionary stages. While power-law indices around 2 on the small scales as observed by \citet{gieser2021,gieser2022,gieser2023} correspond well to classical collapse solutions (e.g., \citealt{larson1969,shu1977,stahler2005,bhandare2020}), the flattening on larger scales does to first order not fit into that picture. However, high-mass star-forming regions are typically parts of larger cloud structures as also evident in our 1.2\,mm IRAM 30\,m continuum maps (Figs.~\ref{nika2_1}, \ref{nika2_2} and \ref{nika2_3}). Therefore, the fitted structures of the star-forming regions cannot resemble infinite power-law profiles, but they may flatten out when they merge with the environmental molecular clouds. Such a merging of collapsing star-forming clumps with the environmental cloud could then result in a flattening of the observable density structures on large spatial scales. In addition to this, as shown in analytical and numerical studies, even starting with initially flatter density distributions, during the collapse process, the density structures should approach $p=2$ (e.g., \citealt{naranjo-romero2015,gomez2021}). Assuming that the flatter large-scale profiles found in our study resemble more the initial conditions of the region, the steeper interferometrically derived inner density profiles would then correspond to the collapse-induced density structure.

Regarding the finding that we do not see a correlation between the large-scale intensity distribution (and hence density distribution, see Sect.~\ref{density}) and the number of fragments (see also \citealt{palau2014}), \citet{girichidis2011} argued that the initial density profile of a region should be related to fragmentation of a region in a way that flatter density profiles would result in more fragments whereas steeper density profiles would result in less fragmentation. Considering the millimeter continuum intensity distribution as a proxy of the density distribution, our data do not allow such conclusions. However, one has to keep in mind that the theoretical studies (e.g., \citealt{girichidis2011}) consider the density distribution of the initial conditions of the parental gas clump whereas we observe the gas clump now at a later time with already ongoing star formation and fragmentation. Furthermore, in the simulations by \citet{girichidis2011}, the density distribution is the only parameter that varies. If other parameters, e.g., the magnetic field, vary as well, less clear trends in the observational data can be expected. Following the chemical analysis by \citet{gieser2021}, the mean approximate ages of our regions are roughly $6\times 10^4$\,yr. Hence, we are not observing the initial intensity distribution but one that has already evolved over several $10^4$ years. Nevertheless, if the main collapse were largely progressing in an inside-out fashion, the larger-scale density distribution may not be affected that much during the early evolution. Independent of that, while our data do not allow a conclusion about a potential relation between the initial density structure and fragmentation properties, the number of fragments apparently does not directly correlate with the present day corresponding observed large-scale intensity and density structures.

To investigate previously suggested relations between column density and density versus the level of fragmentation (e.g., \citealt{lombardi2013,palau2014,palau2015}), we checked whether our derived column densities and mean densities (Table \ref{obs}) could be correlated with the number of cores in the respective regions, but we did not find any such correlation. This is consistent with the findings of \citet{palau2013}, where the mean densities were derived following the same approach as in this work. 

A future step to address a potential relation between the density distribution versus fragmentation properties will be to conduct similar mm dust continuum observations toward younger regions, e.g., infrared dark clouds. These should resemble the initial conditions better and hence allow a clearer association of the initial density structure with the early fragmentation of high-mass star-forming regions.

\subsection{Magnetic field structures and fragmentation}
\label{discussion_magn_field}

While the angle dispersion versus number of cores plot in Fig.~\ref{num_cores_disp} suggests that there may be a relation between the polarization angle dispersion with the number of fragments, after conversion to magnetic field strength in Fig.~\ref{num_cores_magn_field} such a trend is not obvious. While for a low number of fragments (below 10), there is a broad spread of angle dispersion and magnetic field strength, the one region with 20 cores, for which we can also estimate a magnetic field strength\footnote{The other highly fragmented region G100.3779 has not strong enough polarized signal to estimate a magnetic field strength (Table \ref{polarization_parameters}, Fig.~\ref{sma_pol2})}, exhibits an intermediate magnetic field strength around $\sim$1\,mG. Hence, the strength of the magnetic field itself does not allow us to predict the number of fragments.

Additional important magnetic field parameters can be estimated from the data, in particular the Alfvenic velocity, the ratio of turbulent to magnetic energy and the mass-to-flux ratio.

The 1D Alfvenic velocity $\sigma_A$ can be estimated from the magnetic field strength $B$ and the density $\rho$ via  $\sigma_A=B/\sqrt{4\pi\rho}$. Using a mean $B(\sigma_{\psi})\approx 1.0$\,mG (Table \ref{polarization_parameters} and Fig.~\ref{num_cores_magn_field}) and a mean density of $\rho\approx 0.9\times 10^6$\,cm$^{-3}$, the 1D Alfvenic velocity is $\sigma_A\approx 1.37$\,km\,s$^{-1}$. With the uncertainties in $B$ and $\rho$, the $\sigma_A$ uncertainty is roughly a factor 2. For comparison, based on the mean line width of $\approx 3.2$\,km\,s$^{-1}$ (Table \ref{polarization_parameters}, based on single-dish C$^{18}$O(2--1) data at $11''$ resolution, \citealt{gieser2021}\footnote{The data are available at \url{https://www.mpia.de/core}.}), the mean 1D velocity dispersion is $\sigma_v\approx 1.4$\,km\,s$^{-1}$. Considering the given uncertainties, these values are surprisingly close and indicate that Alfvenic and turbulent velocities are roughly on the same order of magnitude. 

In a similar direction, one can estimate the ratio of turbulent to magnetic energy $\beta\sim 3(\sigma_v/\sigma_A)^2$ (e.g., \citealt{girart2009}). Using the 1D Alfvenic and turbulent velocity dispersion values discussed before, one finds a $\beta\approx 3.1$. With the velocities being squared for the estimate of $\beta$, the uncertainties on $\beta$ are around a factor of 4. Nevertheless, even such rough estimates indicate similar importance of magnetic and turbulent energies in these regions with a tendency for slightly larger turbulent energies. This is different to turbulent-to-magnetic field energy ratios found in younger still starless regions where $\beta$ was estimated $<<1$ (e.g., \citealt{beuther2018}), indicative of a bigger importance of the magnetic field at earlier evolutionary stages. 

\begin{figure}[htb]
\includegraphics[width=0.49\textwidth]{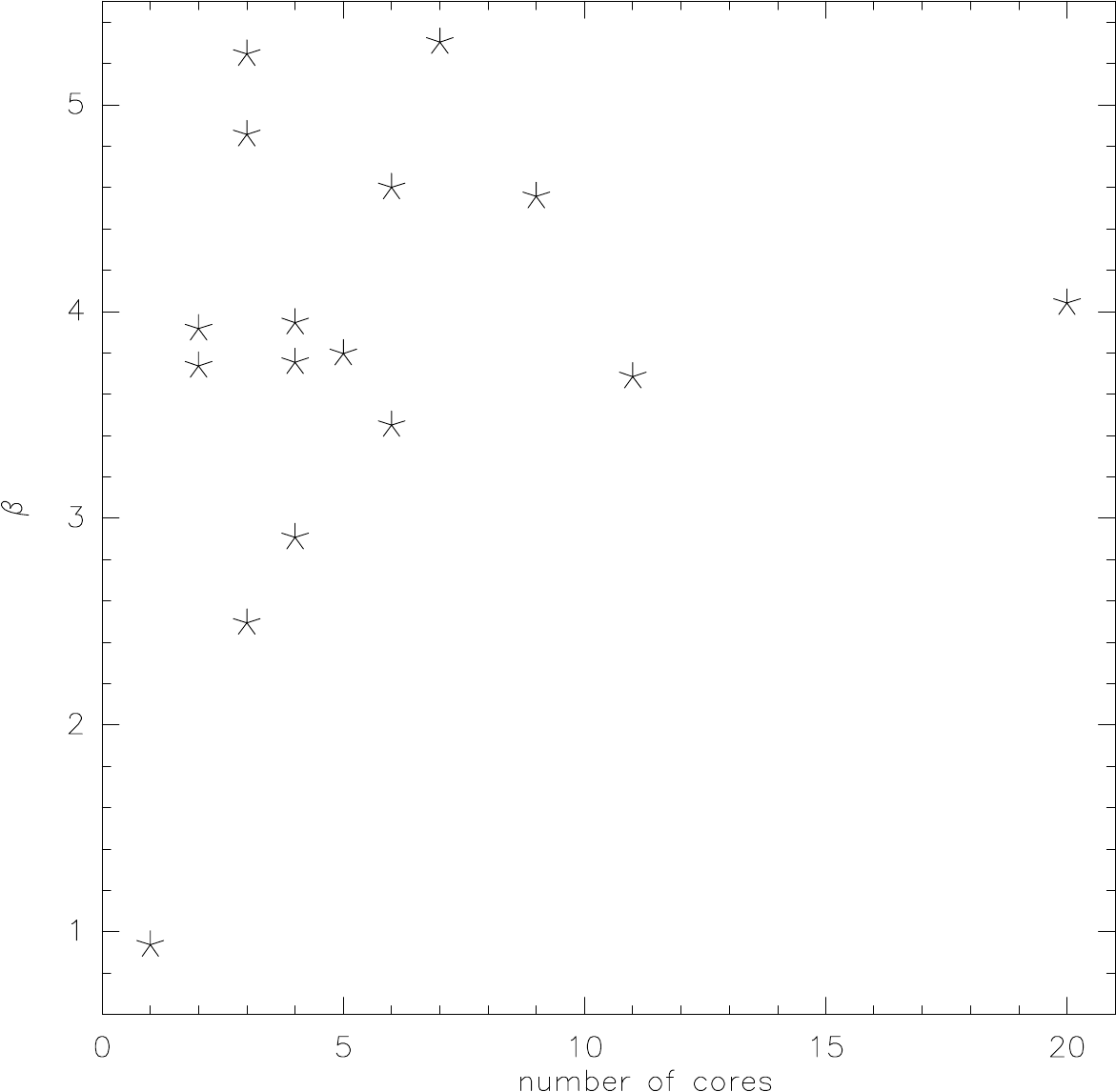}
\caption{Plot of the ratio of turbulent to magnetic energy $\beta$ versus number of cores. Approximate uncertainties on $\beta$ are around a factor 4.}
\label{beta} 
\end{figure} 

In addition to sample estimates, we can also derive $\beta$ for each region and plot that against the number of fragments which is shown in Figure \ref{beta}. While there is a scatter in $\beta$ between $\sim$1 and $\sim$5.4 for fragmented regions with less than 10 cores, the two regions with greater 10 cores, for which we again also have a $\beta$ measurement, have both turbulent to magnetic energy ratios $\beta$ around 4. Such a trend could be interpreted in a direction that more kinetic energy may favor more fragments. However, this trend remains inconclusive with the given uncertainties.

\begin{figure}[htb]
\includegraphics[width=0.49\textwidth]{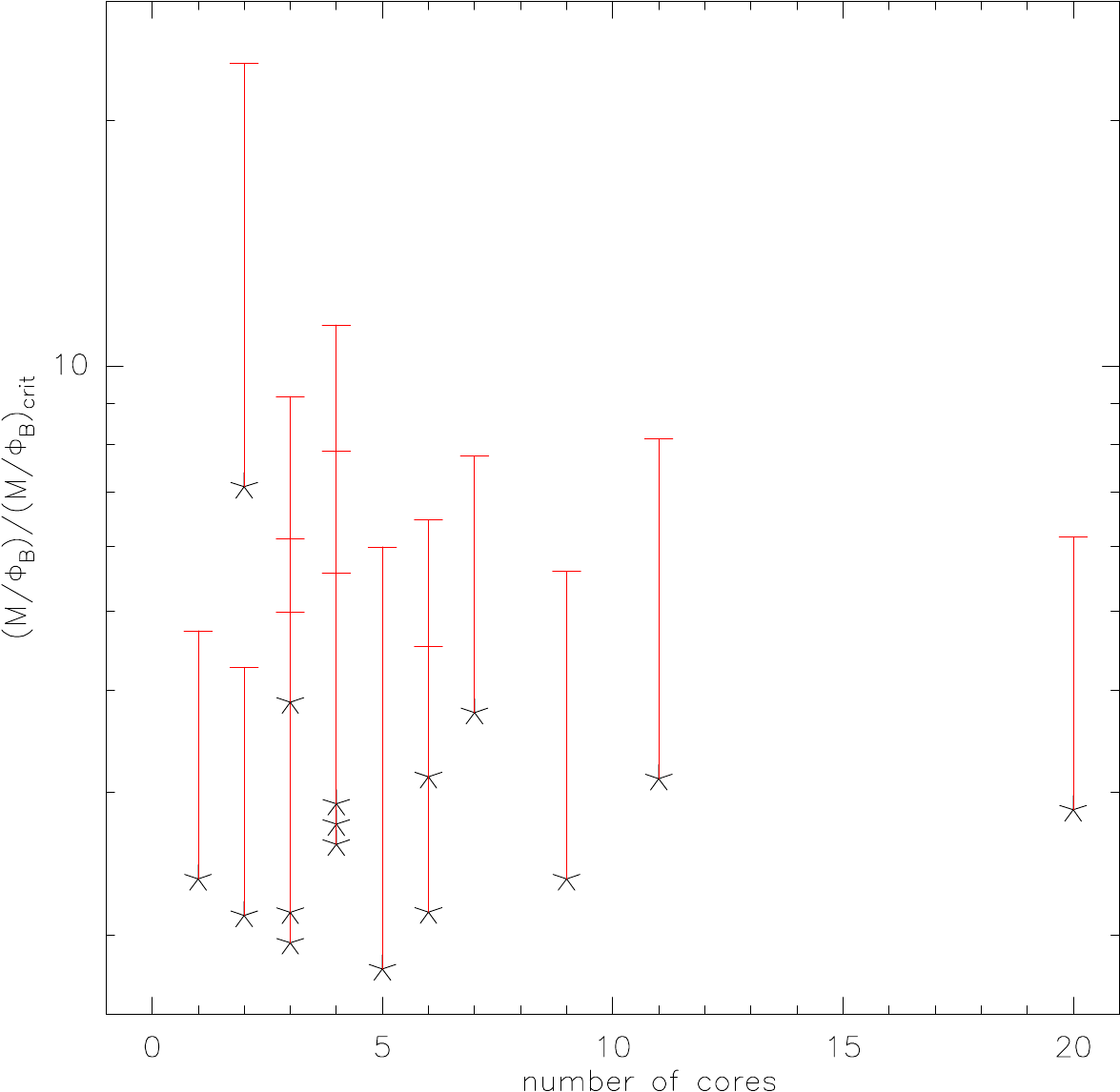}
\caption{Plot of the mass-to-flux ratio versus number of cores. For clarity in the logarithmic plotting, the red errorbars are only shown one-sided to higher values.}
\label{mass_flux} 
\end{figure} 

A different way to assess the stability of the regions against collapse is estimating the mass-to-flux ratio $M/\Phi_B\sim 7.6\times 10^{-24}\frac{N_{\rm H2}}{B}$ with the mass $M$, the magnetic flux $\Phi_B$, the column density $N_{\rm H2}$ and the magnetic field strength $B$. The latter two are in units of cm$^{-2}$ and mG, respectively (e.g., \citealt{crutcher1999,troland2008}). The mass-to-flux ratio $M/\Phi_B$ is then given in units of the critical mass-to-flux ratio $(M/\Phi_B)_{\rm crit}$ where values greater/smaller than 1 correspond to collapsing or more stable configurations, respectively. For the magnetic field $B$, we use $B_{\rm pos}(\sigma_{\psi})$ (Table \ref{polarization_parameters}). For the column density $N_{\rm{H2}}$, we use the mean column densities derived from the 1.2\,mm single-dish data in Sect.~\ref{density} (Table \ref{obs}).
To get an estimate for the uncertainties of $M/\Phi_B$, we apply Gaussian error propagation assuming a mean error on $B$ of $\sim$0.63\,mG (Fig.~\ref{num_cores_magn_field}) and an uncertainty on $N_{\rm{H2}}$ of same order as the measured mean column density.
The estimated mass-to-flux ratios $M/\Phi_B$ are listed in Table \ref{polarization_parameters}. All mass-to-flux ratios are greater than 1, ranging between $\sim$ 2 and $\sim$ 7. Figure \ref{mass_flux} plots $M/\Phi_B$ versus the number of cores within each region. Even considering the given uncertainties on $M/\Phi_B$, for all regions the mass-to-flux ratio is $\geq 2$, consistent with all regions already collapsing and actively forming stars.  Recently, \citet{palau2021} proposed a tentative positive correlation between the mass-to-flux ratio and the fragmentation level. For comparison, we do not see any clear trend between $M/\Phi_B$ and the fragmentation of the regions, and hence can neither confirm nor reject that tentative relation found before.

We also checked whether there could be any correlation between the magnetic field strength and the density structure measured on large and small spatial scales with the IRAM 30\,m and NOEMA, respectively (Fig.~\ref{p_pi}). However, no correlation could be identified, which indicates that the clump and core density structures of the regions are largely independent of the magnetic field strength.

\section{Conclusions and summary}

With the goal to identify and characterize the main physical processes determining the fragmentation properties of high-mass star-forming regions, we have investigated the large-scale density distributions and the magnetic field properties of the 20 high-mass star-forming regions constituting the CORE sample by means of IRAM 30\,m continuum and SMA polarization observations. Although the data allow diverse other investigations, e.g., large-scale environmental studies with the IRAM 30\,m data or chemical and physical analyses of the corresponding SMA spectral line data, here we concentrate on the Stokes $I$ and linearly polarized continuum emission to characterize the density and magnetic field structure and set that into context with the fragmentation properties previously derived by the high-angular-resolution PdBI/NOEMA data of the CORE program \citep{beuther2018b}.

The IRAM\,30\,m 1.2\,mm dust continuum data allow us to infer the density structure of the regions. While the measured intensity profiles on larger scales ($>35''$) exhibit a steepening, this is an observational artifact caused by the continuum mapping approach filtering out emission beyond those scales. However, analysis of the inner intensity profiles reveals that they should be accurate within $\Delta m_i\sim 0.3$, resulting in similar uncertainties for the inner density structure $p_i$. To first order, we find no correlation between the estimated density structures and the number of fragments (see also \citealt{palau2014}). This may partially be due to the evolutionary stage because the regions are typically in the high-mass protostellar object stage and may not reflect the initial conditions anymore. However, it remains to be investigated whether the density structures do change significantly on the given timescales of $\sim$10$^4$\,yrs, or whether they stay relatively constant in the framework of an inside-out collapse. In the latter cases, the given observations could still resemble the initial conditions comparably well.

Additionally, we compare the large-scale density structures with the small-scale density distributions previously derived from the interferometer data by \citet{gieser2021}. Interestingly, we find that the large-scale density distributions are typically flatter (power-law slopes around $\sim$1.5) than the small-scale density structure (typical power-law slopes around $\sim$2.0). A similar steepening of the density structure toward smaller scales was recently also identified for other regions with ALMA observations by \citet{gieser2023}. Possible reasons for this behaviour are that the star-forming regions are still embedded in larger cloud structures, causing a density profile flattening on larger scales. Furthermore, as discussed for simulations by, e.g., \citet{gomez2021}, even starting with initially flatter density distributions, the collapse of the regions steepens the profiles typically approaching slopes around 2.0. Hence, the measured large-scale flatter intensity profiles may resemble a bit more the initial structures, where the smaller-scale profiles are caused by the collapse motions. 

Regarding magnetic field structure, we detect significant polarization signal in 16 out of 20 regions. This is one of the largest high-resolution magnetic field studies today, for comparable SMA samples see \citet{zhang2014b} or \citet{palau2021}. With ALMA, now larger samples with high angular resolution are started to be studied as well (e.g., \citealt{sanhueza2021,cortes2021}). Within our CORE sample we find magnetic field structures appearing aligned with filaments leading toward the dense central cores. However, because of the still relatively low number of individual polarization measurements within individual regions, a statistical quantification of this behaviour, e.g., with the histogram of oriented gradients, is still limited. 

More than half of the observed polarization fractions are below 4\%, with 88 percent being below 10\%. While some regions exhibit polarization holes toward the Stokes $I$ peak positions, as previously discussed in the literature, others show polarized emission also towards the intensity peaks. Nevertheless, the polarized intensities are inversely related to the Stokes $I$ intensities and roughly follow a power law slope $pol_{\rm frac} \propto S_I^{-0.62}$.

Estimating the magnetic field strength via a modified DCF method \citep{skalidis2021a}, we find a range of magnetic field strengths between $\sim$0.2 and $\sim$4.5\,mG. While the original dispersion of polarization angles may indicate a weak trend between angle dispersion and number of fragments, the derived magnetic field strength does not exhibit a clear trend with the number of fragments.
The mass-to-flux ratio varies between $\sim$2 and $\sim$7, consistent with all regions being collapsing and actively forming stars.
Comparing the estimated 1D turbulent and Alfvenic velocity dispersion as well as the turbulent to magnetic energy ratio, we find that the turbulent and magnetic energies in this sample are of similar importance. 

Coming back to our original question whether the level of fragmentation depends on the density and/or magnetic field structure, for this CORE sample we find no clear correlation between these parameters. Such a result may allow the tentative conclusion that apparently high-mass star-forming regions fragment almost independent of their initial density distribution or magnetic field properties.

However, the presented density and magnetic field analysis of the CORE sample is obviously limited by several parameters, e.g., evolutionary stage or spatial scales. Hence, this points to several different study approaches for forthcoming projects. Regarding the density structure analysis, similar studies of younger regions reflecting potentially better the initial conditions, e.g., infrared dark clouds, are needed to better study the connection of the initial conditions with the magnetic field properties. With respect to the magnetic field studies, the SMA data presented here investigate intermediate spatial scales for a source sample covering only a comparatively small range in evolutionary stages. Extending the sample size to younger and older regions will allow an even better investigation of evolutionary changes. Furthermore, it will be important to connect these intermediate spatial scales to larger cloud scales as well as smaller core and disk scales. For the larger cloud scale, the forthcoming NIKA2POL polarization capability at the IRAM\,30\,m will allow us to directly connect the SMA magnetic field measurements with the larger cloud-scale magnetic field structure. Regarding smaller spatial scales, higher resolution SMA observations as well as potential polarization at NOEMA are possible in the northern hemisphere. While the CORE sample is only accessible from the northern hemisphere, in the south, obviously ALMA allows much higher angular resolution studies of the innermost regions. All these opportunities outline the exciting prospects on the horizon.

\begin{acknowledgements}
The authors are grateful to the staff at Pico Veleta and Submillimeter Array observatories for their support of these observations. The Submillimeter Array is a joint project between the Smithsonian Astrophysical Observatory and the Academia Sinica Institute of Astronomy and Astrophysics and is funded by the Smithsonian Institution and the Academia Sinica. This work is based on observations carried out under project number 143-19 with the IRAM 30\,m telescope. We particularly like to thank Stefano Barta and Robert Zylka with helping in the NIKA2 data reduction process. RK acknowledges financial support via the Heisenberg Research Grant funded by the German Research Foundation (DFG) under grant no.~KU 2849/9. A.P.~is grateful to Gilberto Gómez and Enrique Vázquez-Semadeni for very insightful discussions. A.P.~acknowledges financial support from the UNAM-PAPIIT IG100223 grant, the Sistema Nacional de Investigadores of CONAHCyT, and from the CONAHCyT project number 86372 of the `Ciencia de Frontera 2019’ program, entitled `Citlalc\'oatl: A multiscale study at the new frontier of the formation and early evolution of stars and planetary systems’, M\'exico. ASM acknowledges support from the RyC2021-032892-I and PID2020-117710GB-I00 grants funded by MCIN/AEI/10.13039/501100011033 and by the European Union ‘Next GenerationEU’/PRTR, as well as the program Unidad de Excelencia María de Maeztu CEX2020-001058-M. JDS acknowledges funding by the European Research Council via the ERC Synergy Grant ``ECOGAL -- Understanding our Galactic ecosystem: From the disk of the Milky Way to the formation sites of stars and planets'' (project ID 855130).
\end{acknowledgements}

\bibliographystyle{aa}    

\begin{appendix}
\section{IRAM 30\,m data}

\begin{figure*}[htb]
\includegraphics[width=0.99\textwidth]{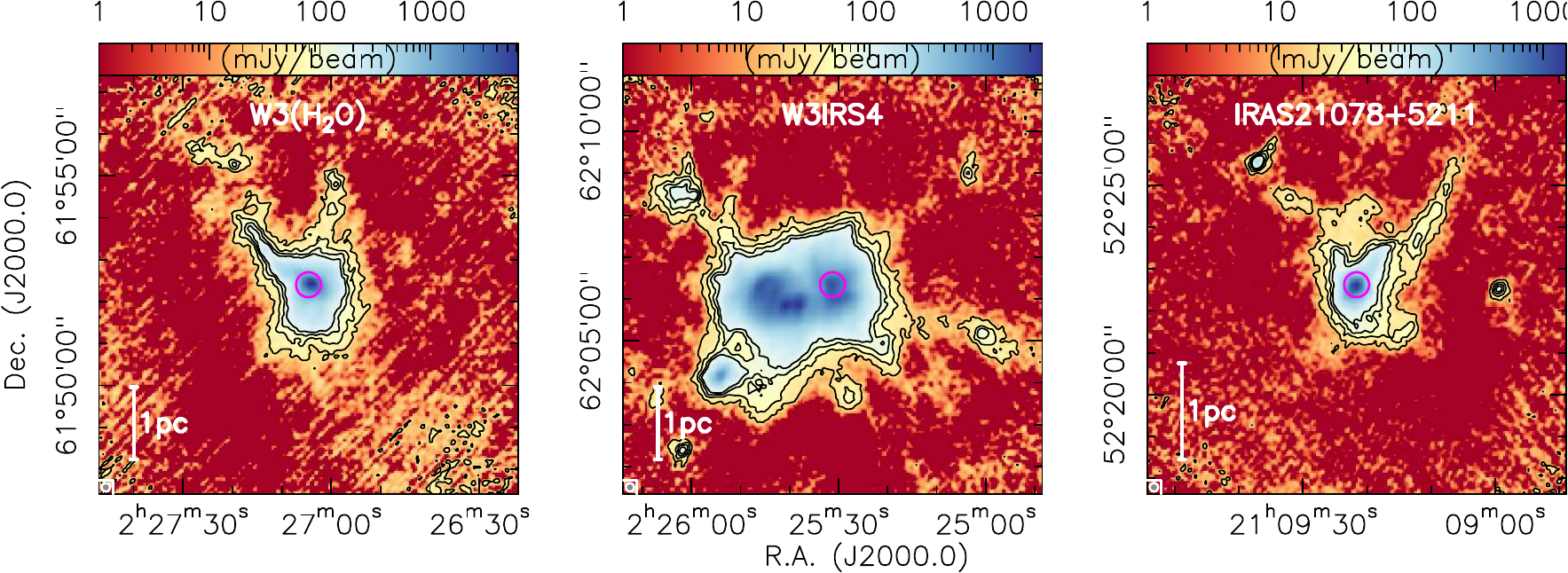}
\includegraphics[width=0.99\textwidth]{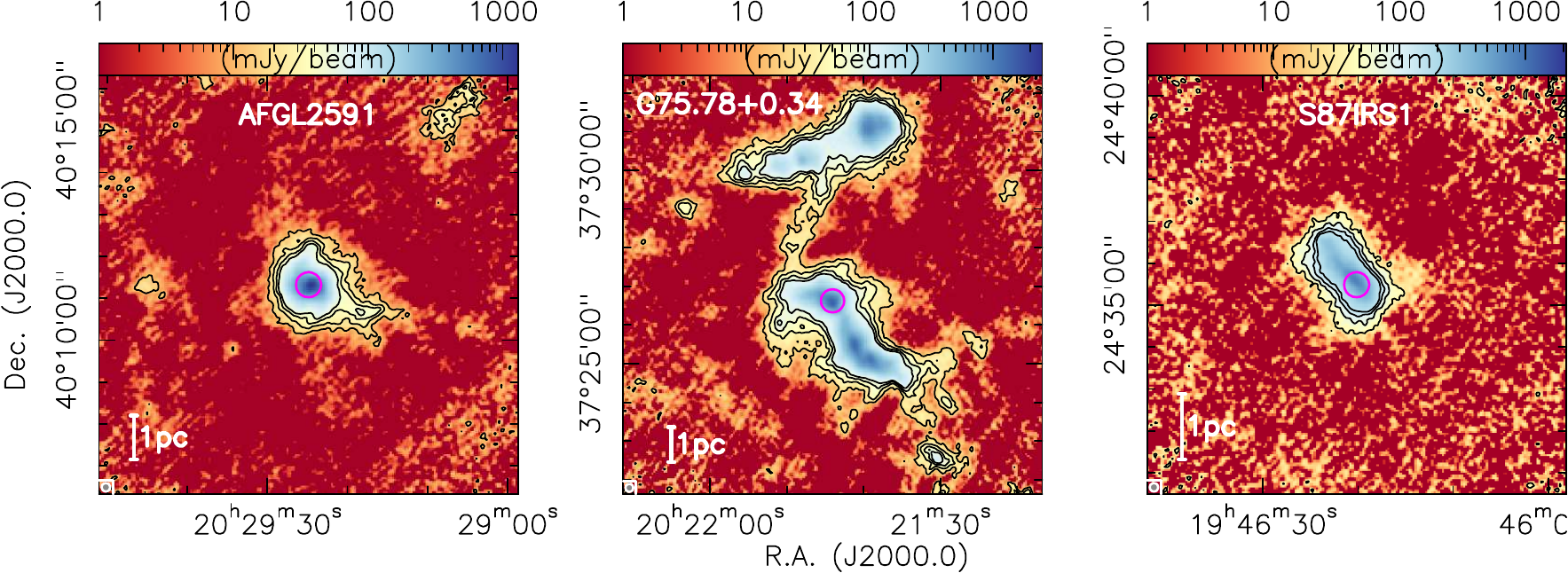}
\includegraphics[width=0.99\textwidth]{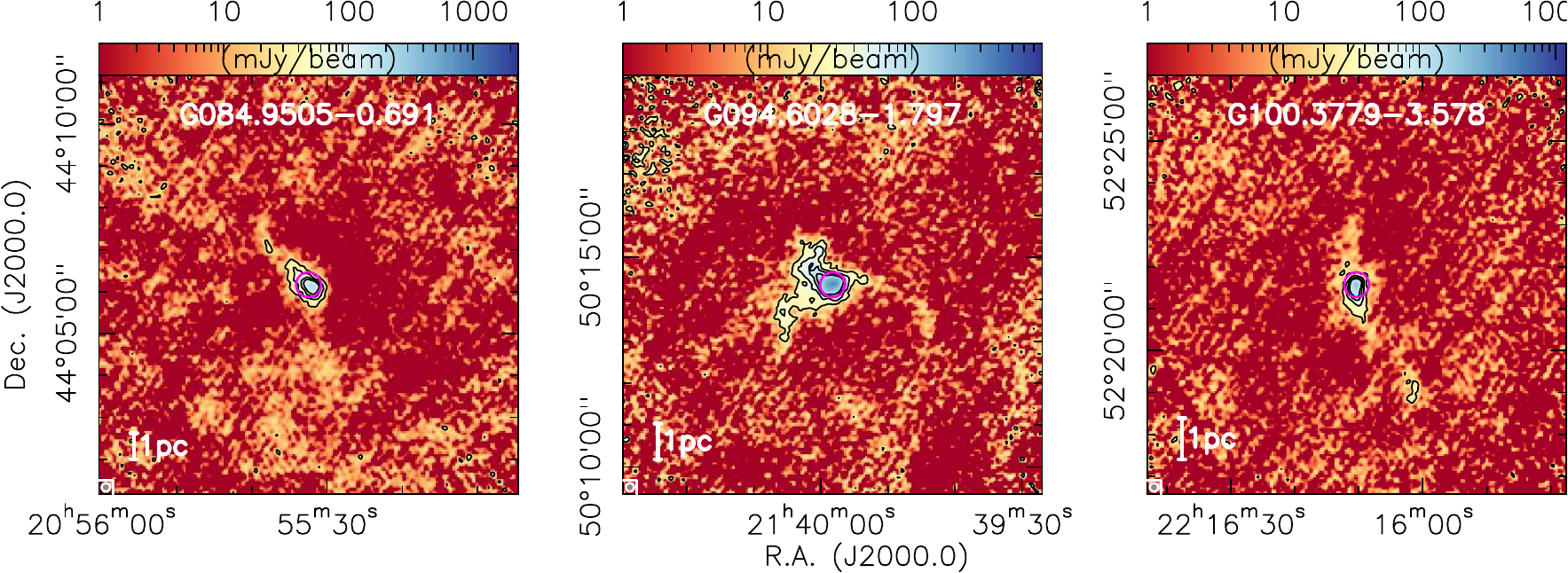}
\caption{NIKA2 1.2\,mm dust continuum images toward the CORE sample. The color-scale shows the flux densities, and the contours are always from $3\sigma$ to $12\sigma$ in $3\sigma$ steps. The $12''$ beam and a 1\,pc scale-bar are shown at the bottom left of each panel. The magenta circles outline the $\sim 36''$ primary beam size of the corresponding SMA observations.}
\label{nika2_2} 
\end{figure*} 

\begin{figure*}[htb]
\includegraphics[width=0.99\textwidth]{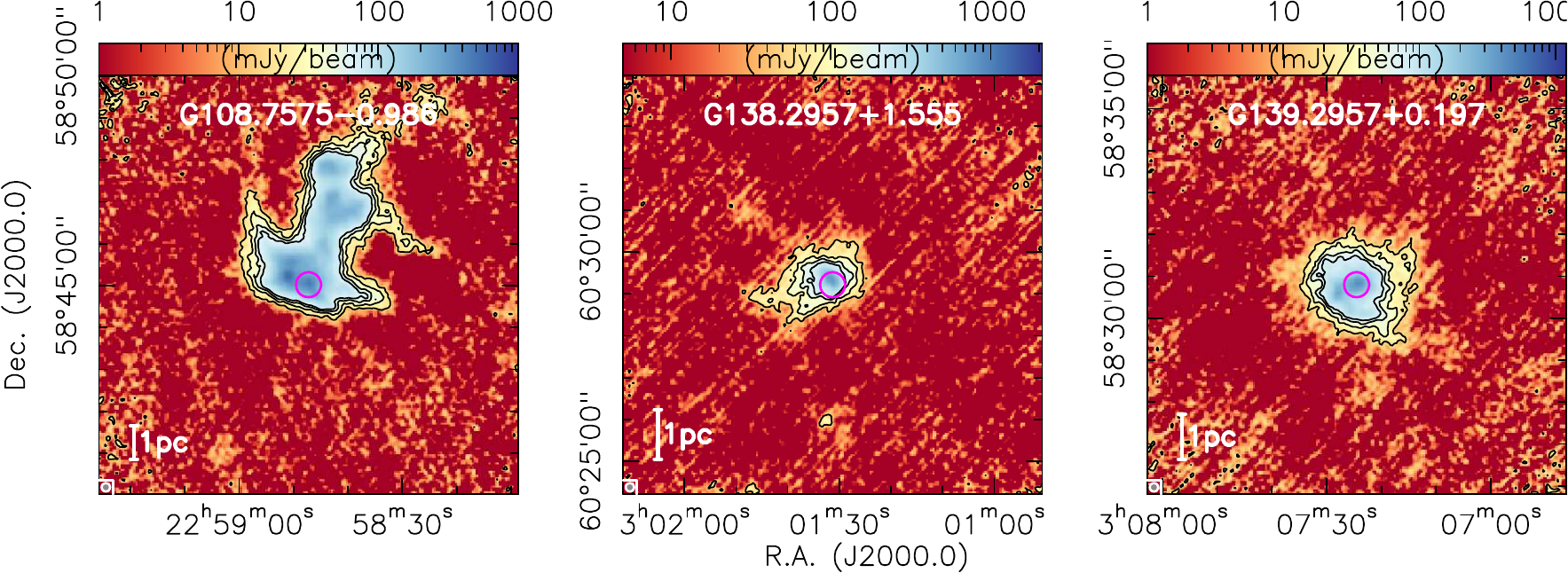}
\includegraphics[width=0.99\textwidth]{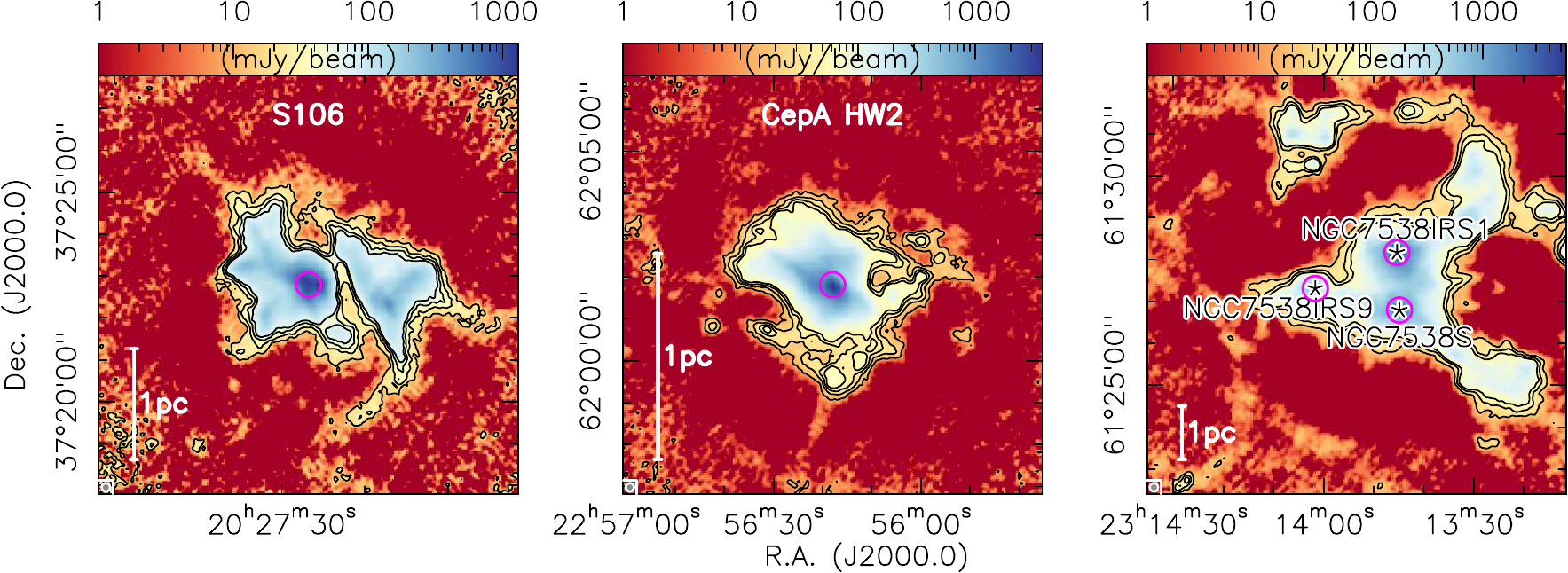}
\caption{NIKA2 1.2\,mm dust continuum images toward the CORE sample. The color-scale shows the flux densities, and the contours are always from $3\sigma$ to $12\sigma$ in $3\sigma$ steps. The $12''$ beam and a 1\,pc scale-bar are shown at the bottom left of each panel. The magenta circles outline the $\sim 36''$ primary beam size of the corresponding SMA observations.}
\label{nika2_3} 
\end{figure*} 

\begin{figure*}[htb]
\includegraphics[width=0.33\textwidth]{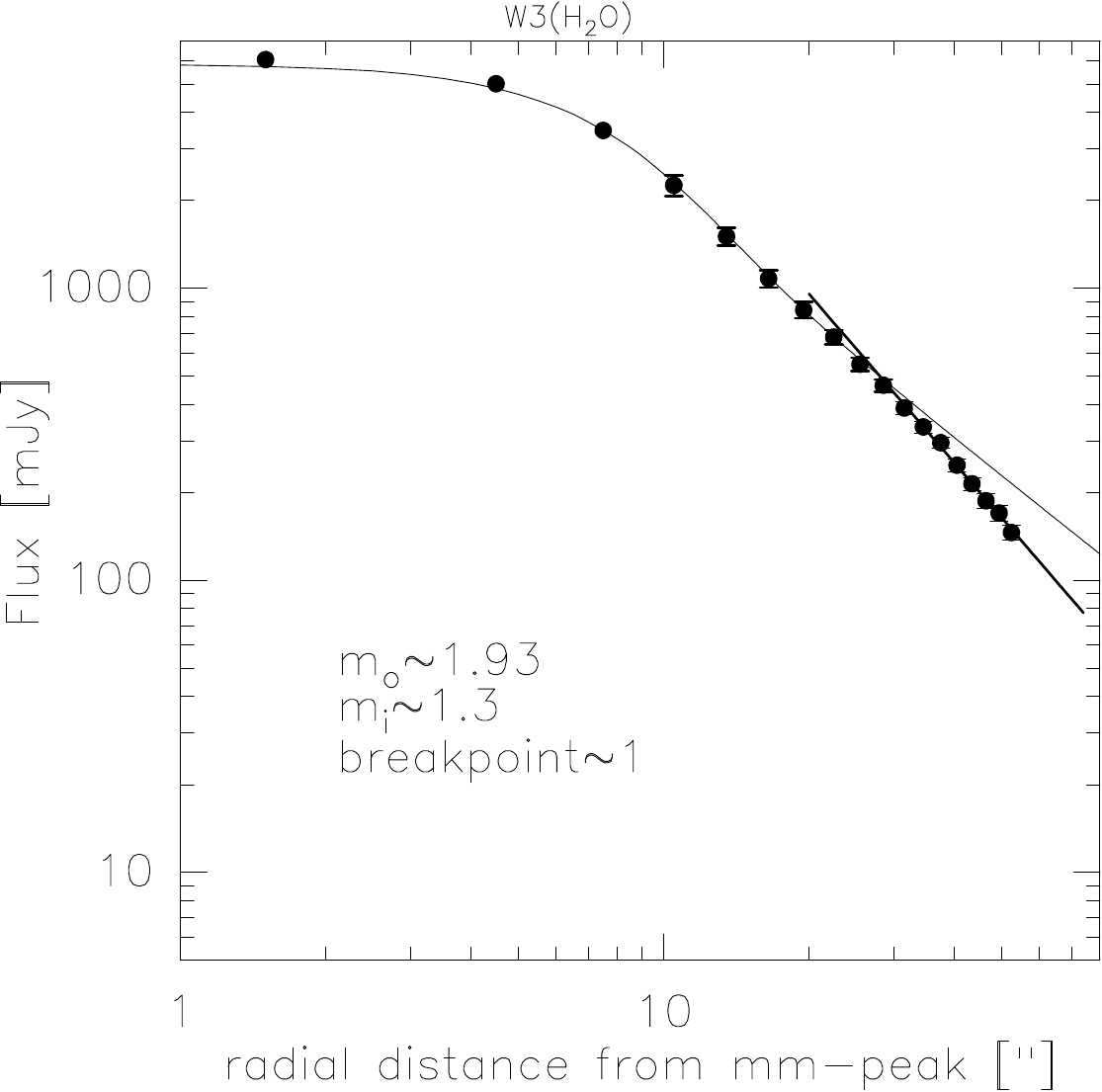}
\includegraphics[width=0.33\textwidth]{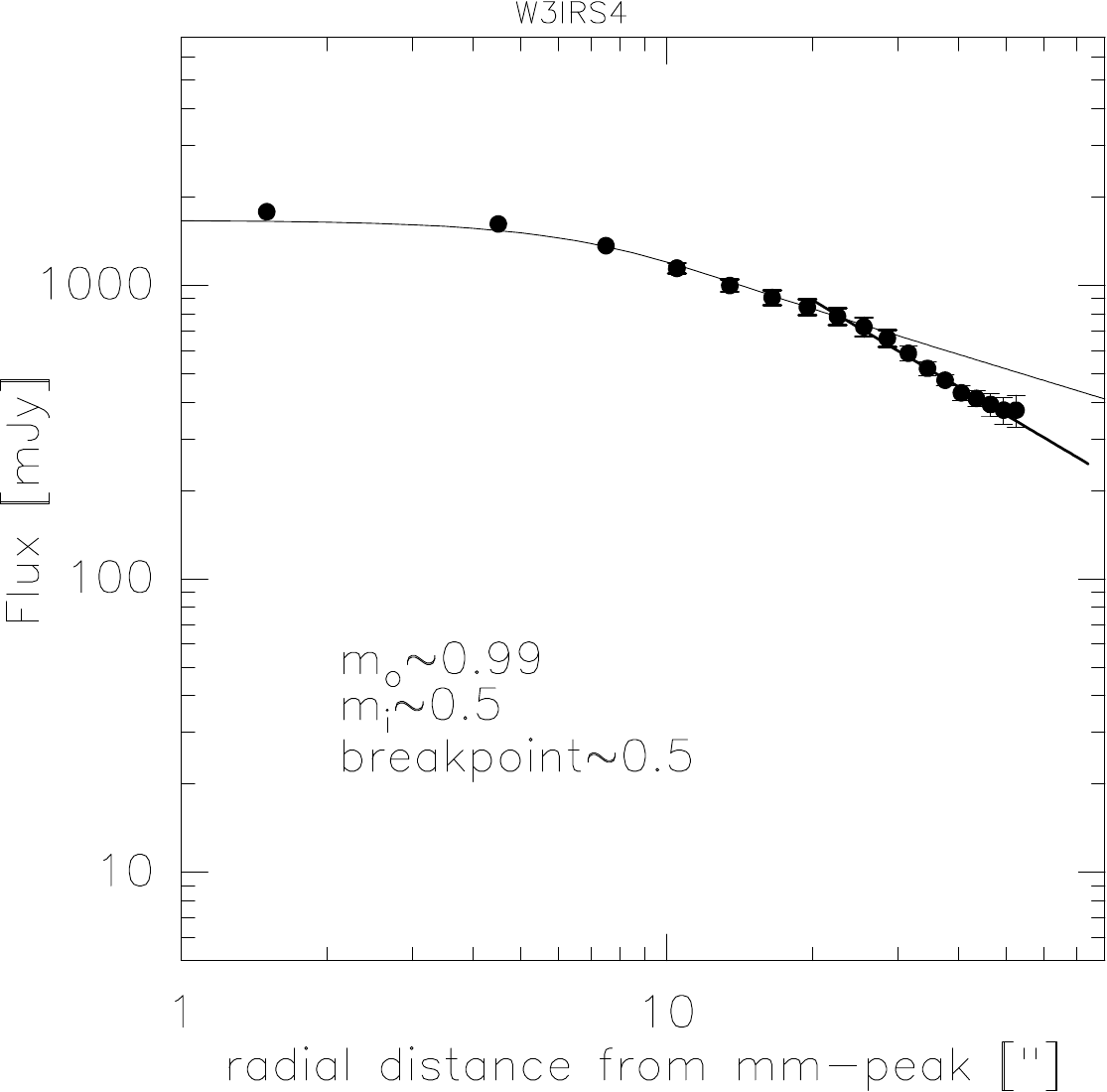}
\includegraphics[width=0.33\textwidth]{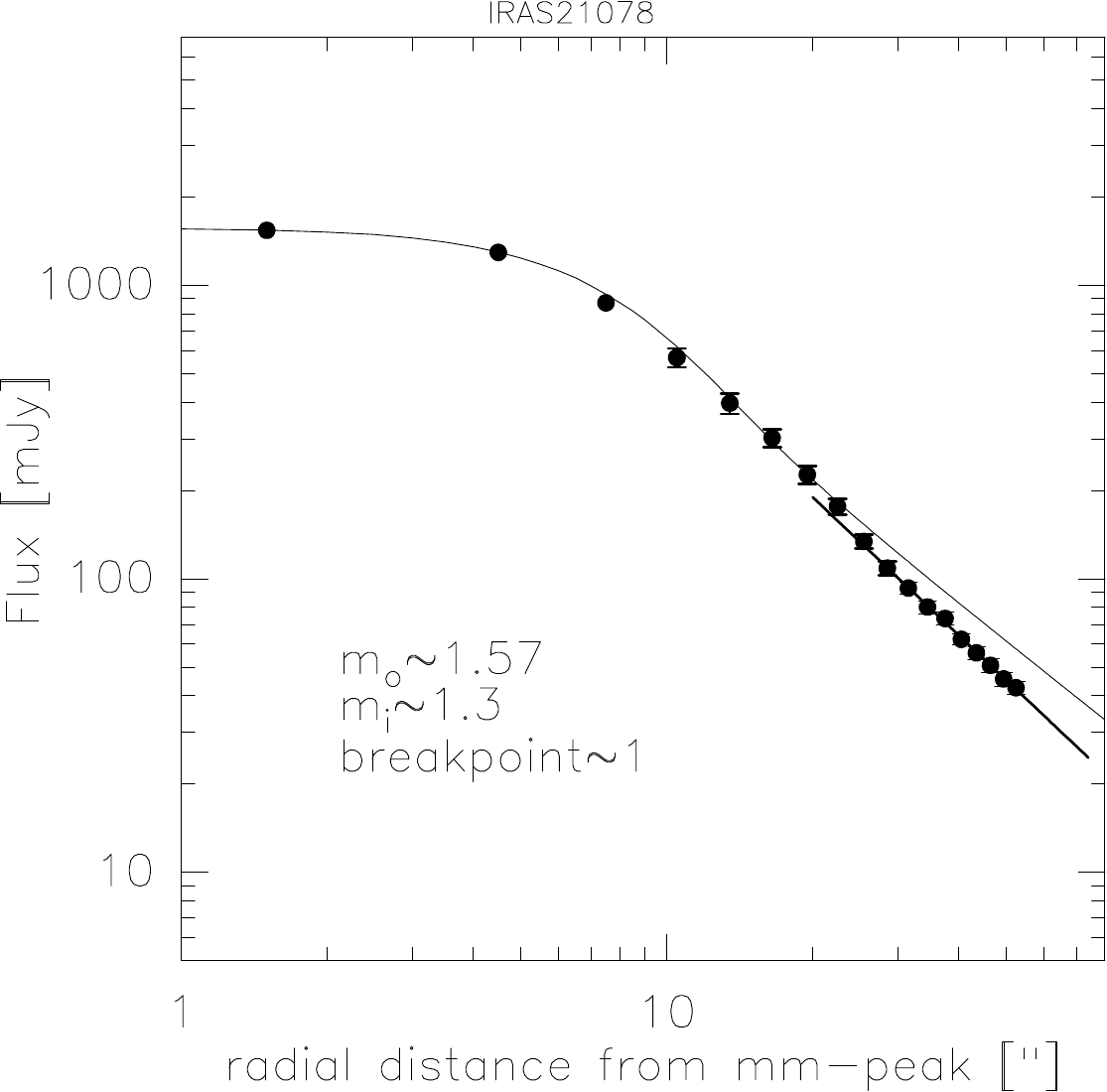}\\
\includegraphics[width=0.33\textwidth]{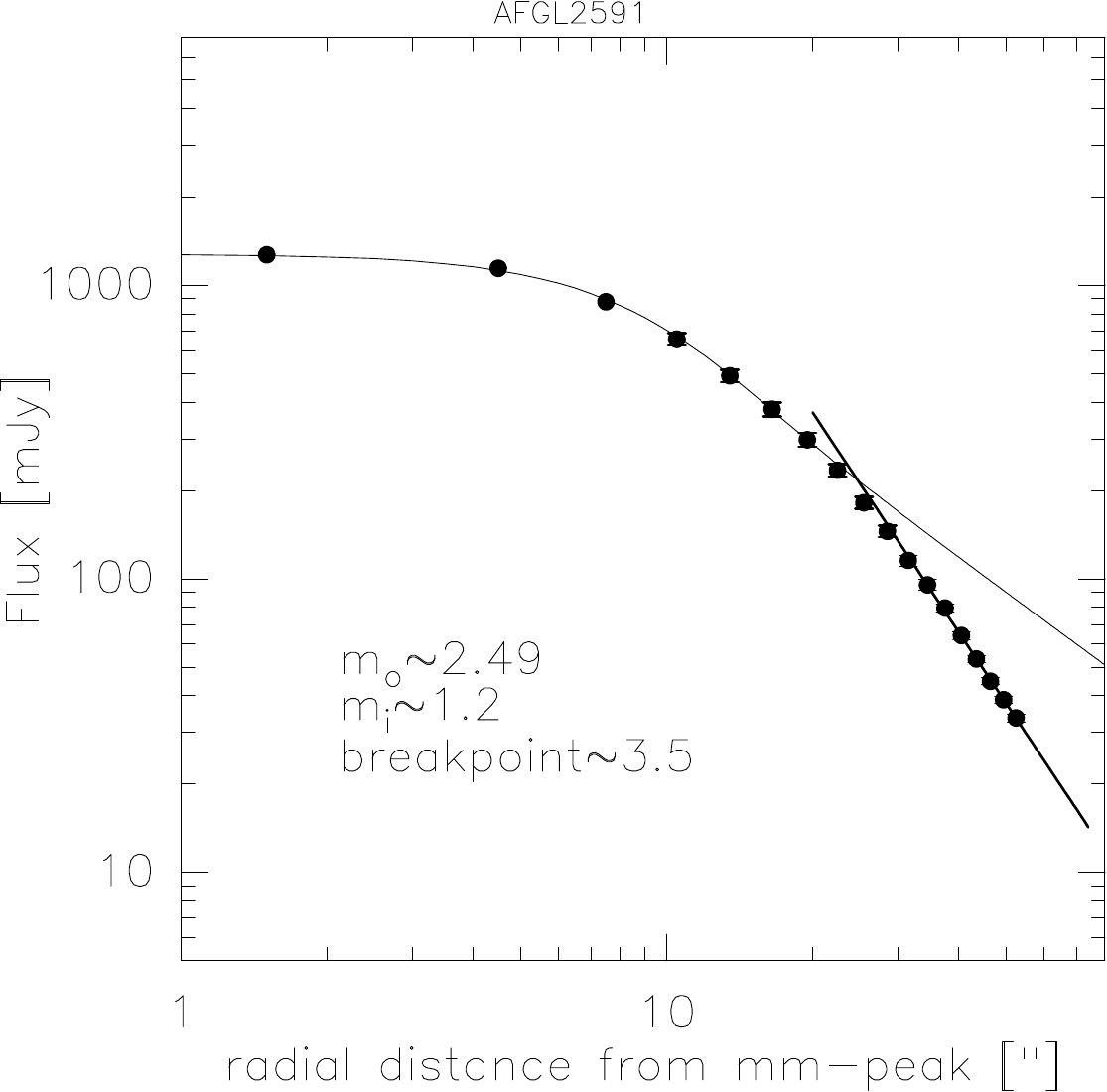}
\includegraphics[width=0.33\textwidth]{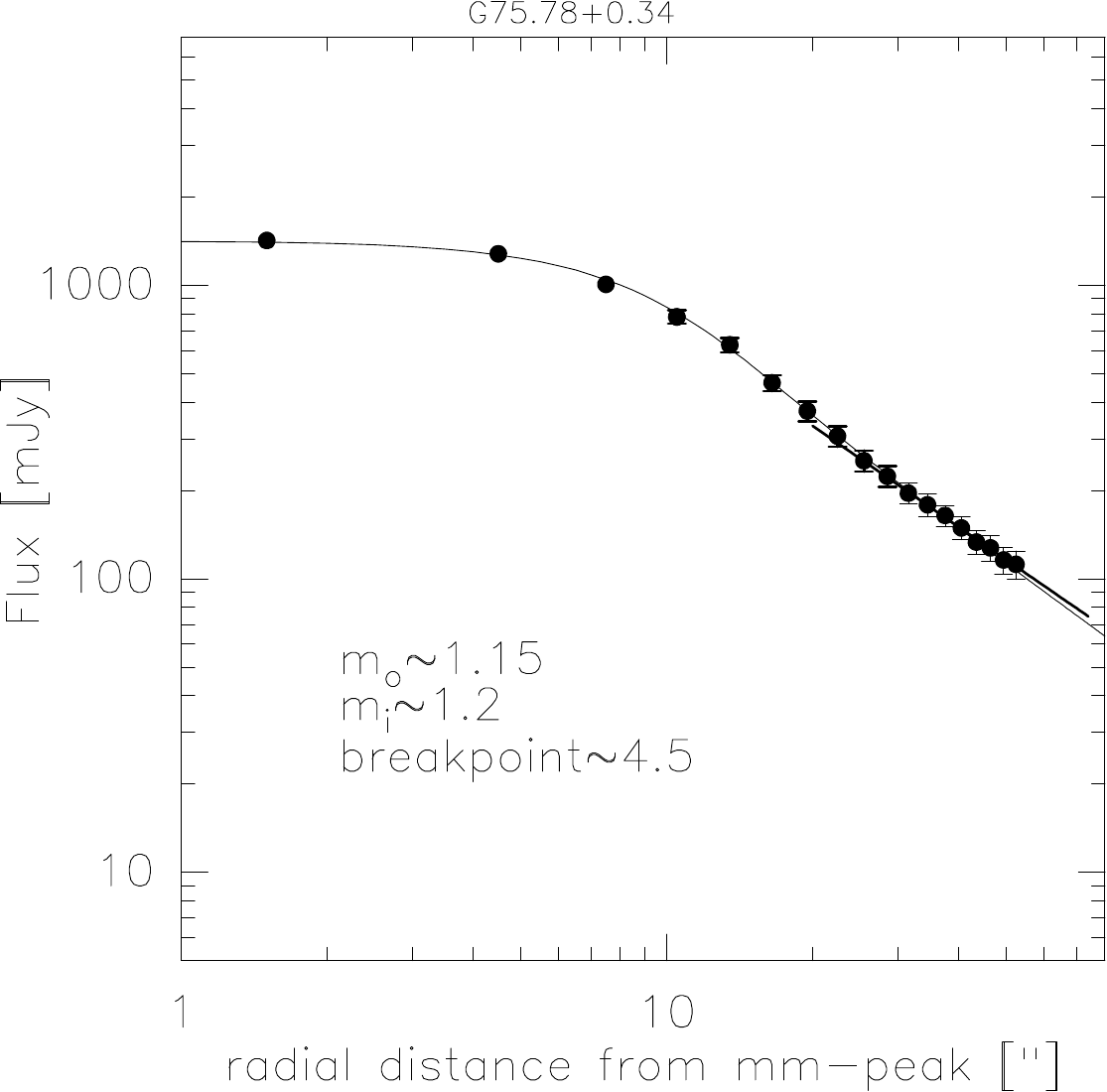}
\includegraphics[width=0.33\textwidth]{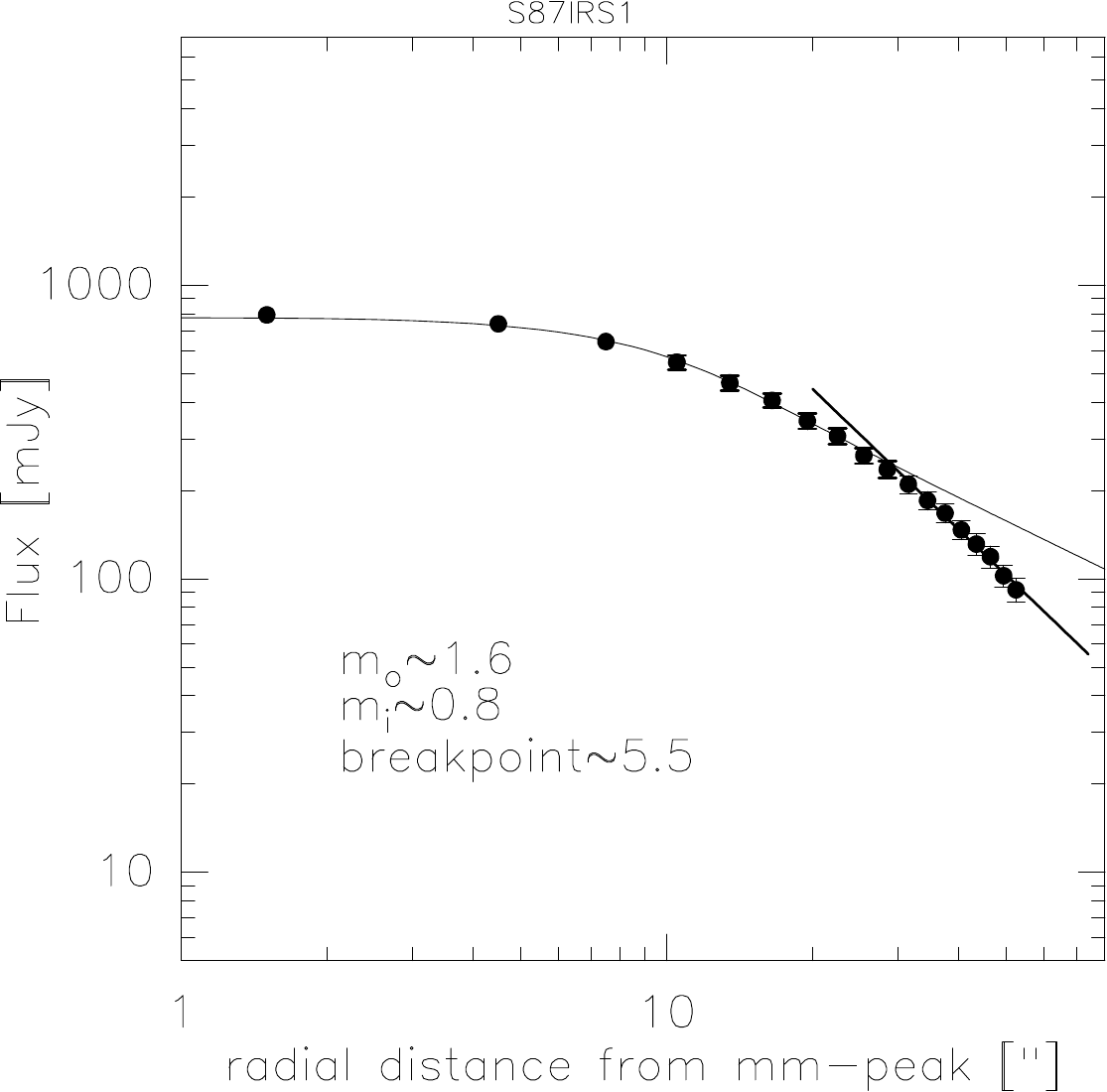}\\
\includegraphics[width=0.33\textwidth]{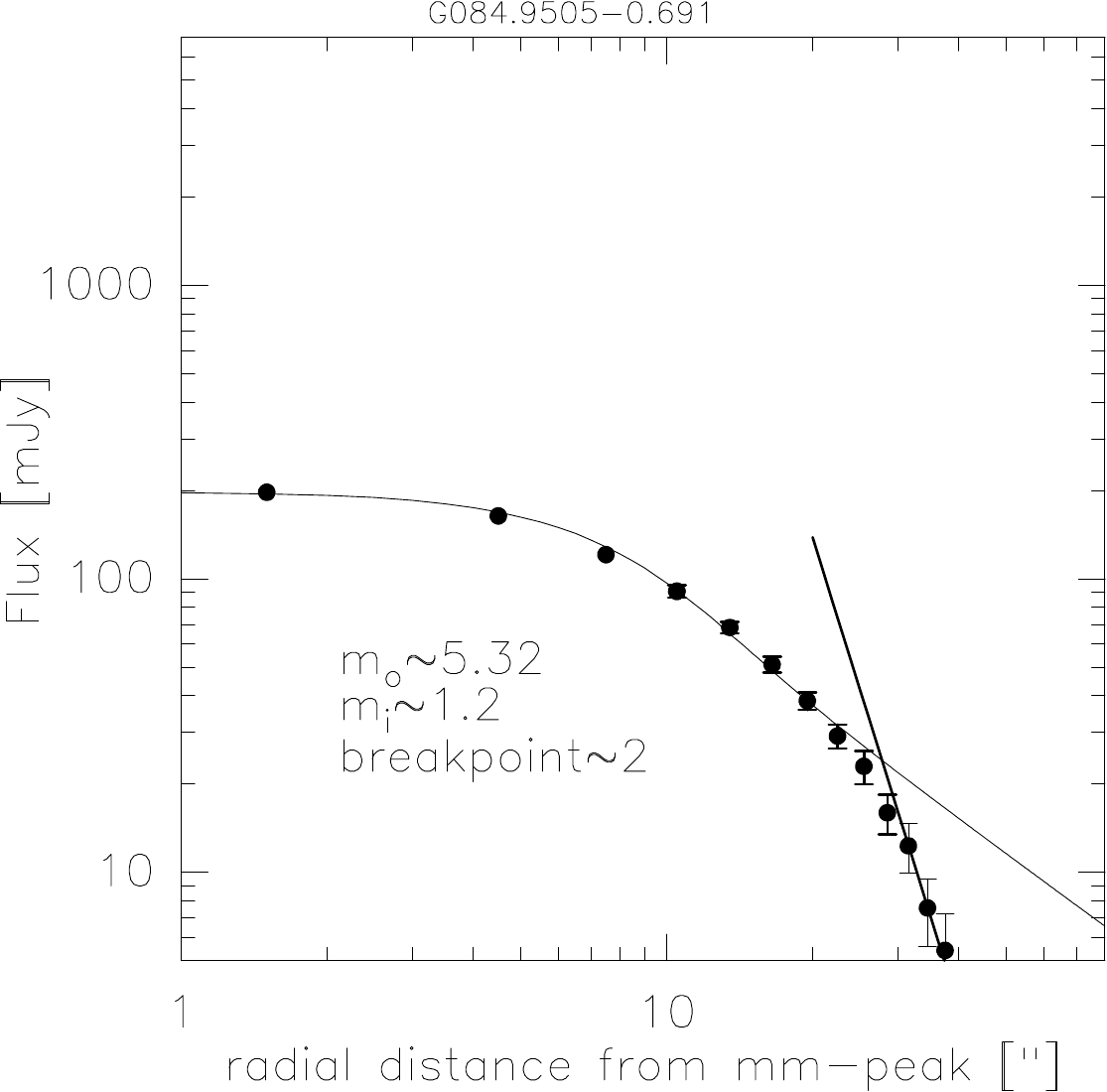}
\includegraphics[width=0.33\textwidth]{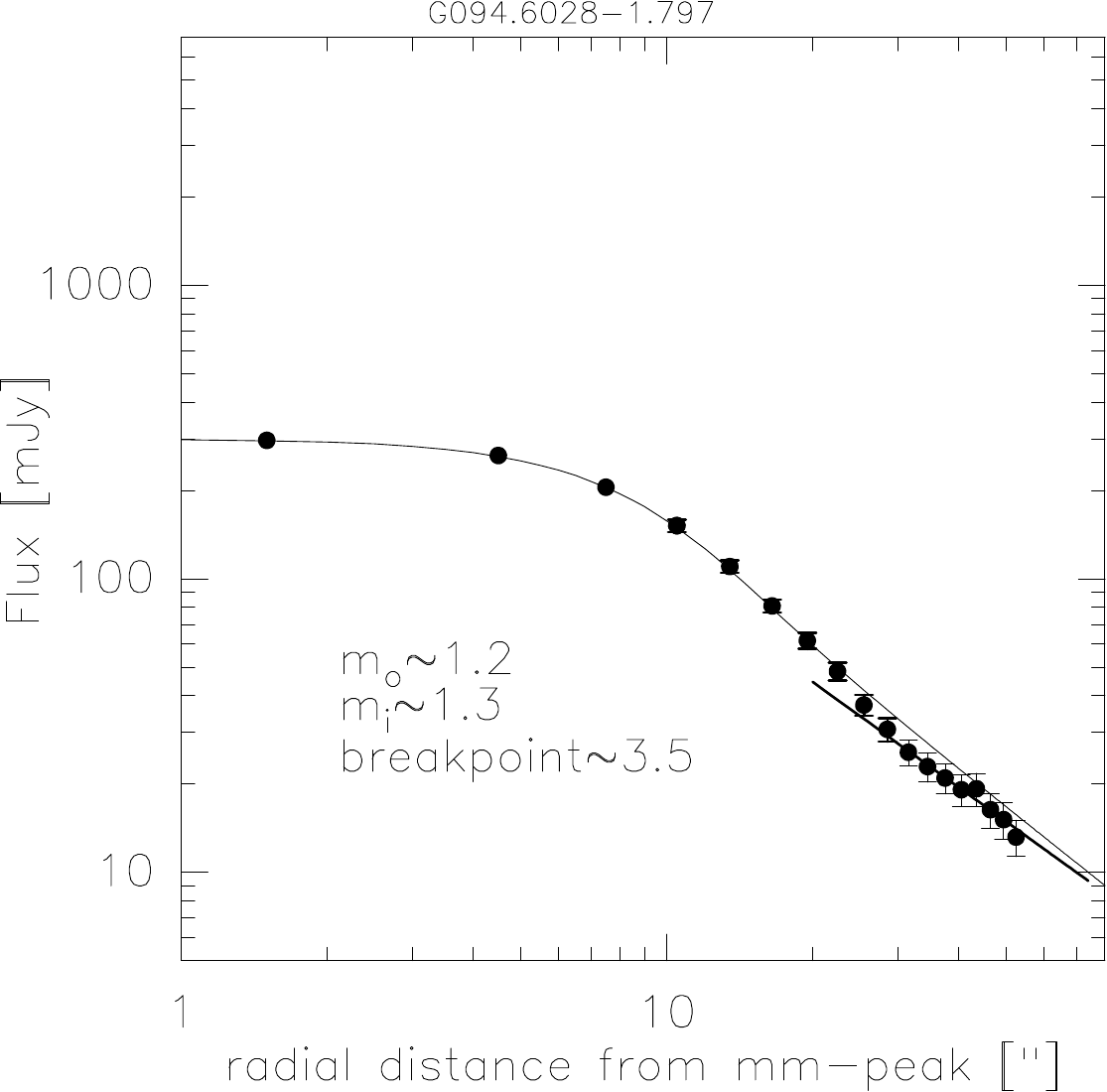}
\includegraphics[width=0.33\textwidth]{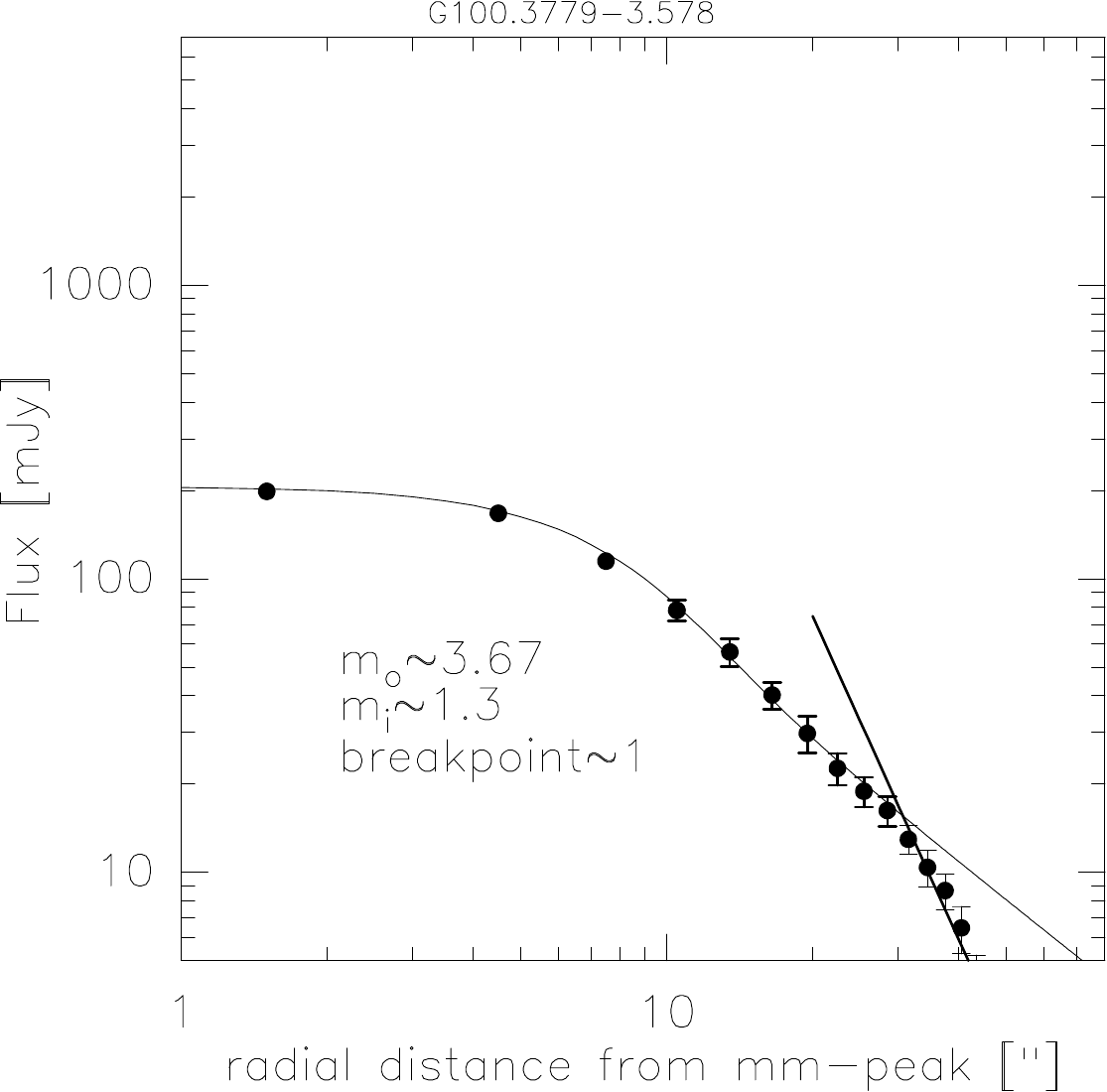}
\caption{Radial intensity profiles derived for the main 1.2\,mm dust continuum sources in the CORE sample. The power-law slopes of the inner and outer fits ($m_i$ and $m_o$) as well as the inner breakpoint  in arcsecond are labeled in each panel. See main text for fitting details. }
\label{profiles_2} 
\end{figure*} 

\begin{figure*}[htb]
\includegraphics[width=0.33\textwidth]{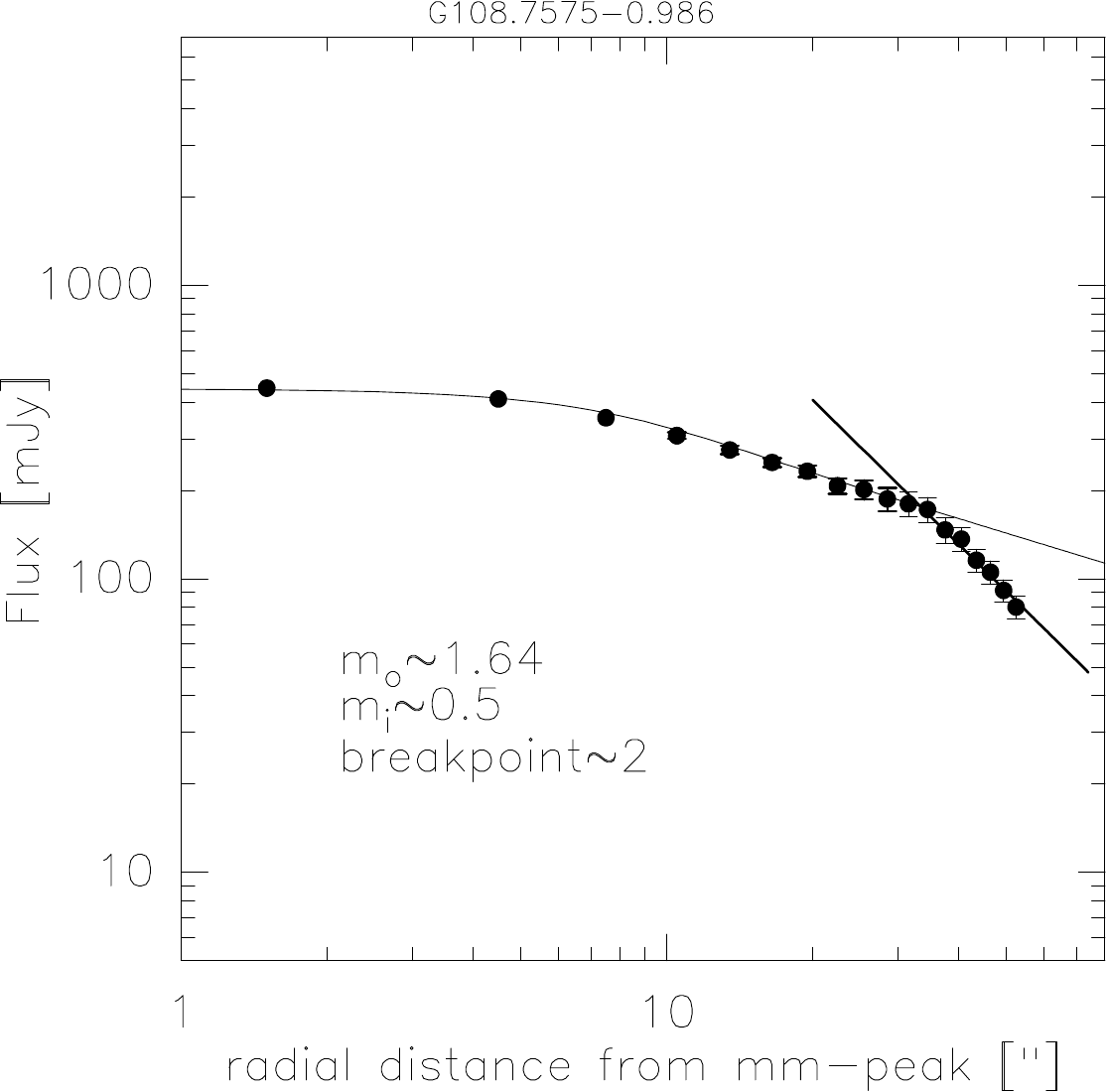}
\includegraphics[width=0.33\textwidth]{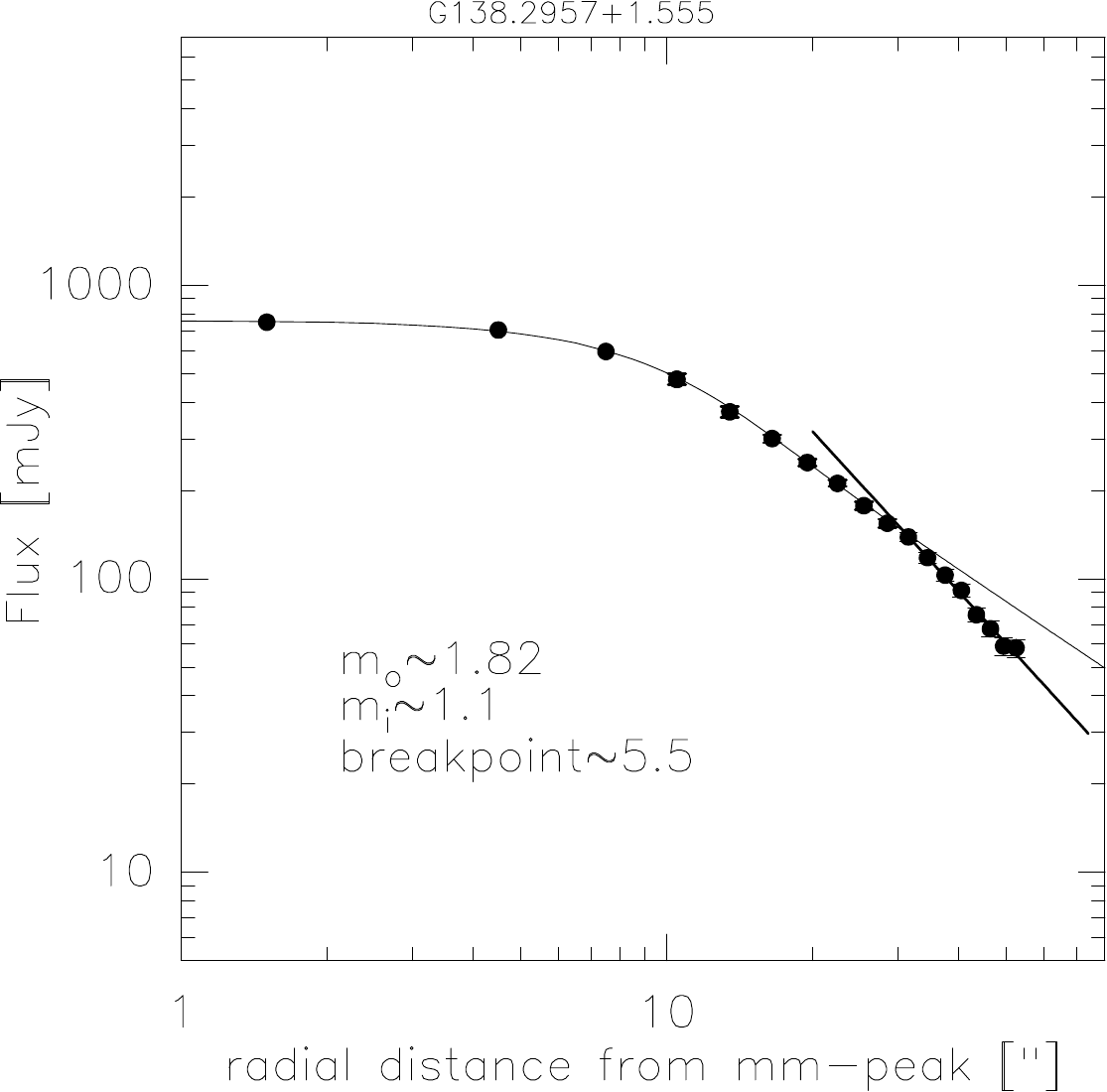}
\includegraphics[width=0.33\textwidth]{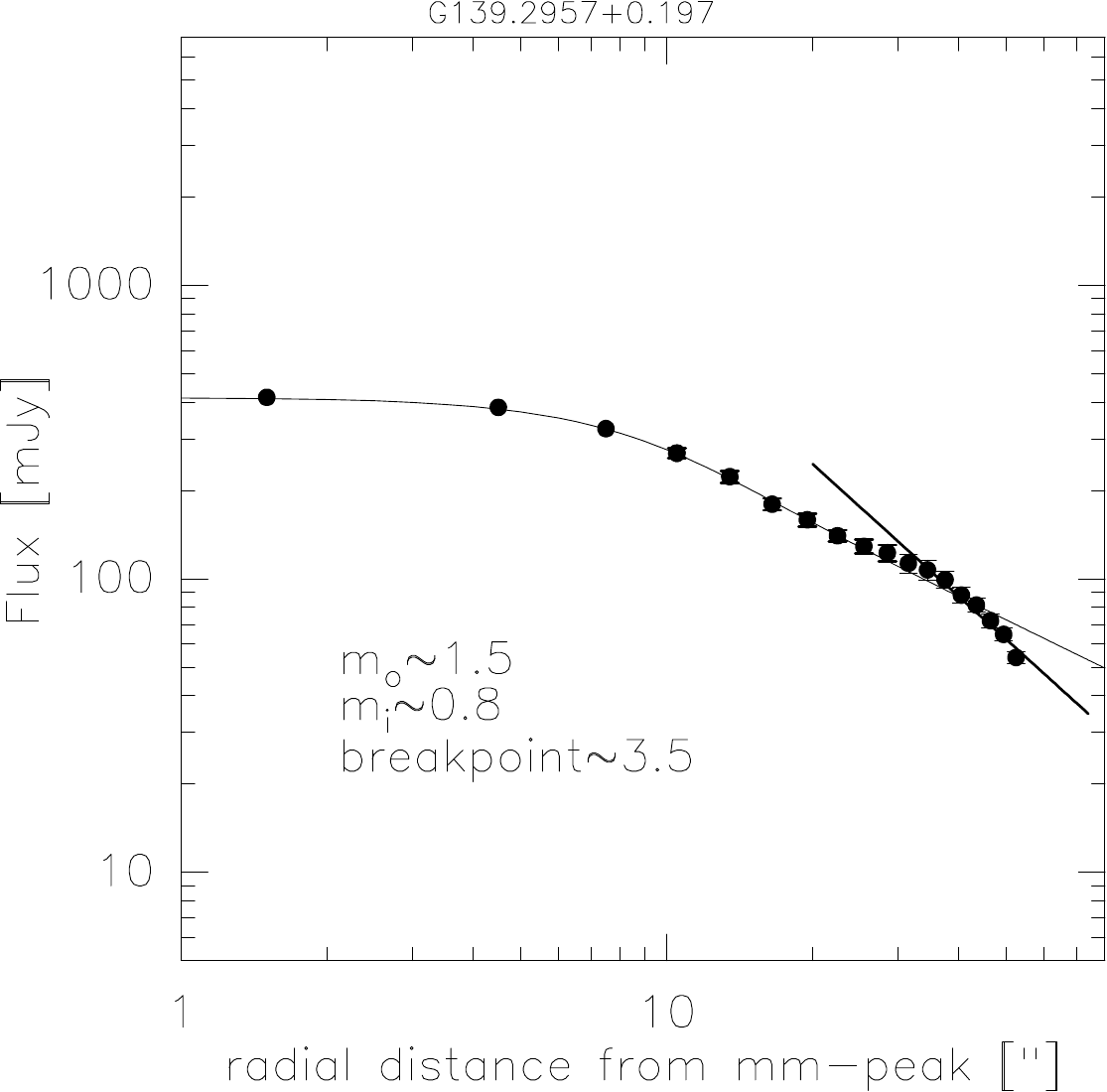}\\
\includegraphics[width=0.33\textwidth]{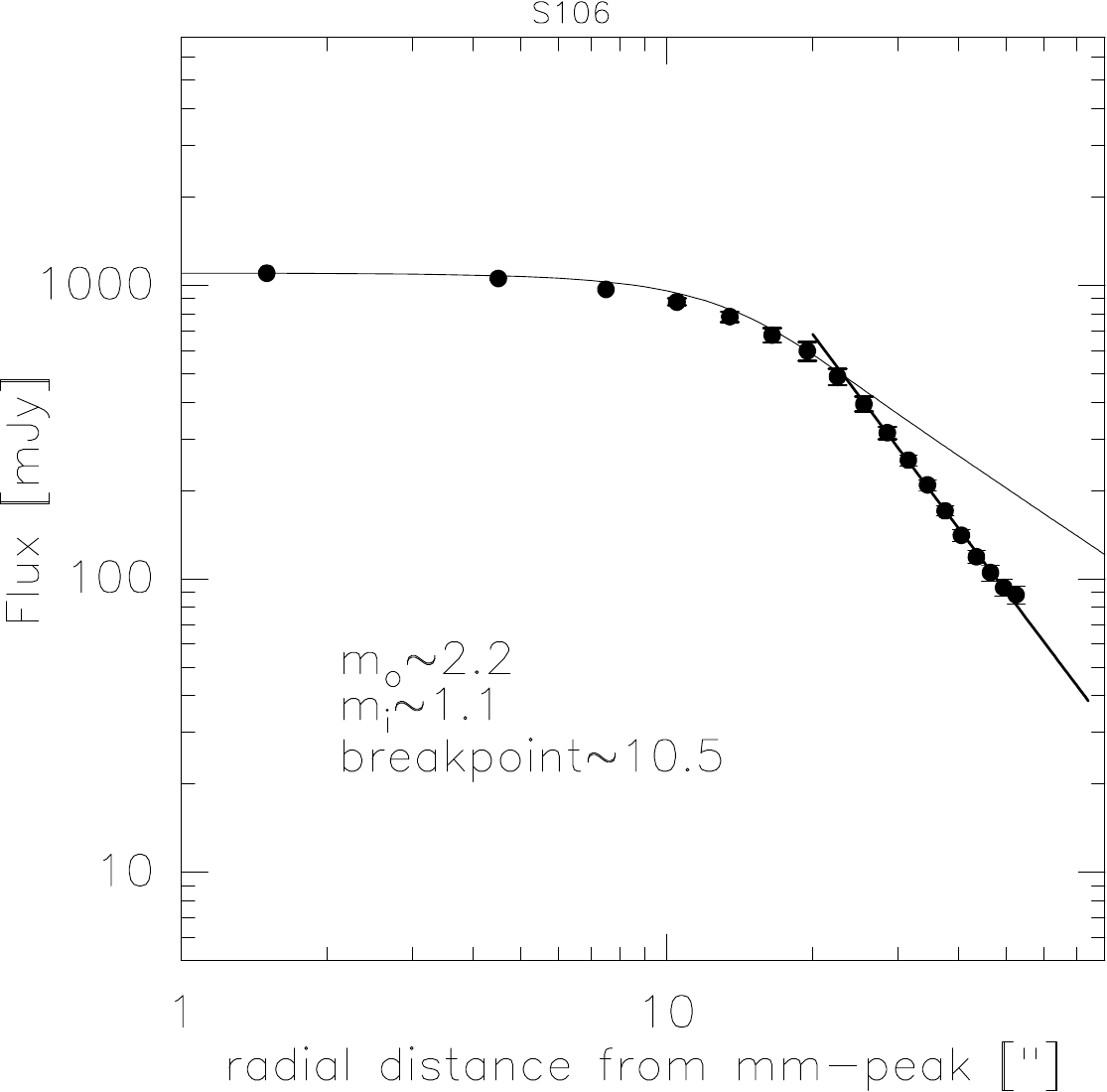}
\includegraphics[width=0.33\textwidth]{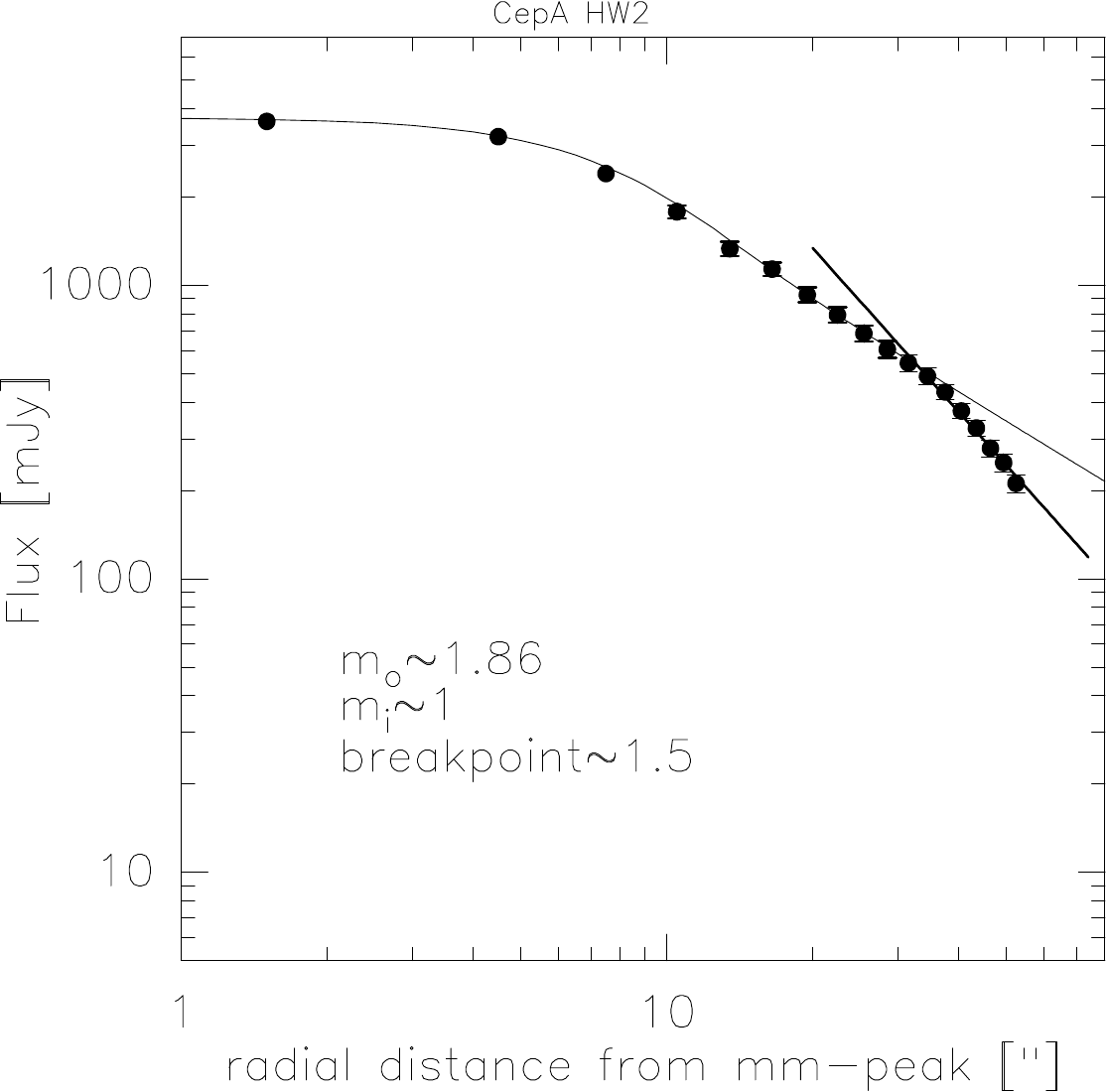}
\includegraphics[width=0.33\textwidth]{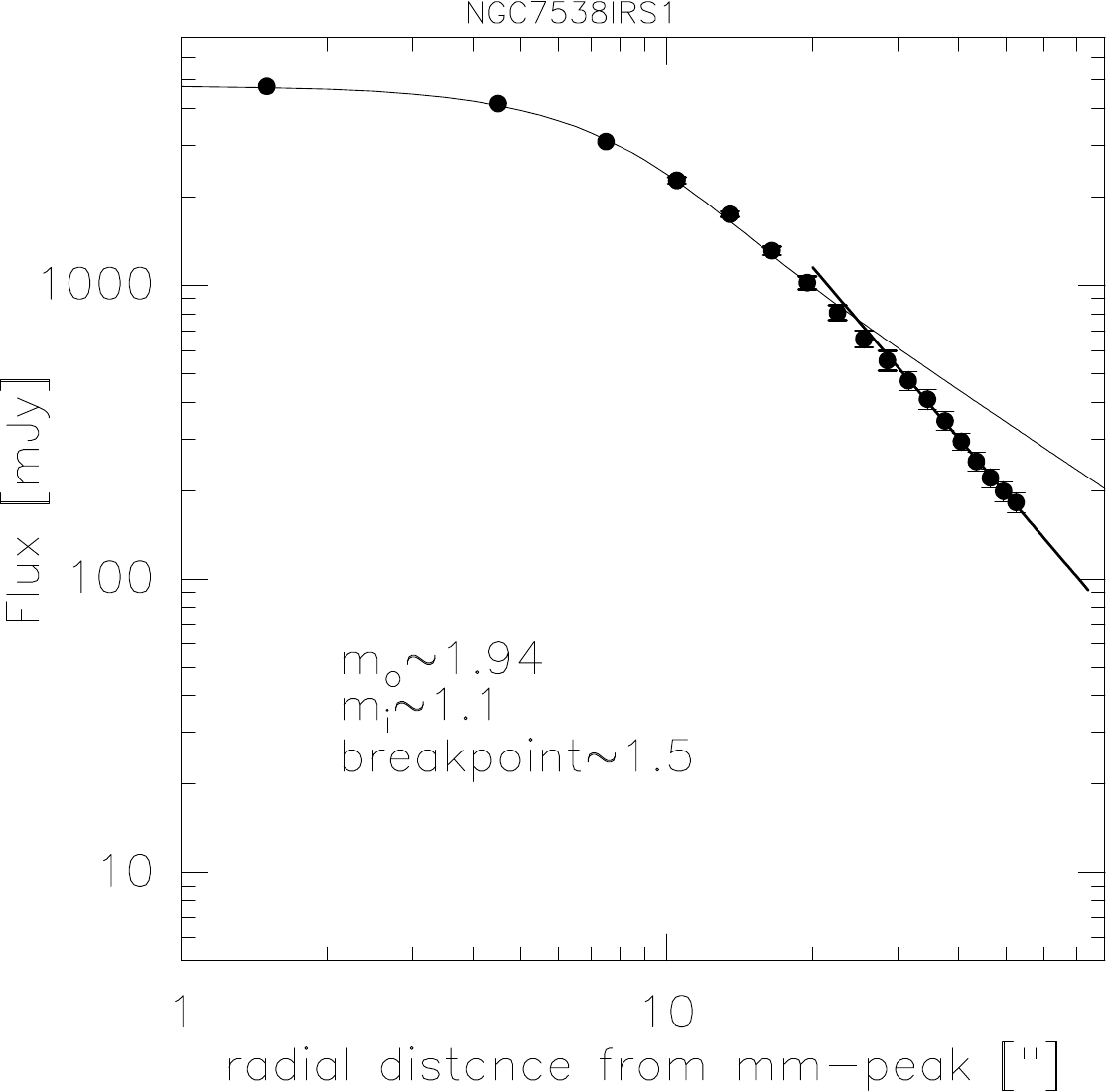}\\
\includegraphics[width=0.33\textwidth]{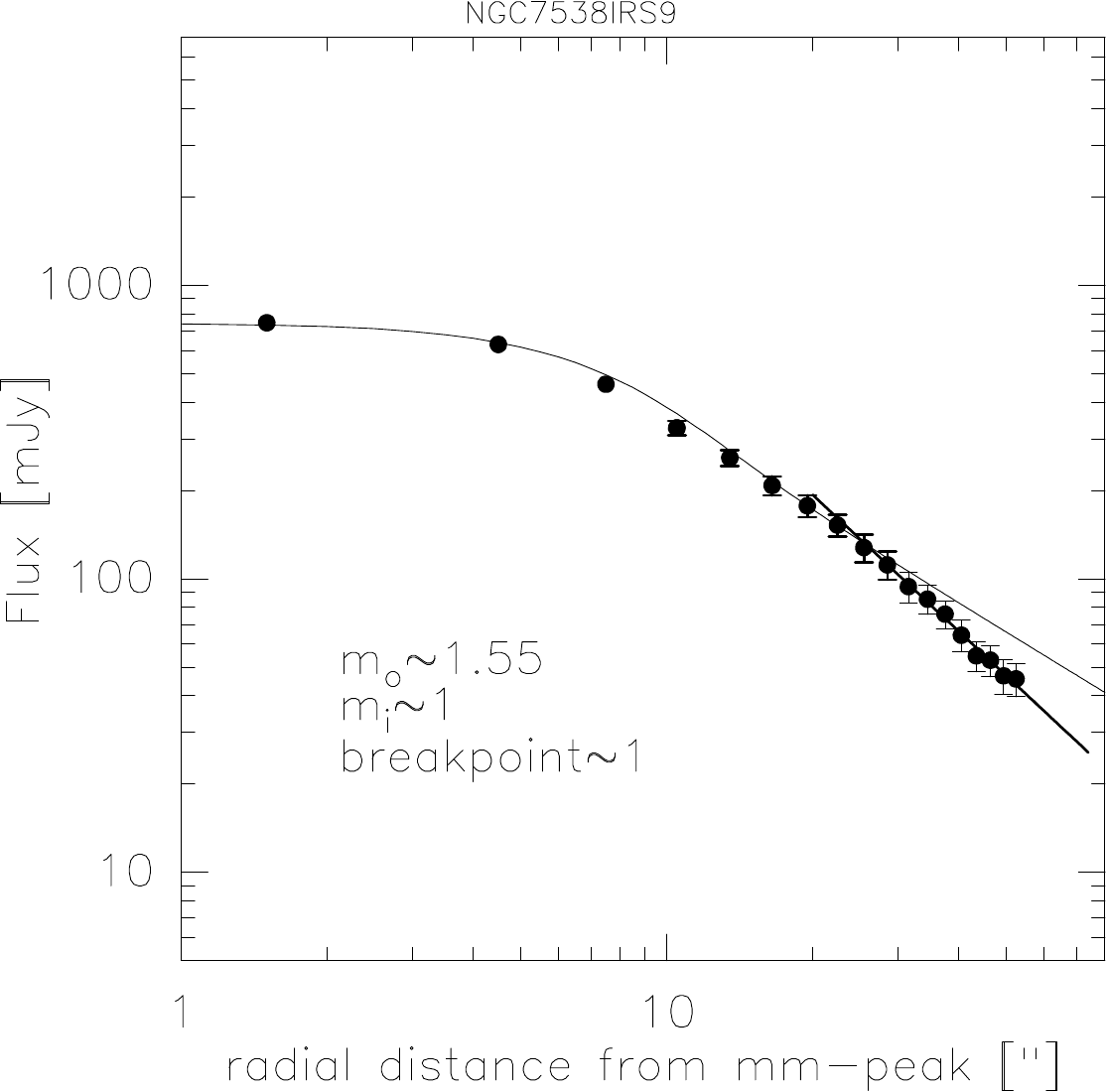}
\includegraphics[width=0.33\textwidth]{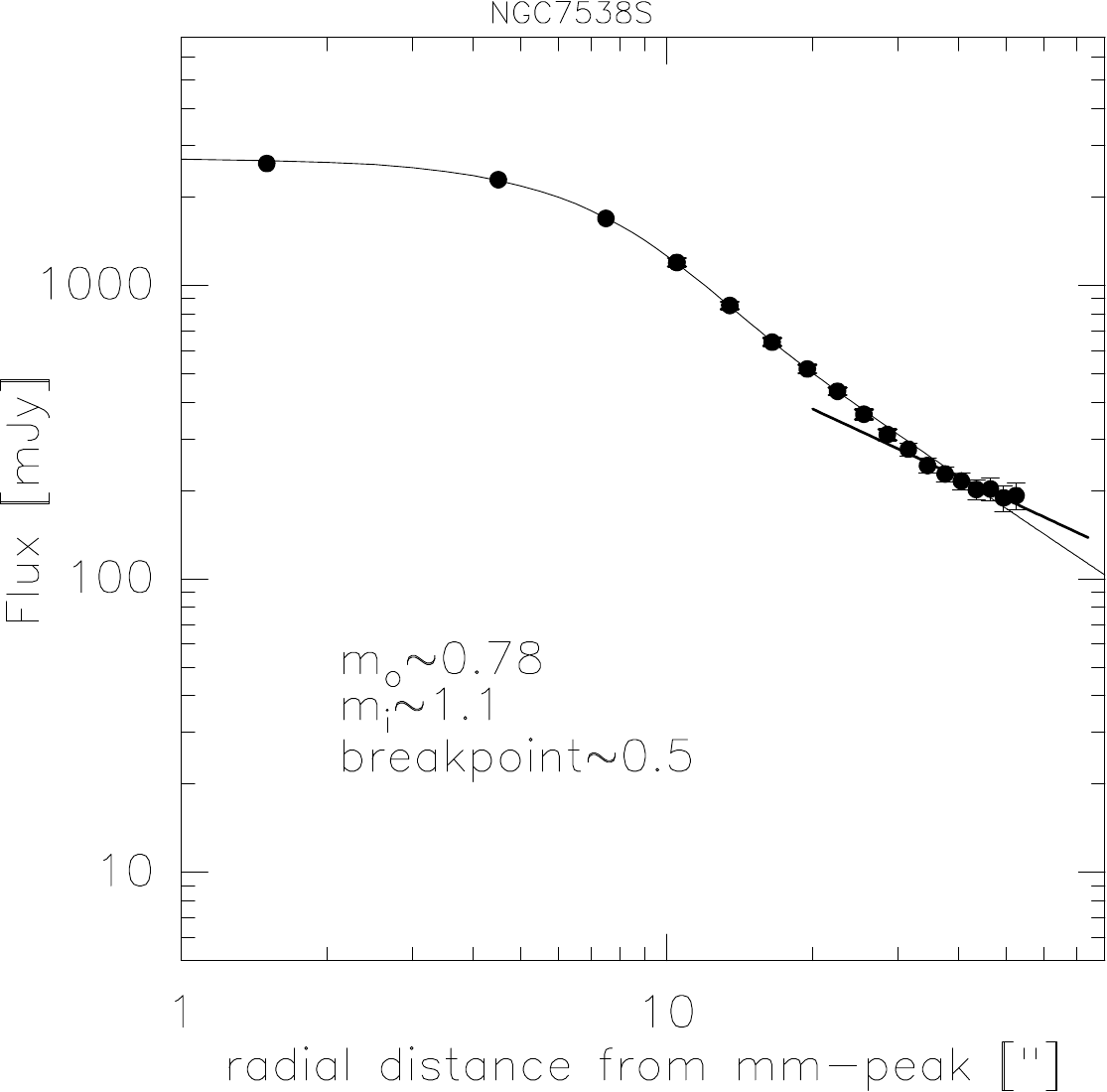}
\caption{Radial intensity profiles derived for the main 1.2\,mm dust continuum sources in the CORE sample. The power-law slopes of the inner and outer fits ($m_i$ and $m_o$) as well as the inner breakpoint  in arcsecond are labeled in each panel. See main text for fitting details. }
\label{profiles_3} 
\end{figure*} 

\section{SMA data}

\begin{figure*}[htb]
\includegraphics[width=0.33\textwidth]{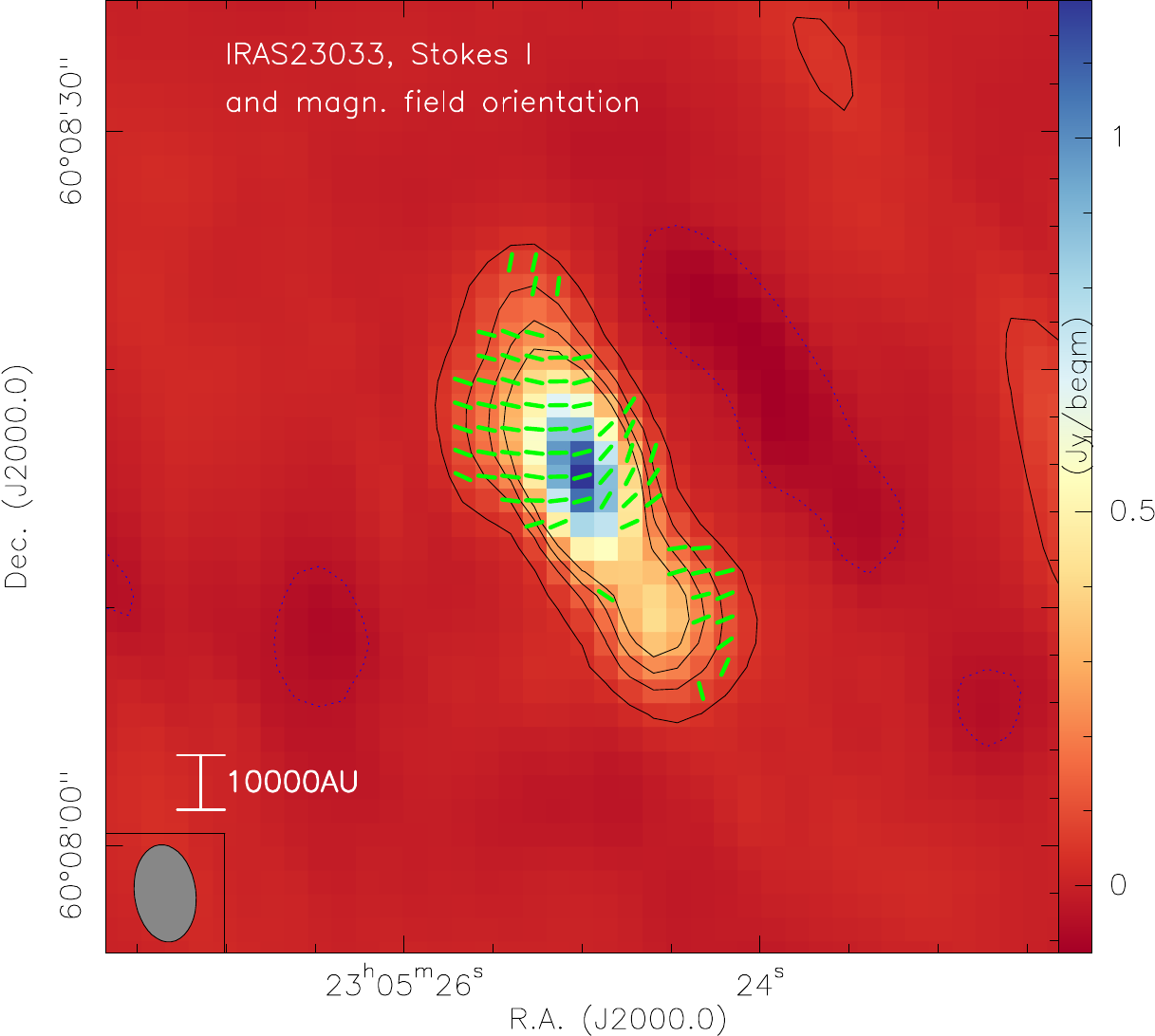}
\includegraphics[width=0.33\textwidth]{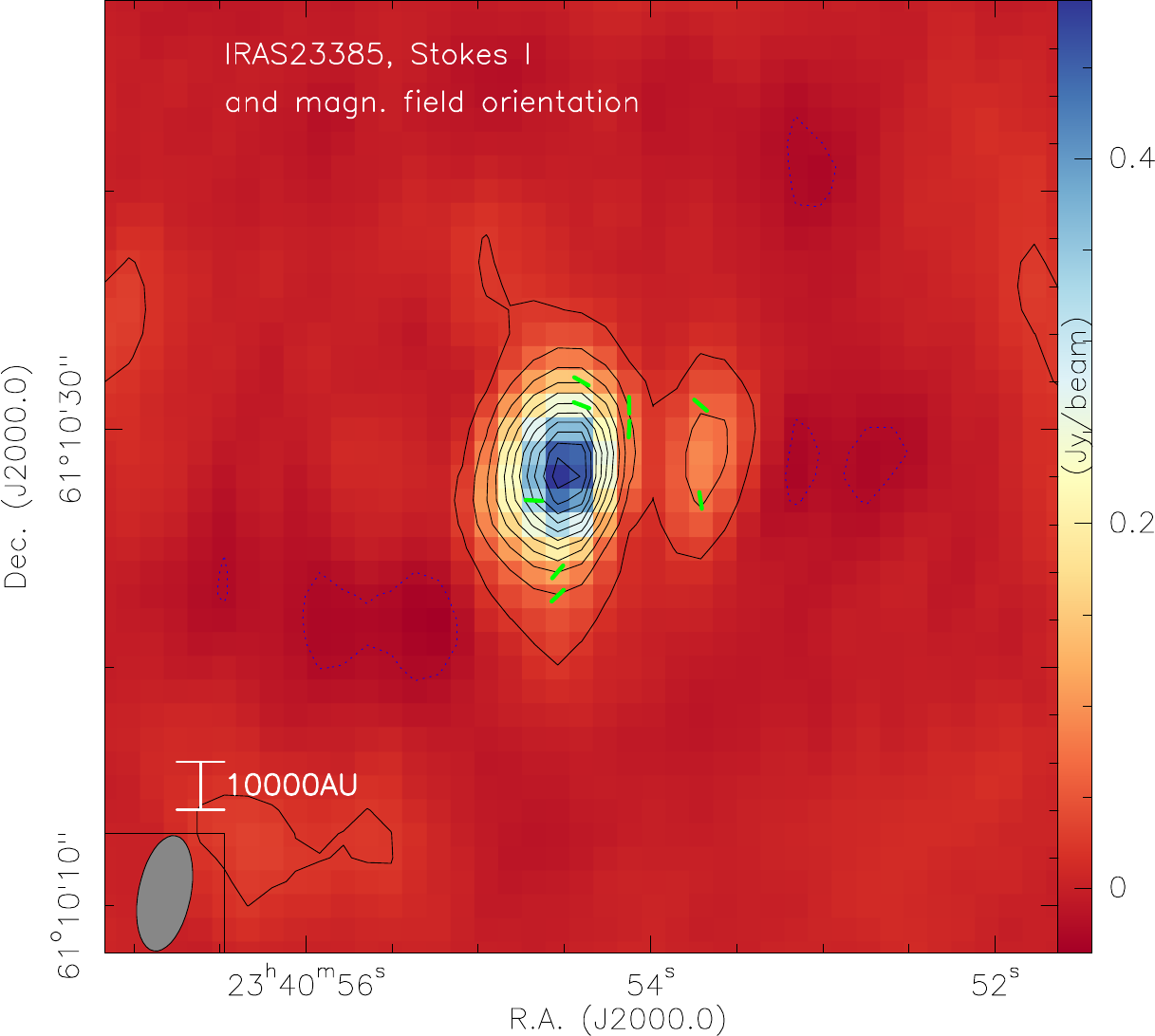}
\includegraphics[width=0.33\textwidth]{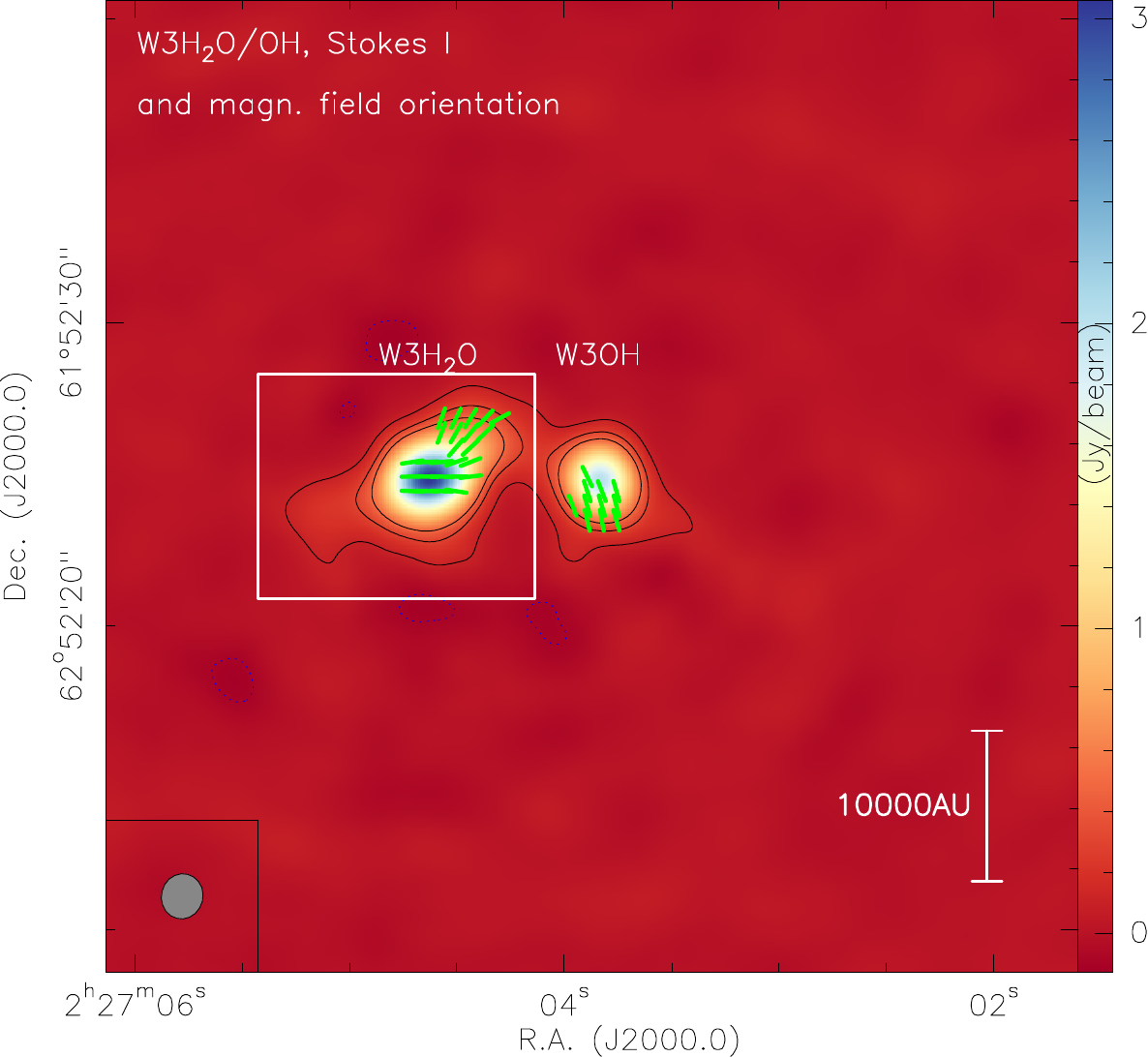}\\
\includegraphics[width=0.33\textwidth]{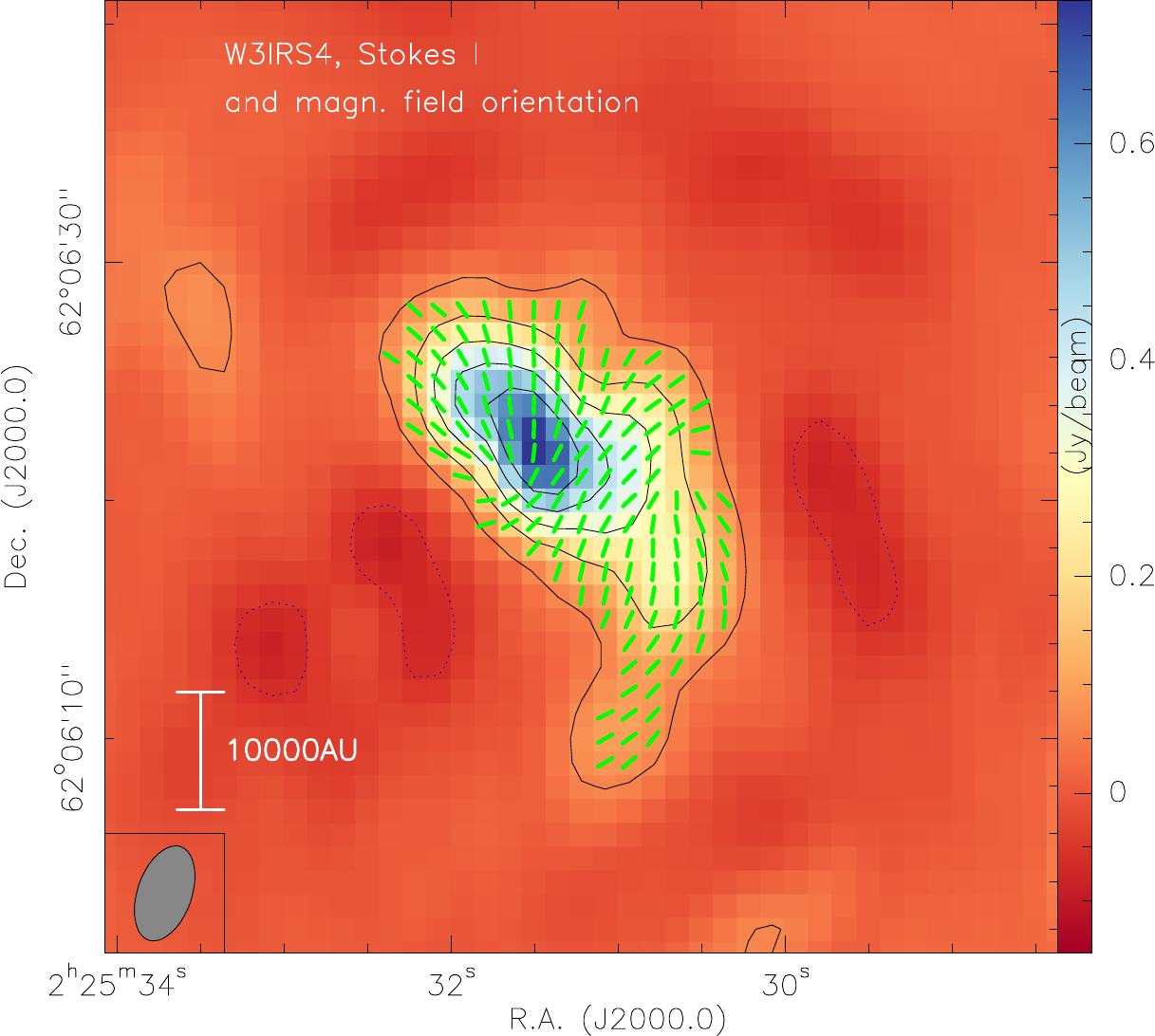}
\includegraphics[width=0.33\textwidth]{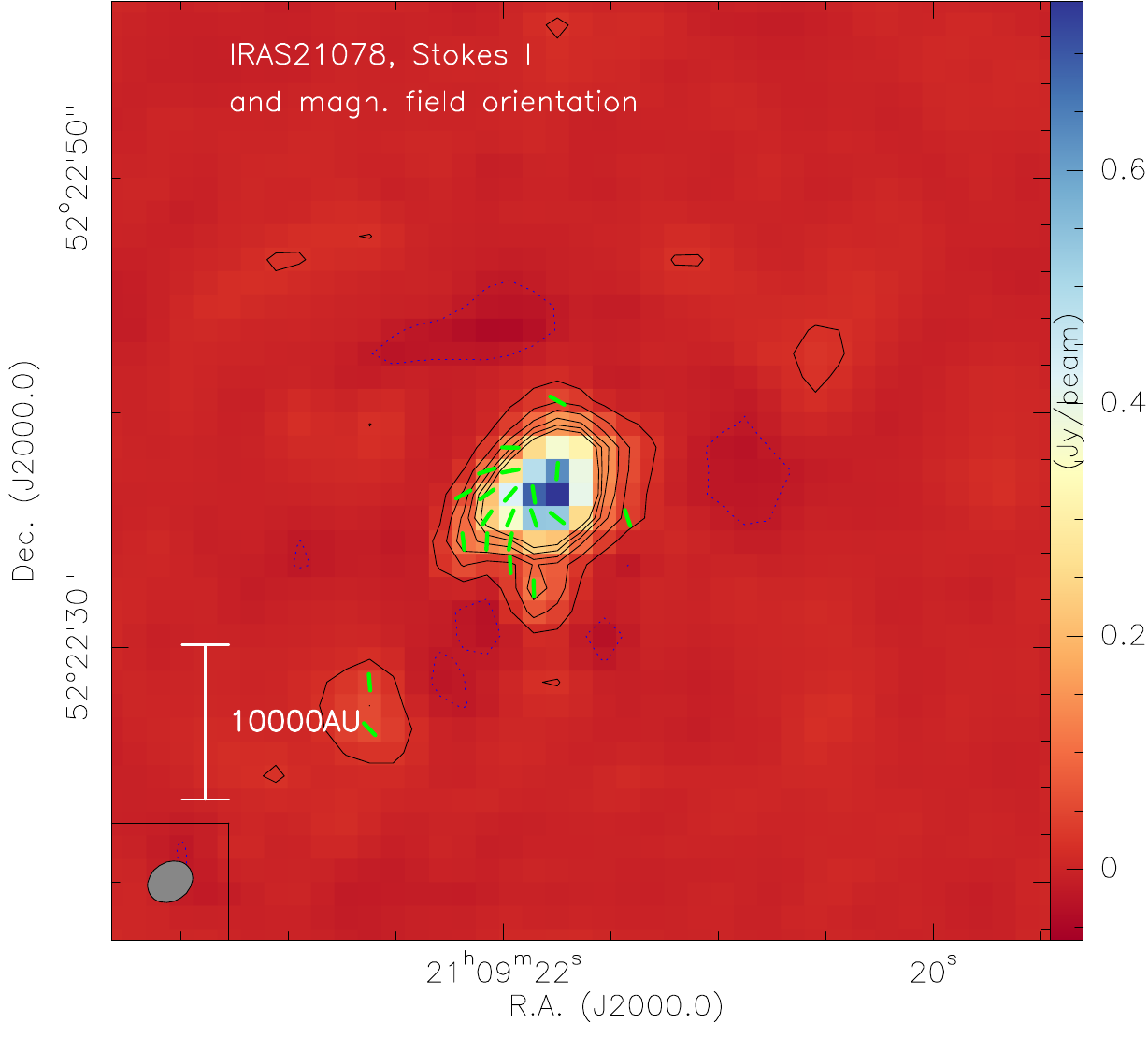}
\includegraphics[width=0.33\textwidth]{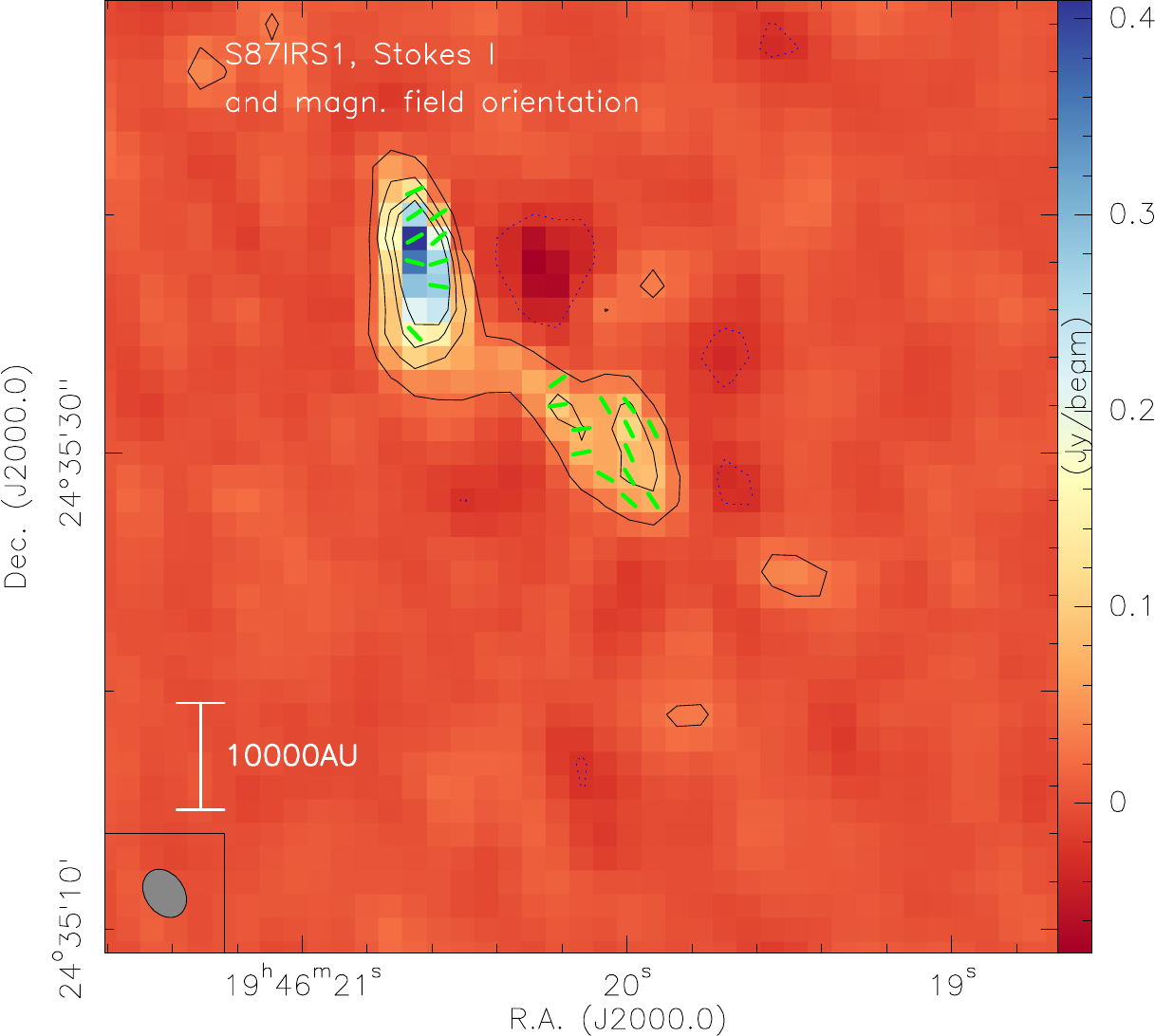}\\
\includegraphics[width=0.33\textwidth]{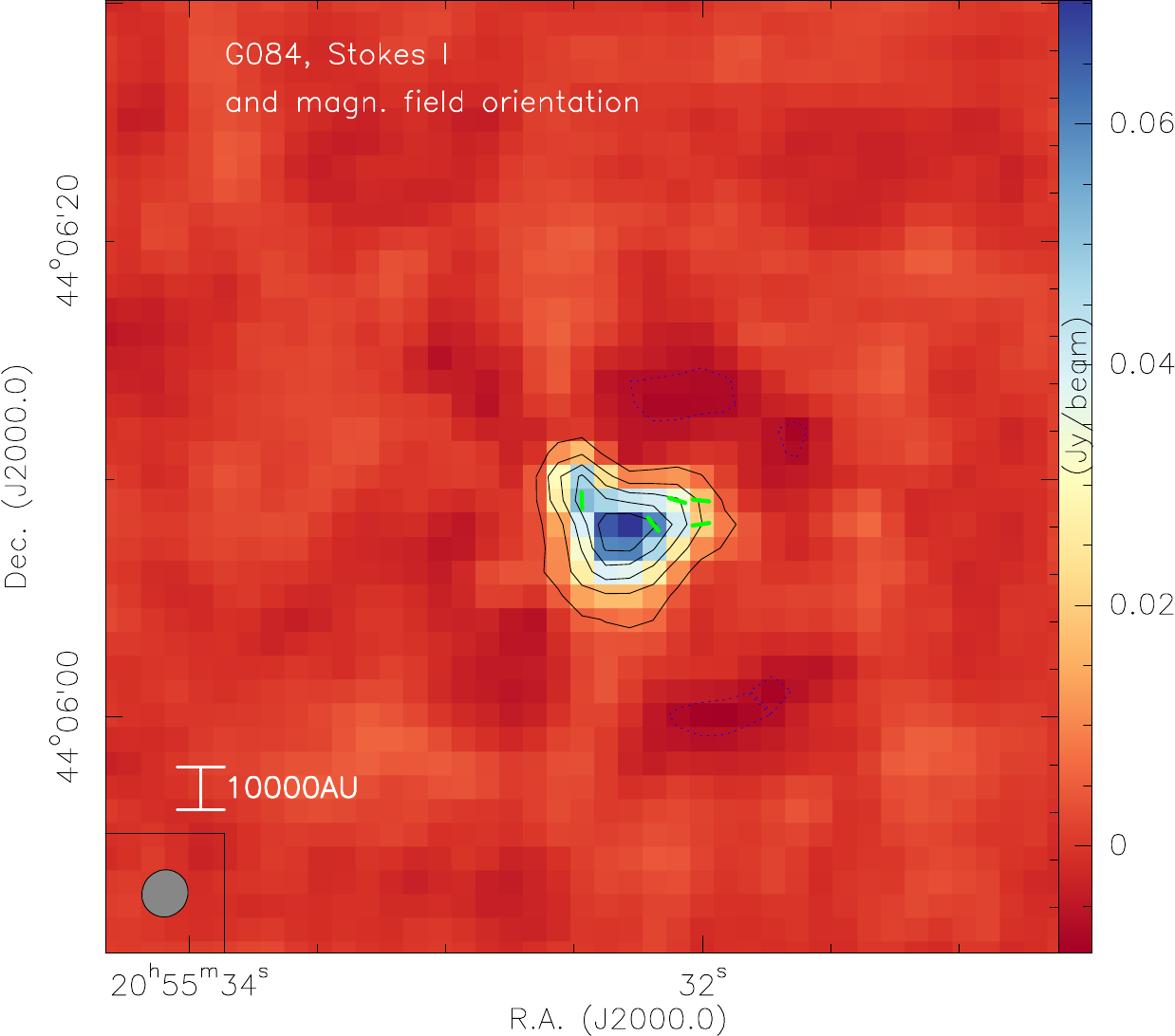}
\includegraphics[width=0.33\textwidth]{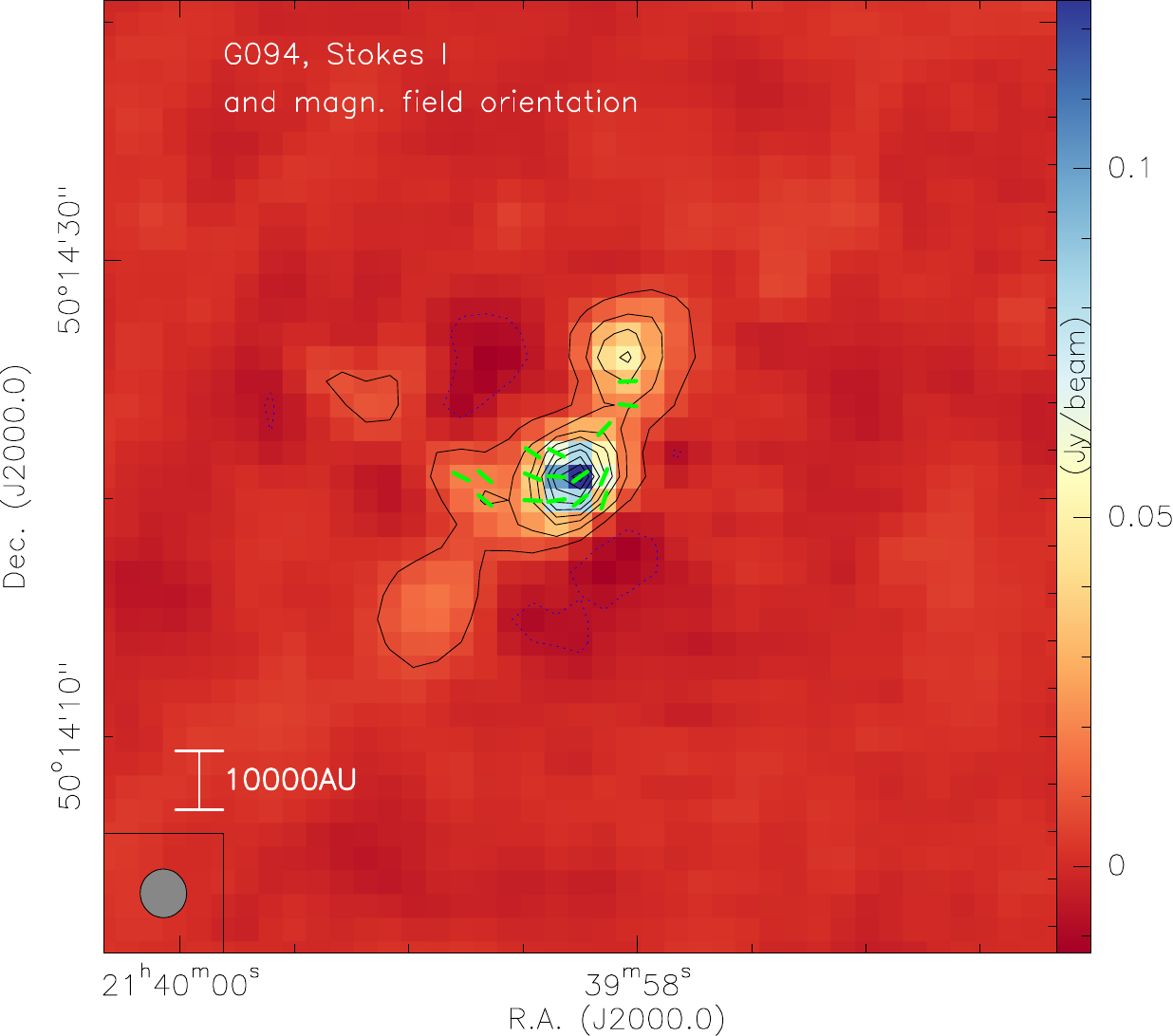}
\includegraphics[width=0.33\textwidth]{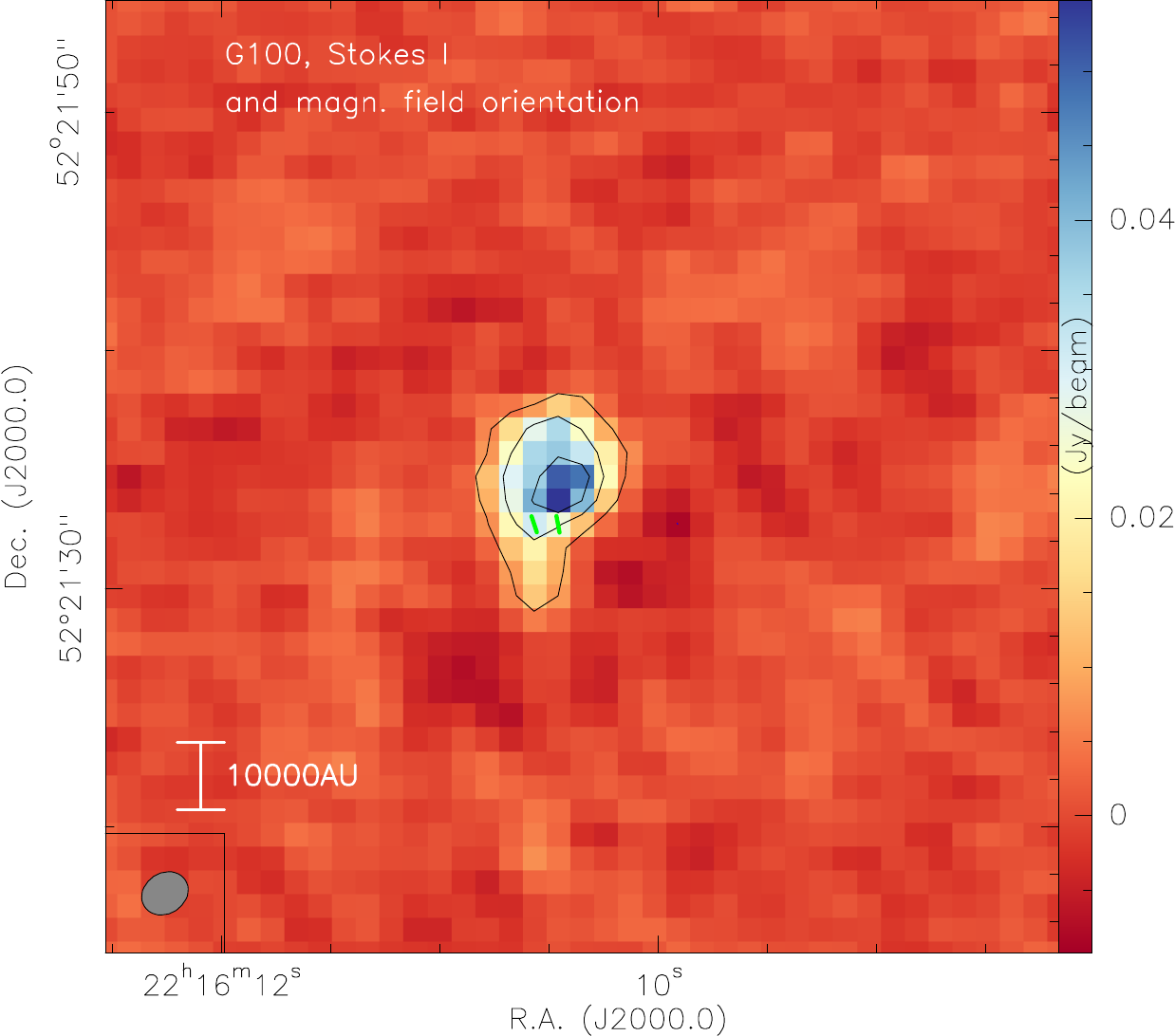}\\
\includegraphics[width=0.33\textwidth]{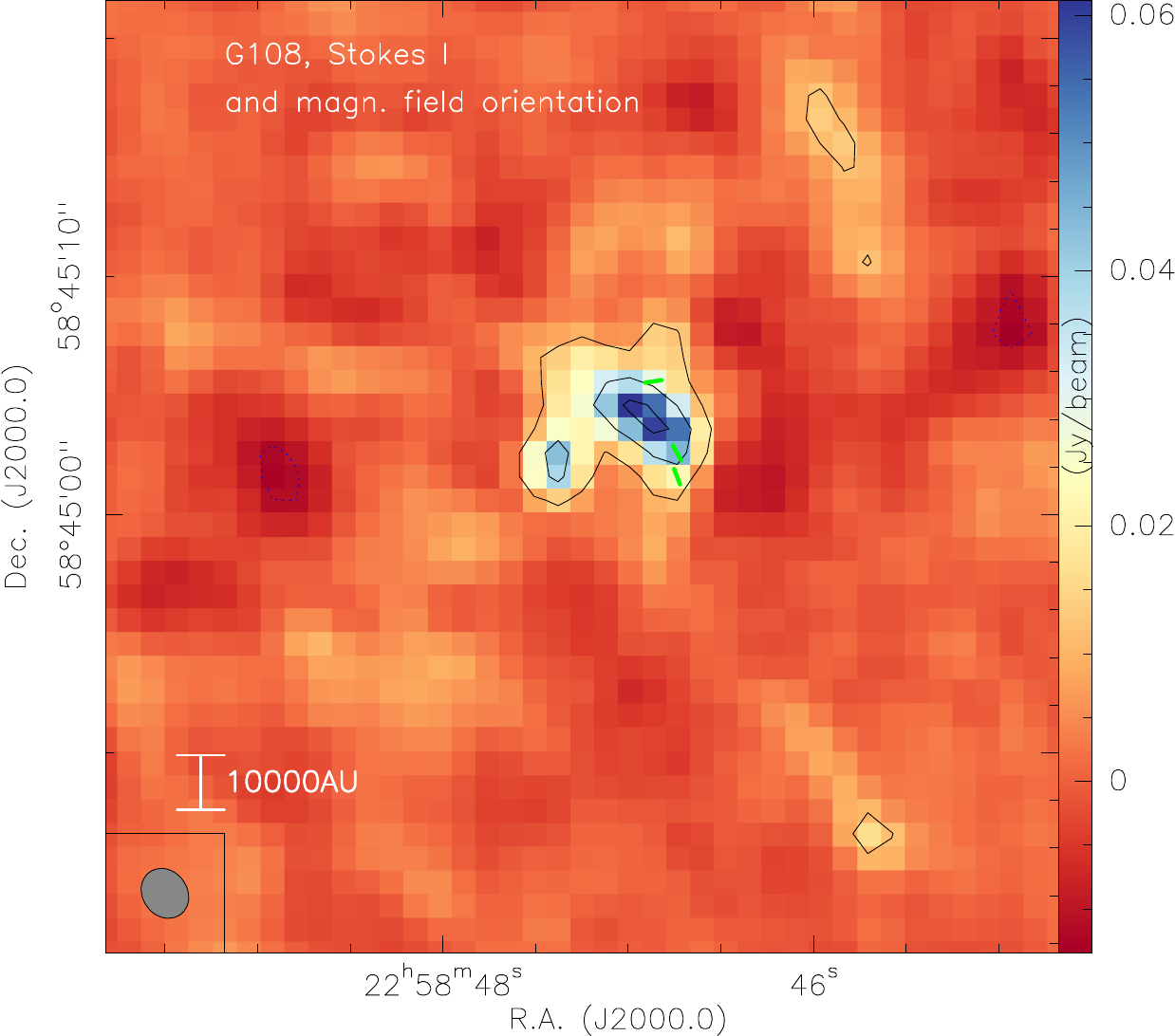}
\includegraphics[width=0.33\textwidth]{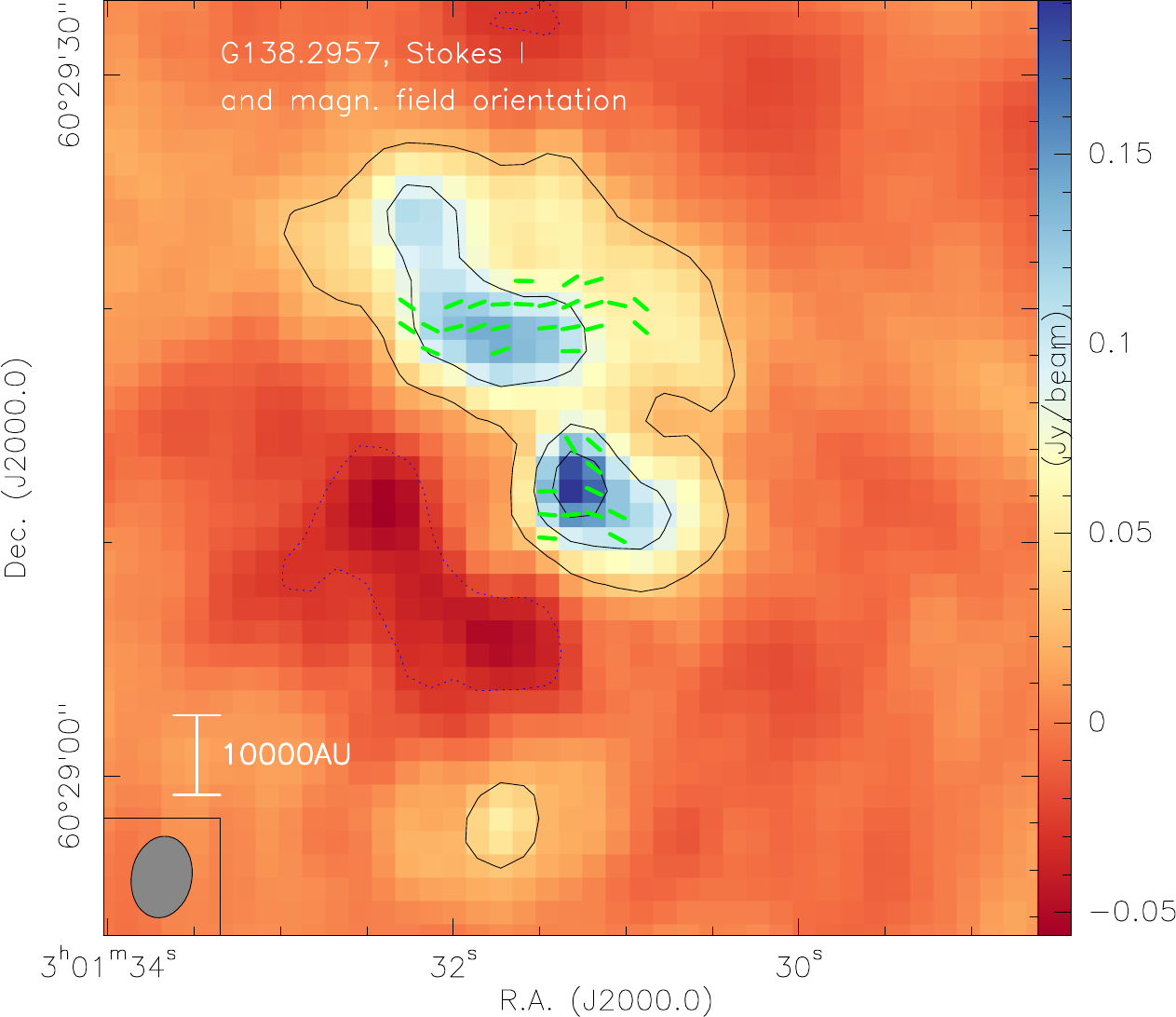}
\includegraphics[width=0.33\textwidth]{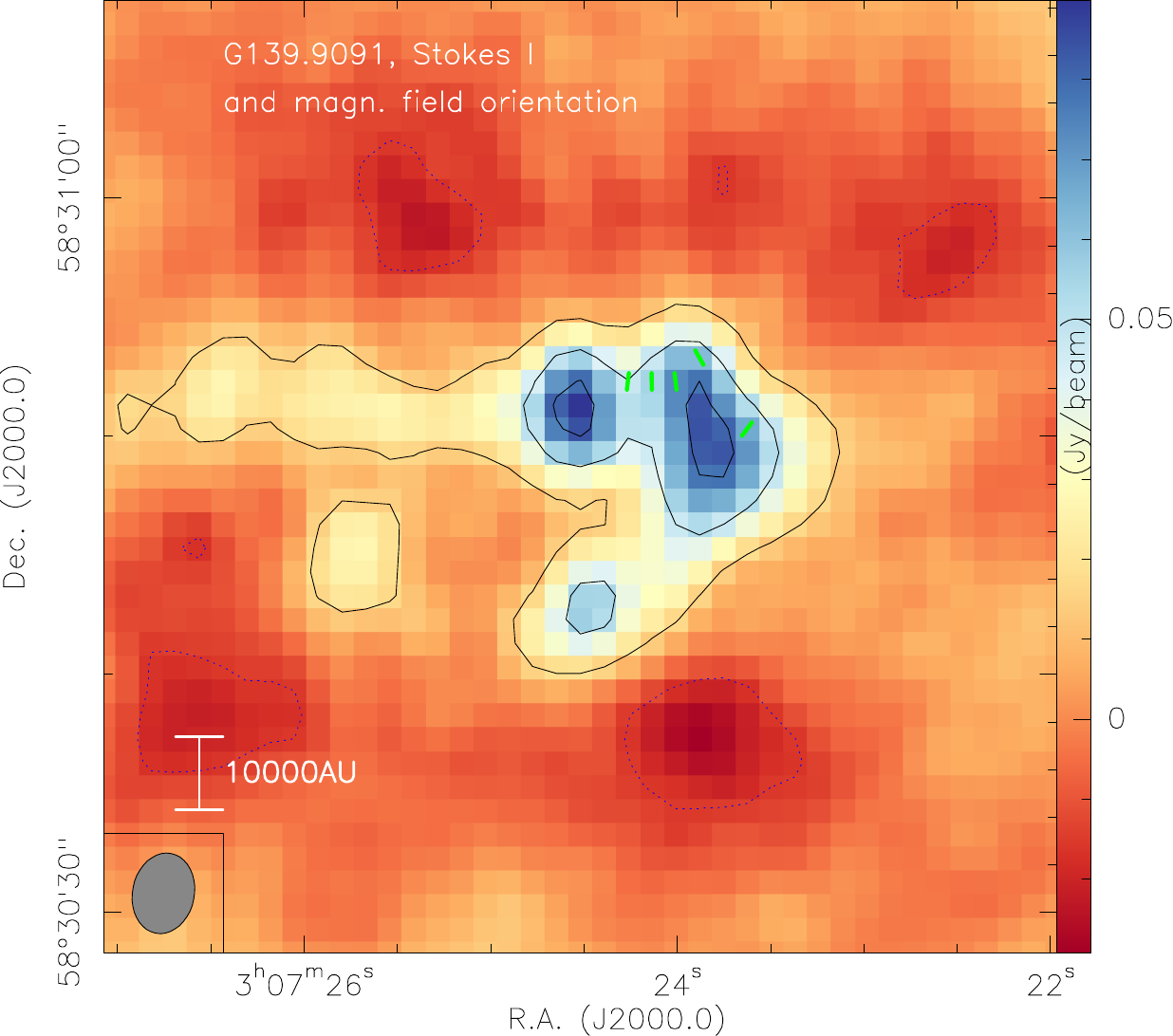}\\
\caption{Remaining SMA polarization maps. The color-scale presents the Stokes $I$ total intensity data. The contours show the same data starting at the $4\sigma$ level and continue in $8\sigma$ steps. The green constant-length line segments present the magnetic field orientation (polarization angles rotated by 90\,deg) derived from the linearly polarized continuum data above the $2\sigma$ level (independent of the polarization fraction). The synthesized beam and a linear scale bar are shown at the bottom of each panel.}
\label{sma_pol2} 
\end{figure*} 

\begin{figure*}[htb]
\includegraphics[width=0.33\textwidth]{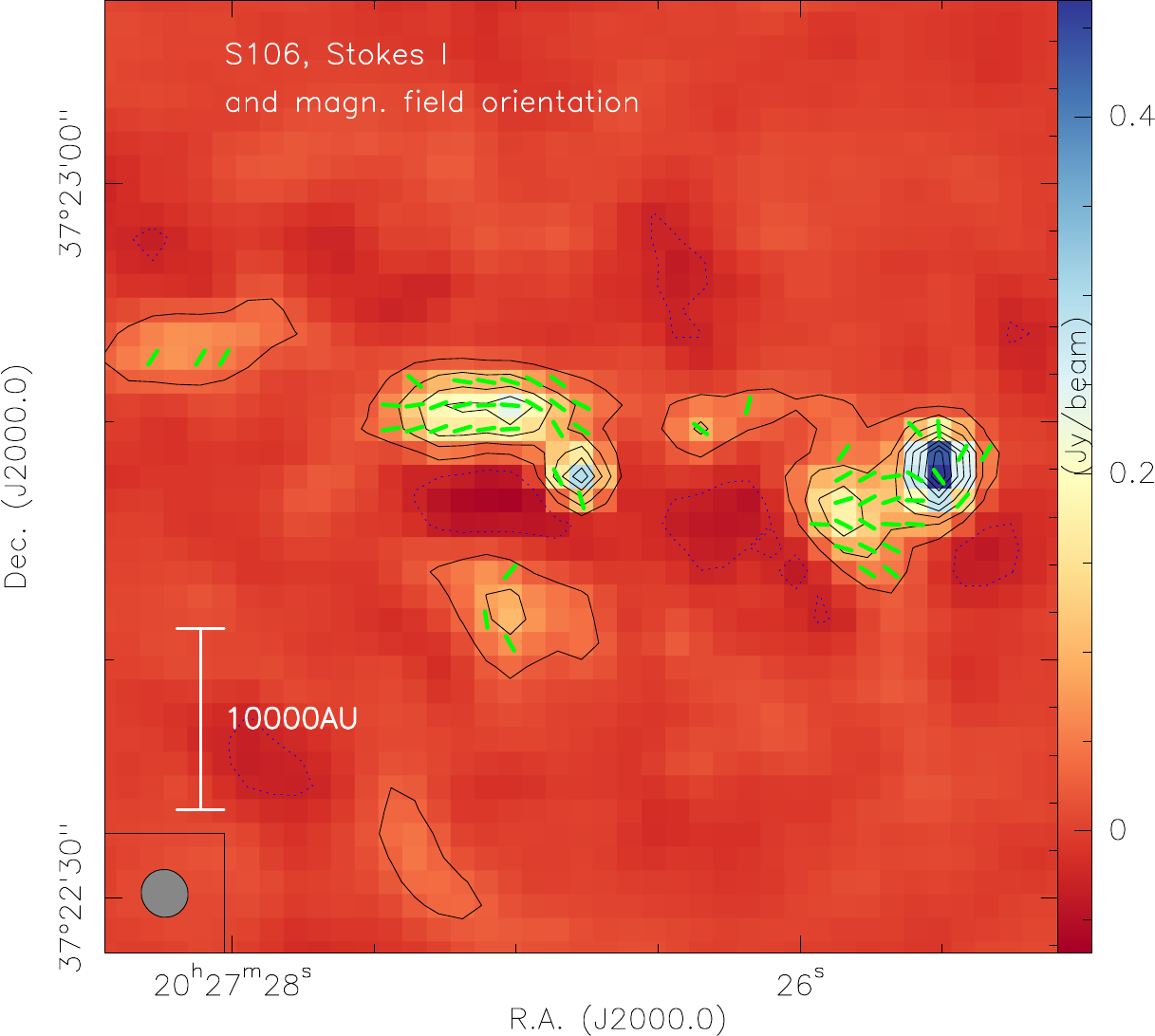}
\includegraphics[width=0.33\textwidth]{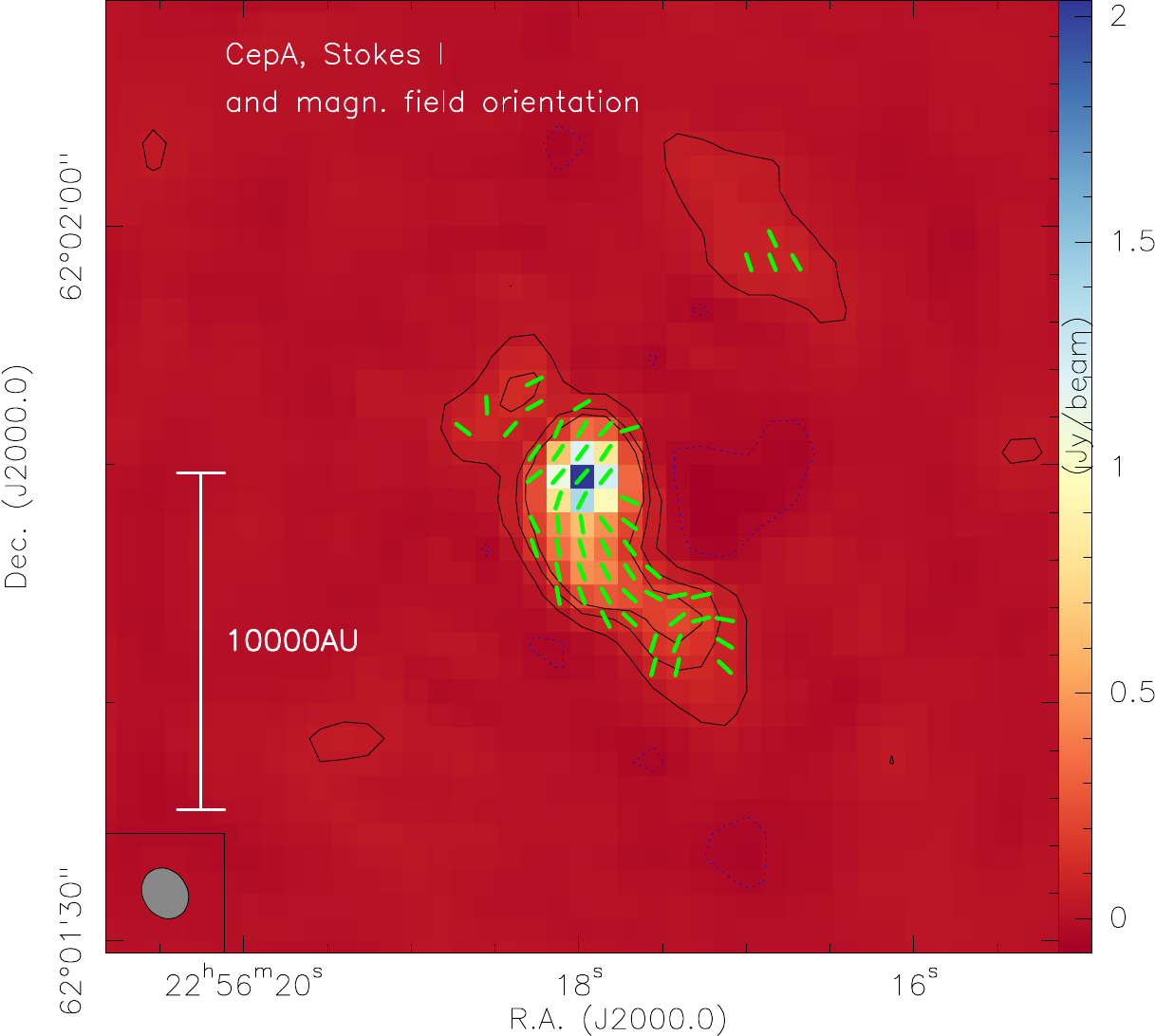}
\includegraphics[width=0.33\textwidth]{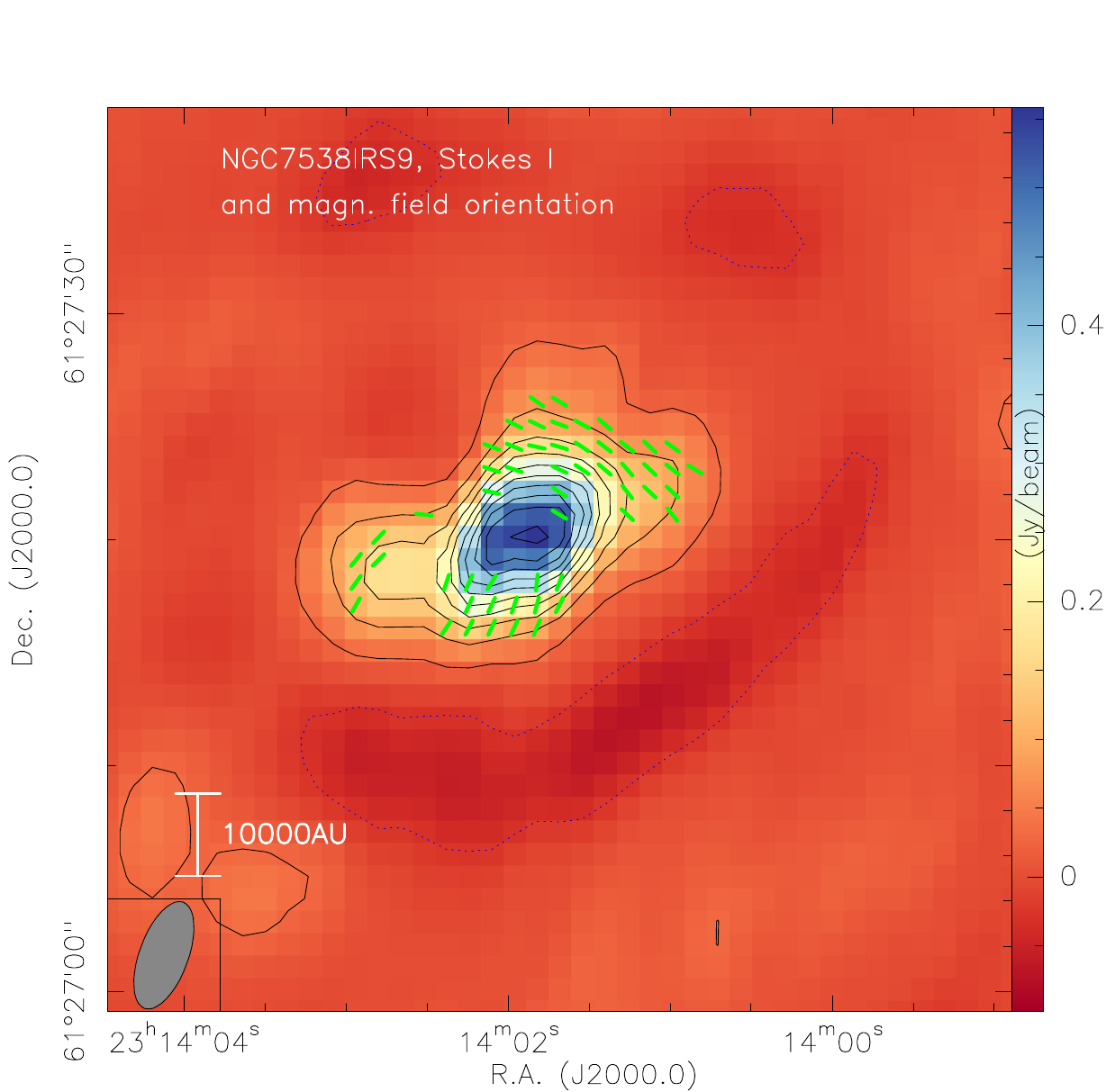}\\
\includegraphics[width=0.33\textwidth]{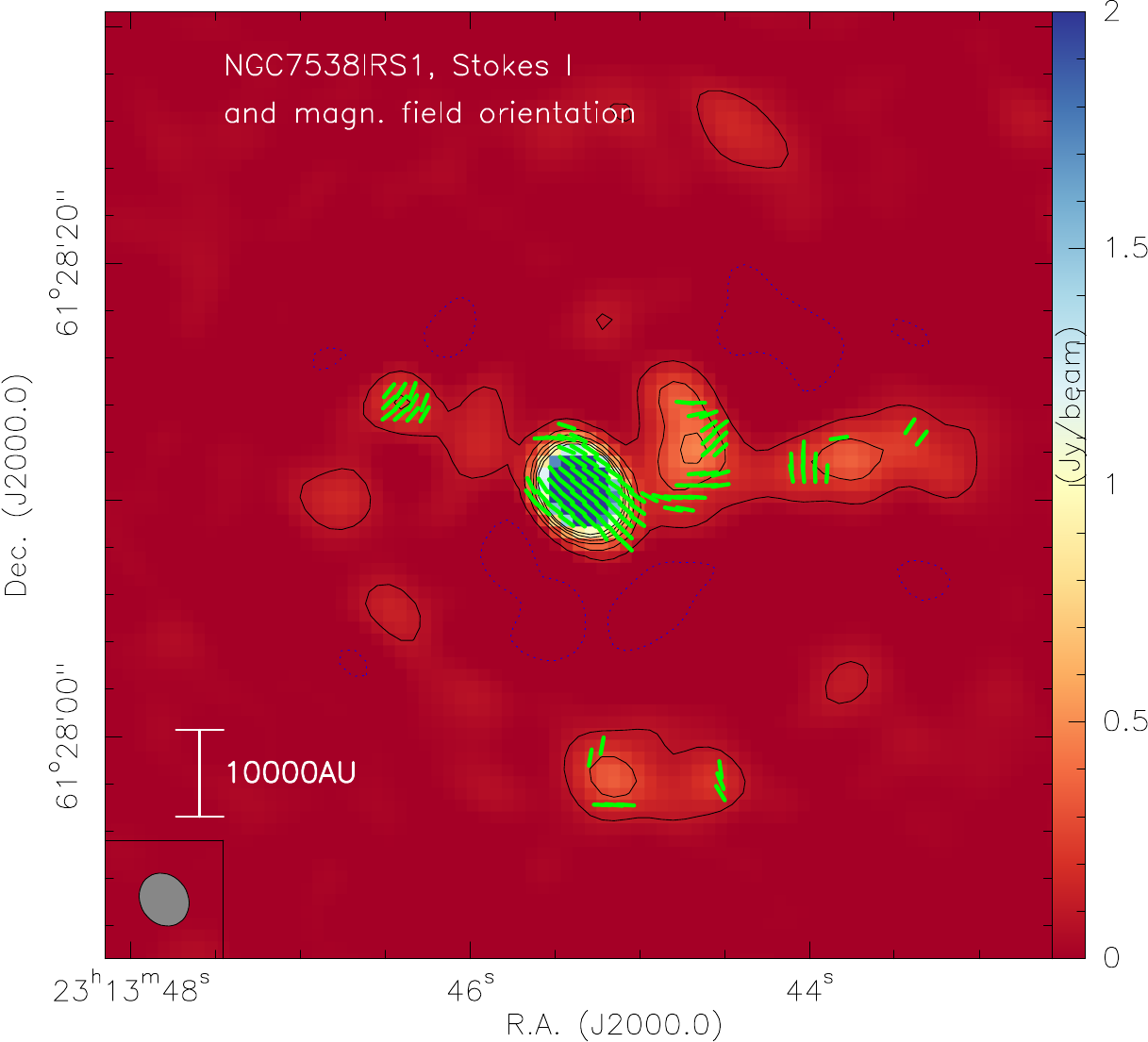}
\includegraphics[width=0.33\textwidth]{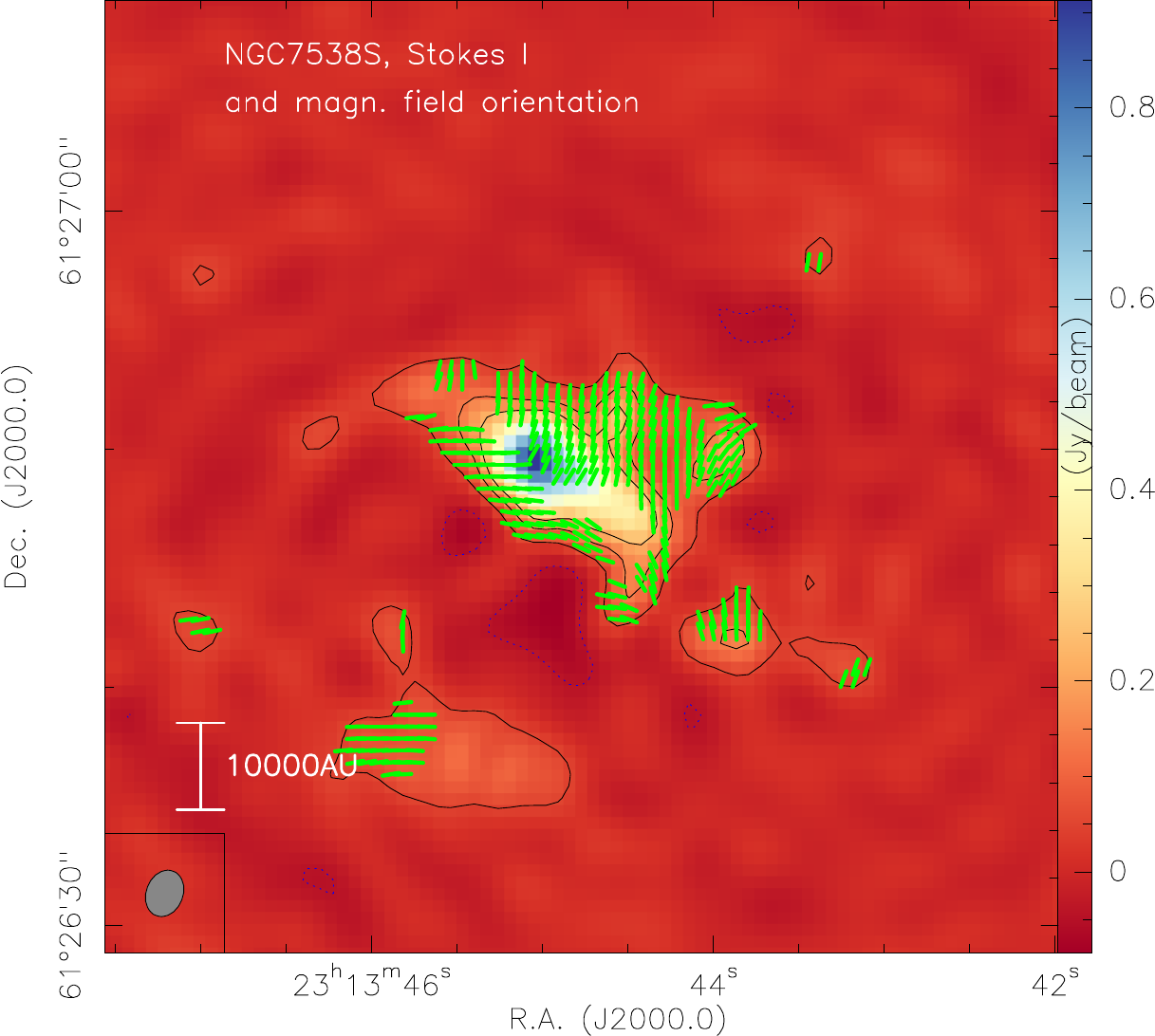}
\caption{Remaining SMA polarization maps continued. The color-scale presents the Stokes $I$ total intensity data. The contours show the same data starting at the $4\sigma$ level and continue in $8\sigma$ steps. The green constant-length line segments present the magnetic field orientation (polarization angles rotated by 90\,deg) derived from the linearly polarized continuum data above the $2\sigma$ level (independent of the polarization fraction). The synthesized beam and a linear scale bar are shown at the bottom of each panel.}
\label{sma_pol3} 
\end{figure*} 

\begin{figure*}[htb]
 \includegraphics[width=0.99\textwidth]{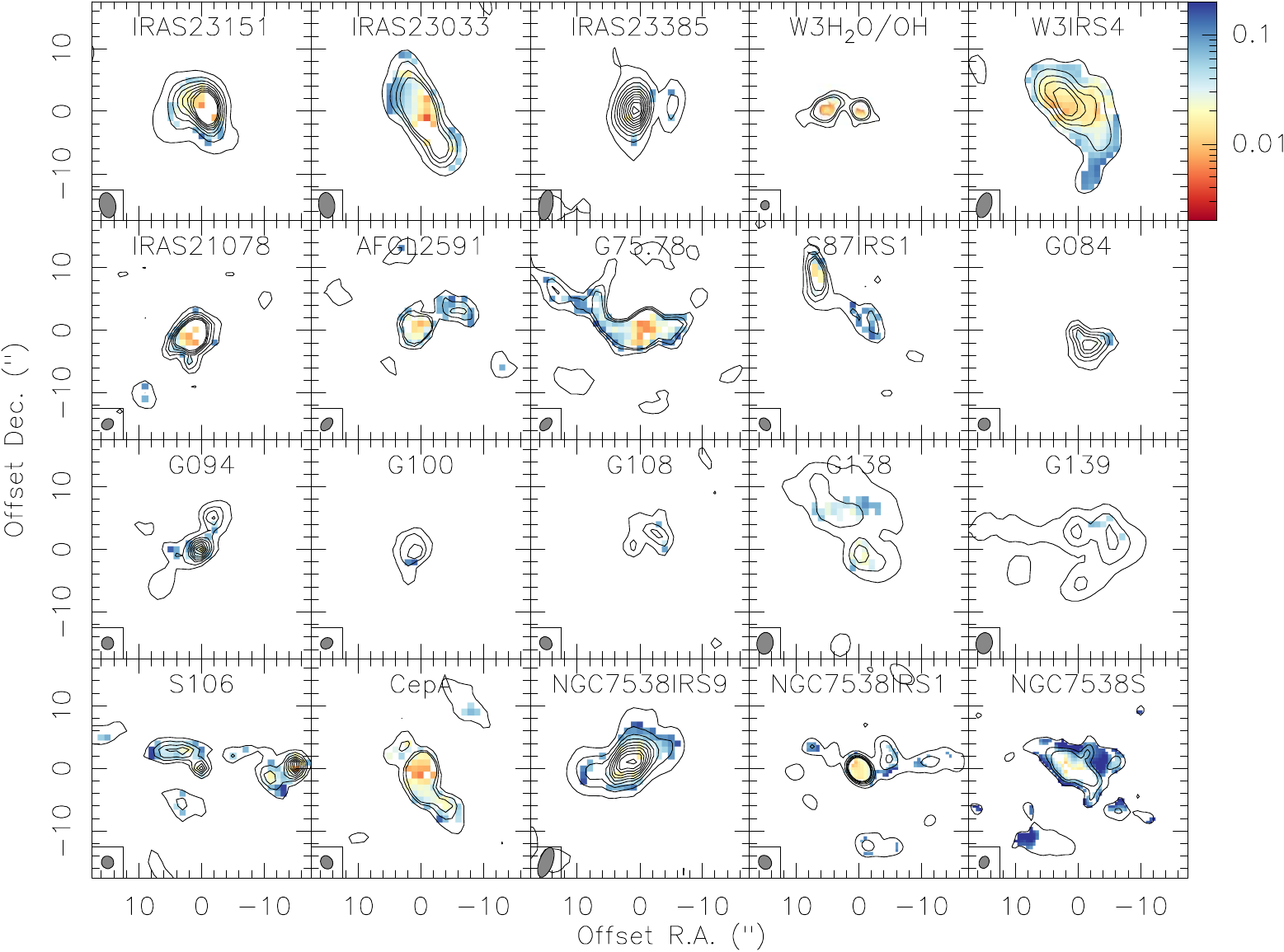}
\caption{Polarization fraction maps for all regions are shown in color-scale. The color-bar is the same for all and shown next to the top-right panel. The contours show the Stokes $I$ total intensity data starting at the $4\sigma$ level and continue in $8\sigma$ steps. }
\label{poli} 
\end{figure*}

\end{appendix}

\end{document}